\documentclass[a4paper,12pt,colorlinks=true,urlcolor=blue,anchorcolor=blue,citecolor=blue,filecolor=blue,
               linkcolor=blue,menucolor=blue,linktocpage=true,pdfproducer=medialab,pagebackref=true]{article}
\pdfoutput=1
\usepackage{jheppub_X}
\usepackage[T1]{fontenc}
\usepackage{multirow}
\usepackage{amsmath,mathrsfs,booktabs}
\usepackage{slashed}
\usepackage{physics}
\usepackage{setspace}
\usepackage[dvipsnames,table]{xcolor}
\usepackage[normalem]{ulem}
\usepackage{cleveref}
\usepackage{anyfontsize}
\usepackage{float} 
\usepackage{subcaption}
\usepackage{ifpdf}
\usepackage{slashed}
\usepackage{bbold}
\usepackage{graphicx}
\usepackage{cancel}
\usepackage{dsfont}
\usepackage[most]{tcolorbox}
\definecolor{light-gray}{gray}{0.9}
\usepackage{mathtools}
\usepackage[normalem]{ulem}
\usepackage{orcidlink}

\hypersetup{
  colorlinks   = true, 
  urlcolor     = mygreen, 
  linkcolor    = blue, 
  citecolor   = blue 
}

\crefname{table}{Table}{Tables}
\crefname{equation}{Eq.}{Eqs.}
\crefname{appendix}{App.}{Apps.}
\crefname{section}{Sec.}{Secs.}
\crefname{figure}{Fig.}{Figs.}

\def\eg{\textit{e.g.}}
\def\ie{\textit{i.e.}}
\def\etc{\textit{etc.}}
\def\cf{\textit{cf.}}

\newcommand{\Hcal}{\mathcal{H}}
\newcommand{\Den}{D}
\newcommand{\Num}{N}
\definecolor{brightpink}{rgb}{1.0, 0.0, 0.5}
\newcommand{\Sat}[1]{{\color{brightpink}{#1}}}
\newcommand{\Ge}{\text{trivial}}
\newcommand{\BB}{\Phi}
\newcommand{\HS}{\Hcal}
\newcommand{\RBB}{{R_\BB}}
\newcommand{\I}{\text{Inv}}
\newcommand{\Ibf}{\textbf{\I}}
\newcommand{\PI}{{\text{P}\I}}
\newcommand{\ring}{\mathbb{r}}
\newcommand{\rBB}{{\ring}}
\newcommand{\rI}{{\ring_\I}}
\newcommand{\rPI}{{\ring_\PI}}
\newcommand{\Mod}{\mathcal{M}}

\newcommand{\eR}{r}
\newcommand{\eM}{v}
\newcommand{\tu}{u}
\newcommand{\td}{d}
\newcommand{\trless}[1]{{\overline{#1}}}
\newcommand{\hc}{\text{h.c.}}
\newcommand{\mm}{M}

\renewcommand{\L}{\mathcal{L}}
\renewcommand{\O}{\mathcal{O}}
\renewcommand{\H}{\mathcal{H}}

\definecolor{mygreen}{rgb}{0.0, 0.5, 0.0}
\definecolor{azuremist}{rgb}{0.94, 1.0, 1.0}
\definecolor{mylightgray}{rgb}{0.93, 0.93, 0.93}

\newcommand{\be}{\begin{equation}}
\newcommand{\ee}{\end{equation}}
\newcommand{\MeV}{\ \text{MeV}}

\newcommand{\TeV}{\ \text{TeV}}
\newcommand{\GeV}{\ \text{GeV}}

\newcommand{\cB}{\mathcal{B}}
\newcommand{\cH}{\mathcal{H}}
\newcommand{\cO}{\mathcal{O}}
\newcommand{\cR}{\mathcal{R}}
\newcommand{\ov}[1]{\overline{#1}}
\newcommand{\Heff}{{\cal H}_\text{ eff}}
\def\diag{{\tt diag}}

\preprint{IFT-UAM-CSIC-23-67}

\title{\Large 
Hilbert series for covariants and their applications to Minimal Flavor Violation}

\author[a]{Benjamín Grinstein\orcidlink{0000-0003-2447-4756},}
\author[a]{Xiaochuan Lu\orcidlink{0000-0001-8821-2574},}
\author[b]{Luca Merlo,\orcidlink{0000-0002-5876-4105}}
\author[a]{and Pablo Quílez\orcidlink{0000-0002-4327-2706}}

\affiliation[a]{Department of Physics, University of California, San Diego, La Jolla, CA 92093, USA}
\affiliation[b]{Departamento de F\'isica Te\'orica and Instituto de F\'isica Te\'orica UAM/CSIC,\\
Universidad Aut\'onoma de Madrid, Cantoblanco, 28049, Madrid, Spain}

\emailAdd{bgrinstein@ucsd.edu}
\emailAdd{xil224@ucsd.edu}
\emailAdd{luca.merlo@uam.es}
\emailAdd{pquilezlasanta@ucsd.edu}

\abstract{
We elaborate how to apply the Hilbert series method to enumerating group covariants, which transform under any given representation, including but going beyond group invariants.
Mathematically, group covariants form a module over the ring of the invariants. The number of independent covariants is given by the rank of the module, which can be computed by taking a ratio of two Hilbert series.
In many cases, the rank equals the dimension of the group covariant representation. When this happens, we say that there is a \textit{rank saturation}.
We apply this technology to revisit the hypothesis of Minimal Flavor Violation in constructing Effective Field Theories beyond the Standard Model.
We find that rank saturation is guaranteed in this case, leading to the important consequence that 
the MFV symmetry principle does not impose any restriction on the EFT, \ie\ MFV SMEFT = SMEFT, in the absence of additional assumptions.
}

\begin{document}
\maketitle
\flushbottom
\setcounter{page}{2}
\newpage
\begin{spacing}{1.1}
\parskip=0ex

\section{Introduction}
\label{sec:Introduction}

Flavor physics constitutes a unique window to look for new physics. The fact that many flavor observables are extremely suppressed within the Standard Model (SM) represents an excellent opportunity to explore what lies beyond it. Given that no new particle beyond the SM has been discovered to date, the best tool for studying potential deviations from it at our disposal is the Standard Model Effective Field Theory (SMEFT)~\cite{Buchmuller:1985jz, Grzadkowski:2010es}. Equipped with the SM matter content and symmetries, SMEFT is able to parametrize in a model-independent manner the low-energy consequences of a large class of Ultraviolet (UV) theories with a characteristic energy scale much larger than the electroweak scale. 

However, it can often be impractical to work with the most general SMEFT. For example, it contains 3045 free parameters at dimension 6 (2499 if baryon number conservation is imposed). Therefore, motivated assumptions are often imposed to simplify the analysis. They often concern the flavor sector, where the large part of the SMEFT complexity lies. Moreover, given the strong constraints that flavor observables place on generic BSM theories, introducing flavor assumptions to SMEFT also helps selecting theories that would not suffer from those severe bounds.

One of the most popular flavor hypothesis in constructing EFTs beyond the SM is the Minimal Flavor Violation (MFV)~\cite{DAmbrosio:2002ex, Cirigliano:2005ck}. In this hypothesis, the Yukawa couplings are assumed to be the only source of flavor violation. Operationally, one assigns spurious transformation properties to them such that the SM Lagrangian is formally invariant under the flavor group. Higher dimensional operators beyond the SM are then required to be built out of these Yukawa spurions together with the SM fields such that the flavor group is formally preserved. Compared to a general Beyond the SM (BSM) EFT construction, the MFV criterion requires the Wilson coefficients to be built out of the Yukawa spurions, and to transform as particular representations under the flavor groups. Therefore, it is a common impression that the MFV ansatz places additional constraints on BSM EFTs.

In this paper, we revisit this question carefully. We focus on applying the MFV criterion to the SM quark sector, and check if it would place additional constraints on SMEFT, \ie\ we ask the question
\begin{equation}
\text{MFV SMEFT} \;\;\subsetneq\;\; \text{SMEFT}\;\; ?
\label{eqn:Question}
\end{equation}
Specifically, each higher dimensional term in the SMEFT Lagrangian has the form $C(Y_u, Y_d)\, Q(\phi_\text{SM})$, where $Q(\phi_\text{SM})$ is a usual SMEFT operator, a monomial of SM fields $\phi_{\text{SM}}$ that is generically not an invariant of the quark flavor group, and $C(Y_u,Y_d)$ is a polynomial of the Yukawa coupling constants $Y_u$ and $Y_d$. The MFV criterion then requires the Wilson coefficient $C(Y_u, Y_d)$ to transform accordingly such that the product $C(Y_u, Y_d)\, Q(\phi_\text{SM})$ is a quark flavor invariant. In this language, the question raised in \cref{eqn:Question} is asking whether $C(Y_u, Y_d)$ would have fewer degrees of freedom than a general SMEFT Wilson coefficient for the operator $Q(\phi_\text{SM})$. In this paper, we study this type of questions systematically with the Hilbert series approach.

The Hilbert series is a well-established tool in invariant theory (see \eg\ \cite{Sturmfels:inv, bruns1998cohen, Popov1994}) for counting the number of degrees of freedom. It has been extensively applied in particle physics for enumerating group invariants, including the cases of supersymmetric gauge theories \cite{Benvenuti:2006qr, Feng:2007ur, Gray:2008yu}, flavor invariants \cite{Jenkins:2009dy, Hanany:2010vu}, SMEFT \cite{Lehman:2015via, Henning:2015daa, Lehman:2015coa, Henning:2015alf, Henning:2017fpj, Marinissen:2020jmb, Bonnefoy:2021tbt, Kondo:2022wcw, Bonnefoy:2023bzx}, SMEFT with gravity \cite{Ruhdorfer:2019qmk}, QCD Chiral Lagrangian \cite{Graf:2020yxt}, $O(N)$ nonlinear sigma model \cite{Bijnens:2022zqo}, nonlinearly realized gauge theories like Higgs EFT \cite{Graf:2022rco, Sun:2022aag}, non-relativistic EFTs like NRQED and NRQCD \cite{Kobach:2017xkw, Kobach:2018pie}, general supersymmetric EFTs \cite{Delgado:2022bho, Delgado:2023ivp}, EFTs for axion-like particles \cite{Grojean:2023tsd}, as well as the so-called primary observables at colliders \cite{Chang:2022crb}.

However, applying the Hilbert series to enumerating group covariants such as $C(Y_u, Y_d)$, \ie\ including but going beyond the invariants, is a less explored practice in particle physics. In \cref{sec:Hilbert} of this paper, we elaborate how to carry out such an analysis. To demystify the Hilbert series and its applications in Quantum Field Theories (QFTs), we first provide a pedagogical review in \cref{subsec:Invariants} on how to use it for enumerating group invariants. We then extend it in \cref{subsec:Covariants} to address the group covariants. Mathematically, group covariants form a \emph{module} over the ring formed by the group invariants, and the number of independent group covariants is given by the \emph{rank} of this module. As we will show, properties of this module can be systematically studied with the Hilbert series. In particular, the rank of the module can be obtained by taking a ratio of two Hilbert series.

It is remarkable that in a wide variety of scenarios, the rank of the module $\Mod_R$ is equal to the dimension of the representation $R$ of the group covariants:
\begin{equation}
\rank \left( \Mod_R \right) = \dim R \,.
\label{eqn:rankSaturation}
\end{equation}
When this occurs, we say that there is a \emph{rank saturation} (as the right-hand side is a general upper bound on the rank). It implies that the number of independent group covariants equals the number of its components, \ie\ no further restriction is achieved by requiring these Wilson coefficients to form group covariants. In \cref{subsec:RankSaturation}, we give a detailed discussion on the condition for \cref{eqn:rankSaturation} to occur. In particular, a mathematical theorem by Brion \cite{brion1993modules} (see also \cite{broer1994hilbert}) tells us that
\begin{equation}
\rank \left( \Mod_R \right) = \dim \left( R^H \right) \,,
\label{eqn:TheoremBrionIntro}
\end{equation}
where $H$ denotes the unbroken subgroup when the building blocks (\eg\ $Y_u, Y_d$ in case of the quark-MFV, taken as spurions) adopt a generic vacuum expectation value (vev), and $R^H$ denotes the subspace of $R$ that is invariant under $H$. With this theorem, we know that rank saturation will occur, in particular, when
\begin{equation}
R^H = R \,,
\end{equation}
a condition that one can check without even computing the Hilbert series.

In \cref{sec:MFV}, we apply the extended Hilbert series technique in \cref{subsec:Covariants} to enumerating the quark-MFV covariants in SMEFT. We show that rank saturation is guaranteed in this case. Consequently, we find
\begin{equation}
\text{MFV SMEFT} \;\;=\;\; \text{SMEFT} \,,
\label{eqn:Answer}
\end{equation}
as the answer to the question raised in \cref{eqn:Question}. Therefore, the restricting power of the MFV hypothesis actually comes from the additional assumptions that accompany the MFV.
We clarify that one could alternatively derive the above results with Hilbert series technique for invariants only, by working with the original setup of the problem --- including the Yukawa spurions $Y_u, Y_d$ together with the SM fields $\phi_\text{SM}$ as building blocks for the Lagrangian, and then enumerating only invariants. However, our approach of enumerating the covariant Wilson coefficients factor $C(Y_u, Y_d)$ (enabled by knowing the flavor structures of the operator $Q(\phi_\text{SM})$) is more direct and hence more efficient. It involves much fewer building blocks, and also allows us to obtain the Hilbert series to all orders, which is a lot more informative. Moreover it allows one to access a vast literature of powerful theorems, such as the theorem by Brion in \cref{eqn:TheoremBrionIntro}.

In \cref{sec:Phenomenology}, we explore some example additional assumptions on top of MFV, and check their consequences on flavor phenomenology. Specifically, we will work with a sample of extreme scenarios --- considering only one flavor covariant at a time. Our goal is to demonstrate that once equipped with additional assumptions, MFV can be very predictive, meanwhile different additional assumptions often leads to different predictions.

Finally, we provide conclusions and outlooks in \cref{sec:Conclusions}.

\section{Hilbert series for group covariants}
\label{sec:Hilbert}

One of the crucial tasks in QFTs, especially in EFTs, is to construct the Lagrangian from a given set of fields. Let us collectively denote all these fields by $\BB$, which then has many many (say $n$) components:
\begin{equation}
\BB \equiv \big\{ \BB_i \big\} = \big\{ \BB_1,\, \BB_2,\, \cdots,\, \BB_n \big\} \,.
\label{eqn:BBs}
\end{equation}
We emphasize that here we are taking a \emph{full decomposition} --- each $\BB_i$ denotes a \emph{single} component of $\BB$ with no more internal structure. It could be a Grassmann even or odd number, depending on whether it is bosonic or fermionic, but that is all.\footnote{For example, a Weyl fermion would have two \emph{single} components in this language, they are both Grassmann odd and together they form a doublet under the Lorentz group.} 
With such a full decomposition, the Lagrangian is then constructed out of all the polynomials in these components:\footnote{This paper is concerned with non-derivative interactions in the Lagrangian. But when derivatives of fields are considered, they can also be included as components in the building blocks $\BB$.}
\begin{align}
&\Big\{ \BB_1^{k_1} \BB_2^{k_2} \cdots \BB_n^{k_n} \,,\quad
\text{with} \;\; k_i \;\; \text{non-negative integers} \Big\} \notag\\[5pt]
&\hspace{-10pt}
= \Big( 1 \oplus \BB_1 \oplus \BB_1^2 \oplus \cdots \Big)
\otimes \Big( 1 \oplus \BB_2 \oplus \BB_2^2 \oplus \cdots \Big)
\otimes \cdots \otimes \Big( 1 \oplus \BB_n \oplus \BB_n^2 \oplus \cdots \Big) \,.
\label{eqn:BBpolys}
\end{align}
Clearly, we are considering nothing but \emph{multiplying all field components in all possible ways}. The number of linearly independent terms, $n (k_1, k_2, \cdots, k_n)$, at each given power structure, $(k_1, k_2, \cdots, k_n)$, indicates the degrees of freedom (or the number of free parameters) of the theory. This important piece of information can be encoded in a generating function (or partition function) defined as
\begin{equation}
\HS \left( q_1, q_2, \cdots, q_n \right) \equiv
\sum_{k_1, k_2, \cdots, k_n=0}^\infty n (k_1, k_2, \cdots, k_n)\, q_1^{k_1} q_2^{k_2} \cdots q_n^{k_n} \,.
\label{eqn:HSdef}
\end{equation}
This is the so-called Hilbert series. Here a distinct grading variable $q_i$ has been introduced for each field component $\BB_i$. In our applications, the Hilbert series converges to an analytic function at $|q_i|<1$, which will also tell us about the properties of the operator set that we are enumerating.

Note that for each bosonic field component $\BB_b$, the power $k_b$ in \cref{eqn:BBpolys} runs across all non-negative integers $\mathbb{N}$. On the other hand, a fermionic field component $\BB_f$ has its power $k_f$ truncated at the first power, due to its anti-commuting nature $\BB_f^2=0$. In summary, the power structure has the following domain
\begin{subequations}\label{eqn:kdomain}
\begin{alignat}{2}
k_b &\in \mathbb{N} = \{ 0,\, 1,\, 2,\, \cdots \}
&&\qquad\text{for each bosonic field component}\; \BB_b \,, \\[5pt]
k_f &\in \{ 0,\, 1 \}
&&\qquad\text{for each fermionic field component}\; \BB_f \,.
\end{alignat}
\end{subequations}
Without any further restrictions, we should simply have $n (k_1, k_2, \cdots, k_n) = 1$ for each distinct power structure that is allowed by \cref{eqn:kdomain}. Carrying out the sums in \cref{eqn:HSdef} then leads us to the trivial Hilbert series
\begin{equation}
\HS^\Ge \left( q_1, q_2, \cdots, q_n \right) =
\left[ \prod_{\text{bosonic } \BB_b} \frac{1}{1-q_b^{}} \right]
\left[ \prod_{\text{fermionic } \BB_f} \left( 1+q_f^{} \right) \right] \,.
\label{eqn:HStrivial}
\end{equation}
Here we clarify that each grading variable $q_i$ is always a Grassmann-even number, regardless of the nature of the its corresponding field component $\BB_i$.

The Hilbert series becomes more interesting and informative when symmetry constraints are further considered. This will be our focus for the rest of the section.

Symmetries are typically described by a (generally product) group $G$, under which the field components in \cref{eqn:BBs} form a (linear) representation.\footnote{This is the case in linear realizations of the symmetries. For Lagrangians constructed with nonlinearly realizations, one could first follow the CCWZ prescription \cite{Coleman:1969sm, Callan:1969sn} to find the linearly transforming building blocks, and then apply the Hilbert series technique; see \eg\ Sec. 3.2 in Ref.~\cite{Graf:2022rco} for more details.}
Imposing the symmetries then means to allow only $G$-invariant polynomial combinations in \cref{eqn:BBpolys}. With this additional constraint, the number of linearly independent terms becomes nontrivial and can be difficult to count directly, making the Hilbert series a useful and efficient approach. Because of this, it has been extensively applied in particle physics. In \cref{subsec:Invariants}, we will review the calculation techniques of the Hilbert series for group invariants, as well as some of their general properties relevant for our discussions. In particular, we will see that the set of group invariants always have a nice decomposition called the \emph{Hironaka decomposition}.

In some cases, our interest is not a whole Lagrangian term, but only a partial factor in it, which is then allowed to transform nontrivially under the symmetry group. For example, in spurion analysis, couplings are viewed as nontrivial representations of certain symmetry group $G$. In such scenarios, we are interested in a certain set of $G$-covariant polynomial combinations in \cref{eqn:BBpolys}. In \cref{subsec:Covariants}, we will extend the Hilbert series calculation techniques for enumerating group covariants. Although straightforward, this is a less explored practice in particle physics.\footnote{However, using this approach to select out the spacetime group covariants has been a standard practice to handle the integration by parts redundancies in the studies of EFT operator bases.}
But a lot has been worked out in invariant theory; see \eg\ Ref.~\cite{broer1994hilbert} for a succinct summary. In particular, group covariants typically do not have a Hironaka decomposition, due to redundancies in a generating set. This makes the concept of \emph{rank} a crucial property to reflect the number of ``independent'' group covariants. The rank respects an upper bound, which is saturated in many cases. This turns out to have important phenomenological consequences. In \cref{subsec:RankSaturation}, we give a dedicated discussion on ``independent'' group covariants and the \emph{rank saturation}.

\subsection{Review: Hilbert series for a ring of group invariants}
\label{subsec:Invariants}

Suppose that we are imposing a symmetry group $G$ (a product group in general), under which our Lagrangian building blocks $\BB$ transform as a linear representation (rep) $\RBB$. In this case, not all the polynomials in \cref{eqn:BBpolys} will be allowed in the Lagrangian; only $G$-invariant combinations are, and hence the Hilbert series is no longer the trivial one given in \cref{eqn:HStrivial}. Mathematically, the $G$-invariant polynomials in $\BB$ form a \emph{polynomial ring} (or \emph{polynomial algebra}), whose structure is reflected by the Hilbert series. We will come back to this towards the end of the subsection. For now, let us check what the Hilbert series look like and how to compute them.

\subsubsection{Quick examples}
\label{subsubsec:QuickExamples}

Let us see a quick example. Consider a symmetry group $G=U(1)$, and the building blocks as a pair of bosonic fields $\BB = \{ \phi_1,\, \phi_1^* \}$ with opposite charges $(+1, -1)$, \ie\ a single complex scalar. (Again, we are not considering derivatives of the fields in this paper.) All the $G$-invariant polynomials are built out of the product $I\equiv \phi_1 \phi_1^*$:
\begin{equation}
\text{The set of $G$-invariant polynomials}
= 1 \oplus I \oplus I^2
\oplus I^3 \oplus \cdots \equiv \big\{ I^k \big\}\,.
\label{eqn:IkU11}
\end{equation}
As $I$ is quadratic in the building blocks $\BB$, this leads us to the following Hilbert series
\begin{equation}
\HS_\I^{U(1),\, \left(+1, -1\right)} (q) = 1 + q^2 + q^4 + q^6 + \cdots = \frac{1}{1-q^2} \,,\qquad
|q|<1 \,,
\label{eqn:HSU11}
\end{equation}
which can be trivially summed since it corresponds to the geometric series. Generally, we will be using $\HS_\I^{G,\, \RBB}$ to denote the Hilbert series for $G$ invariants (\ie\ non-transforming singlets) built out of the rep $\RBB$ of $G$. Also, we will typically take all the grading variables in \cref{eqn:HSdef} the same ($q_i=q$) unless otherwise stated, and thus $q$ acts as a label indicating the total power of building block fields in a given invariant. Some explanations about this choice together with comments on more general grading options will be provided later. The grading variables are understood to have modulus smaller than unity, $|q_i|<1$, such that the Hilbert series is convergent to some analytic function, as in \cref{eqn:HSU11}.

Now let us consider a slightly more complicated example --- the same symmetry group, but with two complex scalars of the same charge as the building blocks:
\begin{equation}
G=U(1) \,,\qquad
\BB = \big\{ \phi_1,\, \phi_1^*,\, \phi_2,\, \phi_2^* \big\} \,.
\label{eqn:ExampleU12}
\end{equation}
In this case, all the $G$-invariant polynomials must be built out of the following four basic $G$-invariant products
\begin{equation}
I_1 \equiv \phi_1 \phi_1^* \,,\qquad
I_2 \equiv \phi_2 \phi_2^* \,,\qquad
I_3 \equiv \phi_1 \phi_2^* \,,\qquad
I_4 \equiv \phi_2 \phi_1^* \,.
\label{eqn:IsU12}
\end{equation}
However, naively enumerating all their polynomials
\begin{align}
\hspace{-5pt}
(1 \oplus I_1\oplus I_1^2  \oplus \cdots)
(1 \oplus I_2\oplus I_2^2  \oplus \cdots)
(1 \oplus I_3\oplus I_3^2  \oplus \cdots)
(1 \oplus I_4\oplus I_4^2  \oplus \cdots) \,,
\end{align}
would be over counting, due to the redundancy relation
\begin{equation}
I_1 I_2 = I_3 I_4 \,.
\label{eqn:IRelationU12}
\end{equation}
To factor out the non-redundant set, we can organize them by the powers of $(I_3 I_4)$:
\begin{equation}
\Big\{ I_1^{k_1} I_2^{k_2} I_3^{k_3} I_4^{k_4} \Big\}
= \big\{ \text{non-redundant $G$-invariant polynomials} \big\}
\otimes \left\{ \left( I_3 I_4 \right)^k \right\} \,,
\label{eqn:FactorizationU12}
\end{equation}
which gives us the following Hilbert series
\begin{align}
\frac{1}{(1-q^2)^4} &= \HS_\I^{U(1),\, 2\times\left(+1, -1\right)} \frac{1}{1-q^4} \notag\\[5pt]
&\hspace{50pt}
\quad\implies\quad
\HS_\I^{U(1),\, 2\times\left(+1, -1\right)} = \frac{1-q^4}{(1-q^2)^4}
= \frac{1 +q^2}{\left( 1-q^2 \right)^3} \,.
\label{eqn:HSU12}
\end{align}
An alternative way of constructing this Hilbert series emerges from the second-to-last expression in \cref{eqn:HSU12}, where a negative term $-q^4$ appears in the numerator. One can start from the series of all the polynomials in $I_{i}$, \ie\ ${1}/{(1-q^2)^4}$, and then subtract the number of redundancies to obtain the Hilbert series. For the case at hand, subtracting redundancies gives rise to the term
\begin{equation}
-\frac{q^4}{(1-q^2)^4} \,,
\end{equation}
because the redundancy relation in \cref{eqn:IRelationU12} is of order $q^4$, and multiplying it by any polynomial in $I_i$ gives another redundancy. The lesson we get from this is that negative terms in the numerator originate from subtracting redundancy relations.

\subsubsection*{What can we learn from the Hilbert series?}

Generally, any $G$-invariant polynomial can be decomposed as
\begin{equation}
\sum_{i=0}^{n_S^{}} p_i \big( P_1, \cdots, P_{n_P^{}} \big)\, S_i \,,
\label{eqn:PSdecomposition}
\end{equation}
where $p_i \big( P_1, \cdots, P_{n_P^{}} \big)$ are polynomials of the $n_P^{}$ \emph{primary invariants} $P_a$, which enter as coefficients of the \emph{secondary invariants} $S_i$, including $S_0\equiv 1$. $S_1, \ldots, S_{n_S^{}}$ are characterized by the following property: there is a minimal power $r_i>1$ such that $S_i^{r_i}$ can be written as a polynomial of the primary invariants. Note that this means that $S_i^2, \ldots,S_i^{r_i-1}$ are themselves secondary invariants (and must be included in the sum above). To reiterate, secondary invariants appear linearly, while any power of a primary invariant can appear in \cref{eqn:PSdecomposition}. This is reflected in the \emph{summed} Hilbert series. For each primary invariant of degree $k$, there is a factor of $(1-q^k)^{-1}$. If the degree of each $S_i$ is $d_i$ then there is also a factor of $(1+q^{d_1} +\cdots + q^{d_{n_S^{}}})$. In the example above, the Hilbert series in \cref{eqn:HSU12} indicates 3 primary invariants of degree 2, as well as one secondary invariant of degree 2, whose square can be written in terms of the primary invariants using the redundancy in \cref{eqn:IRelationU12}.

In the above, we have computed the Hilbert series from one complex scalar $\HS_\I^{U(1),\, \left(+1, -1\right)} (q)$, and two complex scalars $\HS_\I^{U(1),\, 2\times\left(+1, -1\right)} (q)$. What if we have more of them in the building blocks? One can imagine that in more complicated cases, there will be more basic $G$-invariant products of the kind in \cref{eqn:IsU12}, and also more nontrivial redundancy relations of the kind in \cref{eqn:IRelationU12}. Then the factorization of the type in \cref{eqn:FactorizationU12} will be a lot more difficult to find, if it exists at all. Fortunately, a more systematic and efficient way of calculating the Hilbert series is available in such cases. From it one can read off the number of primary and secondary invariants, as well as their degrees.

\subsubsection{Grading options}
\label{subsubsec:GradingOptions}

Before moving on to the calculation technique, let us comment on the grading options for the Hilbert series when a symmetry group $G$ is in consideration. Typically, the Lagrangian building blocks $\BB$ form a linear representation $\RBB$ under $G$. Technically, this is identifying each group element $g\in G$ with an $n \times n$ matrix $g_\RBB^{}$ (with $n=\dim(\RBB)$ as in \cref{eqn:BBs}), which is the linear transformation map between the old and the new field components:
\begin{equation}
\left( g \cdot \BB \right)_i \equiv \left( g_\RBB^{} \right)_i{}^j\, \BB_j \,.
\label{eqn:gRBB}
\end{equation}
We see that under the group transformations, the components $\{ \BB_i \}$ generally get mixed, and so will  monomials formed from products of these components. However, since the transformation in \cref{eqn:gRBB} is linear, homogeneous polynomials of degree $k$ mix only among themselves. In other words, all $\BB$ polynomials with a given total degree $k$:
\begin{equation}
\BB^k \equiv \Big\{ \BB_1^{k_1} \BB_2^{k_2} \cdots \BB_n^{k_n} \,,\quad
k_1 + k_2 + \cdots + k_n = k \Big\} \,,
\label{eqn:BBk}
\end{equation}
form a representation of $G$, and one could ask about the number of invariant reps, $n_\I^{}(k)$, that this representation space $\BB^k$ contains. These numbers can be encoded in a generating function in the  spirit of \cref{eqn:HSdef}:
\begin{equation}
\HS_\I^{G,\, \RBB} (q) \equiv \sum_{k=0}^\infty n_\I^{}(k)\, q^k \,.
\label{eqn:HSqdef}
\end{equation}
This defines a Hilbert series graded by the total degree $k$ of $\BB$ polynomials; it is what we have computed in \cref{eqn:HSU11,eqn:HSU12}. In the rest of this paper, we will be mostly adopting this grading option for the Hilbert series, unless otherwise stated. But let us mention here that there are possibly other grading options. Quite often, the rep $\RBB$ is a reducible representation of $G$, which we can decompose as
\begin{equation}
\BB = \Big\{ \phi_1,\, \phi_2,\, \cdots,\, \phi_m \Big\} \,,\qquad
\RBB = \bigoplus_i R_{\phi_i} \,.
\label{eqn:RBBDecompose}
\end{equation}
Here, each $\phi_i$ is a collection of $\BB$ components that form an irreducible representation (irrep) of $G$. Under this decomposition, the representation matrices $g_\RBB^{}$ are block-diagonal, and different $\phi_i$ subspaces would not get mixed. Therefore, one could ask how many $G$ invariants are contained in the representation space $\phi_1^{k_1} \phi_2^{k_2} \cdots \phi_m^{k_m}$ (where each $\phi_i^{k_i}$ denotes a set of polynomials, same as in \cref{eqn:BBk}). These numbers can be encoded in a similar way as in \cref{eqn:HSdef}, leading to a Hilbert series with multiple grading variables. This is the finest grading that one can take when a symmetry group $G$ is in consideration --- at most, one could assign a distinct grading variable for each irrep of $G$ in the Lagrangian building blocks.

\subsubsection{Molien-Weyl formula}
\label{subsubsec:MolienWeyl}

Now we are ready to introduce the Molien-Weyl formula, which provides a more efficient way for calculating the Hilbert series. Suppose that all the field components $\{\BB_i\}$ are bosonic (the fermionic case will be discussed below), then the Hilbert series defined in \cref{eqn:HSqdef} can be computed by an integral over the group elements
\begin{equation}
\HS_\I^{G,\, \RBB} (q) = \int \dd\mu_G(x)\,
\frac{1}{\det \big[ 1 - q\, g_\RBB^{}(x) \big] }
\qquad\text{with}\qquad
|q|<1 \,.
\label{eqn:Molien}
\end{equation}
This is a standard approach in invariant theory (see \eg\ \cite{Sturmfels:inv}), often referred to as the Molien-Weyl formula. Let us explain this formula:
\begin{itemize}
\item \textbf{Notations in the integrand:} In the integrand above, $q$ is the grading variable in \cref{eqn:HSqdef}, \ie\ a label indicating the total power of building block fields in a given invariant. $g_\RBB^{}(x)$ is the $n\times n$ representation matrix ($n=\dim(\RBB)$) in \cref{eqn:gRBB}. Here, $x$ denotes a set of variables that parameterize the group elements $g(x)\in G$; see \cref{eqn:gxExample} in \cref{appsec:Weyl} for an explicit example of this parameterization. The whole object in the square bracket is understood as an $n\times n$ matrix, and we are taking its determinant.
\item \textbf{Character orthonormality:} The key property that allows to derive the Molien-Weyl formula is the character orthonormality among irreps. In group representation theory, the character is defined as the trace of the representation matrix of a group element. For example, the representation matrix in \cref{eqn:gRBB} leads to the character
\begin{equation}
\chi_\RBB^{} \big( g(x) \big) = \tr \big( g_\RBB^{}(x) \big) \,.
\label{eqn:chiRBB}
\end{equation}
In general, a character is a function that depends both on the group element, and the representation. A crucial property is that characters are orthonormal under a sum over all the group elements. For a finite (discrete) group, this reads
\begin{equation}
\sum_{g\in G} \frac{1}{|G|}\, \chi_{R_1}^*(g)\, \chi_{R_2}^{}(g) = \delta_{R_1 R_2}^{} \,,
\end{equation}
where $R_1$ and $R_2$ are two arbitrary unitary irreps of $G$, and $|G|$ denotes the total number of elements in it. In case of a (compact) continuous group, the sum is performed by an integration over the Haar measure $\dd\mu_G \big( g(x) \big)$:
\begin{equation}
\int \dd\mu_G \big( g(x) \big)\, \chi_{R_1}^*\big( g(x) \big)\, \chi_{R_2}^{}\big( g(x) \big) = \delta_{R_1 R_2}^{} \,.
\label{eqn:CharacterOrthonormality}
\end{equation}
This is extremely useful for selecting out irrep components of a given reducible representation. 

For our case at hand, we are interested in counting all possible invariants made out of the building blocks $\{\BB_i\}$. To this end, we first compute the representation $R_{\BB^k}$. Recall from \cref{eqn:BBk} that $\BB^k$ are all the $\BB$ polynomials with a given total degree $k$. Therefore, $R_{\BB^k}$ is given by the fully symmetric component of the $k$-th tensor product of $R_\BB$:\footnote{Antisymmetric components of this tensor product do not appear in $\BB^k$, because the components $\BB_i$ are bosonic and hence they commute. The fermionic case will be discussed later.}
\begin{align}
R_{\BB^k} = \text{sym} \Big( \underbrace{R_{\BB} \otimes R_{\BB} \otimes \dots \otimes R_{\BB}}_{k}\Big) \,.
\label{eqn:RBBk}
\end{align}
This is a reducible representation of $G$ that can be expressed as a direct sum of irreps. In particular, we want to study the number of invariant irreps, $n_\I^{}(k)$, contained in $R_{\BB^k}$:
\begin{equation}
R_{\BB^k} = n_\I^{}(k)\; \Ibf\, \oplus \text{ other irreps} \,.
\label{eqn:RBBkInvs}
\end{equation}
Making use of \cref{eqn:CharacterOrthonormality}, we can compute the number $n_\I^{}(k)$ as
\begin{equation}
n_\I^{}(k) = \int \dd\mu_G(x)\, \chi_\I^*(x)\, \chi_{R_{\BB^k}}^{}(x)
= \int \dd\mu_G(x)\, \chi_{R_{\BB^k}}^{}(x) \,.
\label{eqn:nkRBBk}
\end{equation}
This is the key to derive the Molien-Weyl formula.
\item \textbf{Constructing the integrand:} When the representation $\RBB$ is reducible, one could decompose it as in \cref{eqn:RBBDecompose}. Since the matrix $g_\RBB^{}$ is block-diagonal in this decomposition, the determinant is given by a product of determinants in the subspaces
\begin{equation}
\frac{1}{\det \big[ 1 - q\, g_\RBB^{}(x) \big]}
= \prod_i \frac{1}{\det \big[ 1 - q\, g_{R_{\phi_i}}^{}(x) \big]} \,.
\label{eqn:detDecompose}
\end{equation}
This is the usual way that one constructs the integrand. At this point, a finer grading option is available if one prefers:
\begin{equation}
\prod_i \frac{1}{\det \big[ 1 - q_i\, g_{R_{\phi_i}}^{}(x) \big]} \,,
\label{eqn:FinerGrading}
\end{equation}
but in this paper we will be mostly using the unified grading by the total degree, which amounts to taking $q_i=q$ in \cref{eqn:FinerGrading}, reproducing \cref{eqn:detDecompose}.
\item \textbf{Nature of the integrand:} The integrand in \cref{eqn:Molien} is a sum of the characters of the rep $R_{\BB^k}$, weighted by $q^k$, \ie\ it is a graded character. To see this, let us note that for any unitary representation $\RBB$, $g_\RBB^{} (x)$ is a unitary matrix; it can be diagonalized as (see \cref{eqn:gxDiagExample} in \cref{appsec:Weyl} for an explicit example)
\begin{equation}
g_\RBB^{}(x) = U_x\, z(x)\, U_x^\dagger
\qquad\text{with}\qquad
z(x) = \diag\, \big( z_1(x), \cdots, z_n(x) \big) \,,
\end{equation}
where each $z_i(x) = e^{i\theta_i(x)}$ is a pure phase variable. With this, the character of the representation $R_{\BB^k}$ can be worked out using the relation in \cref{eqn:RBBk}:
\begin{equation}
\chi_{R_{\BB^k}}^{}(x) = \sum_{k_1+k_2+\cdots+k_n=k} z_1^{k_1} z_2^{k_2} \cdots z_n^{k_n} \,.
\label{eqn:chiRBBk}
\end{equation}
Summing them over with a weight $q^k$ gives us the integrand in \cref{eqn:Molien}:
\begin{equation}
\sum_{k=0}^\infty q^k\, \chi_{R_{\BB^k}}^{}(x)
= \prod_{i=1}^n \frac{1}{1-q\, z_i(x)}
= \frac{1}{\det \big[ 1 - q\, g_\RBB^{}(x) \big]} \,.
\label{eqn:chiqBBb}
\end{equation}
Now using \cref{eqn:nkRBBk}, it is clear that the Hilbert series defined in \cref{eqn:HSqdef} can be computed by the Molien-Weyl formula in \cref{eqn:Molien}.

Note that the above story can be applied to each irrep $\phi_i$ in \cref{eqn:RBBDecompose}. So each individual factor in \cref{eqn:detDecompose} can also be understood as a graded character. This is especially useful when one chooses the finer grading option in \cref{eqn:FinerGrading}.
\item \textbf{Weyl characters and Weyl integration formula:} In principle, the integral in \cref{eqn:CharacterOrthonormality} (and hence in \cref{eqn:Molien}) should be done over all group elements. However, when $G$ is a connected Lie group, it can be simplified into an integral over just the Cartan subgroup of $G$. This is because every group element can be conjugated to an element in the Cartan subgroup, while the character is unaffected by the group conjugation. More explanations on this are provided in \cref{appsec:Weyl}, where we also give a list of characters and Haar measure that are relevant for Hilbert series calculations in this paper. We refer the reader to the appendices in Refs.~\cite{Henning:2017fpj} for a more thorough list and their derivations. Technically, when restricted to the Cartan subgroup, the characters are called \emph{Weyl characters}, and the Haar measure integral is called \emph{Weyl integration formula}. In this paper, we will be sloppy and ignore these terminology distinctions.
\item \textbf{Fermionic case:} When the Lagrangian building blocks $\BB$ are fermionic fields, they anti-commute in multiplications in their polynomial. In this case, $R_{\BB^k}$ is given by the antisymmetric components of the tensor product in \cref{eqn:RBBk}, which effectively further restricts each $k_i$ in \cref{eqn:chiRBBk} to be either $0$ or $1$ (same as in \cref{eqn:kdomain}). So the graded character in the fermionic scenario is given by
\begin{equation}
\sum_{k=0}^\infty q^k\, \chi_{R_{\BB^k}}^{}(x)
= \prod_{i=1}^n \big[ 1+q\, z_i(x) \big]
= \det \big[ 1 + q\, g_\RBB^{}(x) \big] \,.
\label{eqn:chiqBBf}
\end{equation}
One should use this as the integrand in Molien-Weyl formula for the fermionic case. When $\BB$ has mixed statistics, one should first decompose it into irreps as in \cref{eqn:RBBDecompose}; each $\phi_i$ then has definite statistics, based on which one can choose between \cref{eqn:chiqBBb,eqn:chiqBBf}.
\item \textbf{An important condition:} Finally, we emphasize that for \cref{eqn:Molien} to yield the Hilbert series defined in \cref{eqn:HSqdef}, one needs to assume $|q|<1$ while carrying out the integral on the right-hand side. When $|q|=1$ there are poles on the contour of the integration (over the phase variables $z_i(x)$; see \cref{eqn:chiqBBb} and also the example below), and therefore the functions defined by the integral in~\cref{eqn:Molien} for $|q|<1$ and $|q|>1$ are not related by analytic continuation.
\end{itemize}

With all due explanations provided, let us now see how to use the Molien-Weyl formula in practice. Consider again the symmetry group $G=U(1)$, and $m \ge 1$ complex scalars with charge $(+1, -1)$ as the building blocks. In this case, the Haar measure is (see \cref{eqn:HaarU1})
\begin{equation}
\int \dd\mu_{U(1)}^{} (z) = \oint_{|z|=1} \frac{\dd z}{2\pi i} \frac{1}{z} \,,
\end{equation}
and the integrand can be constructed as
\begin{subequations}
\begin{align}
g_\RBB^{} &= \diag \left( z,\, z^{-1},\, \cdots,\, z,\, z^{-1} \right) \,, \\[6pt]
\det \big[ 1 - q\, g_\RBB^{}(x) \big] &= (1-qz)^m\, (1-qz^{-1})^m \,.
\end{align}
\end{subequations}
Plugging these into \cref{eqn:Molien}, we get the Hilbert series
\begin{align}
\HS_\I^{U(1),\, m\times\left(+1, -1\right)}(q)
&= \oint_{|z|=1} \frac{\dd z}{2\pi i}\, \frac{1}{z}\, \frac{1}{(1-qz)^m (1-qz^{-1})^m} \notag\\[8pt]
&= \left[ \frac{1}{(m-1)!} \left( \frac{\dd}{\dd z} \right)^{m-1} \frac{z^{m-1}}{(1-qz)^m} \right]\Bigg|_{z=q} \,.
\label{eqn:HSU1m}
\end{align}
The second line is obtained by evaluating the contour integral with Cauchy's residue theorem. In doing so, it is crucial to assume that $|q|<1$ (as emphasized in \cref{eqn:Molien}) --- it implies that the only pole inside our integration contour $|z|=1$ is $z=q$ (as opposed to $z=q^{-1}$). Taking the other pole gives a wrong result.

One can verify that \cref{eqn:HSU1m} does reproduce \cref{eqn:HSU11,eqn:HSU12} at $m=1, 2$. For higher values, \eg\ $m=3, 4, 5$, it gives more complicated results
\begin{subequations}\label{eqn:HSU1345}
\begin{align}
\HS_\I^{U(1),\, 3\times\left(+1, -1\right)} &=
\frac{1 +4q^2 +q^4}{\left( 1-q^2 \right)^5} \,, \label{eqn:HSU13} \\[10pt]
\HS_\I^{U(1),\, 4\times\left(+1, -1\right)} &=
\frac{1 +9q^2 +9q^4 +q^6}{\left( 1-q^2 \right)^7} \,, \label{eqn:HSU14} \\[10pt]
\HS_\I^{U(1),\, 5\times\left(+1, -1\right)} &=
\frac{1 +16q^2 +36q^4 +16q^6 +q^8}{\left( 1-q^2 \right)^9} \,. \label{eqn:HSU15}
\end{align}
\end{subequations}
More explicit examples of using the Molien-Weyl formula are available in \cref{appsec:Examples}. These results are very hard to obtain by a direct counting like \cref{eqn:IkU11}, or a factorization analysis like \cref{eqn:FactorizationU12}.

\subsubsection{Structure of group invariants: Hironaka decomposition}
\label{subsubsec:HironakaDecomposition}

The Hilbert series is defined to record the number of $G$ invariants, $n_\I^{}(k)$, at each total degree $k$ in the $\BB$ polynomials; see \cref{eqn:HSqdef}. But as we mentioned above, the Hilbert series is actually more informative --- by summing the series at all orders, it also tells us about the structure of this $G$-invariant $\BB$ polynomial set. Below we elaborate this point.

First, let us recall the simple example computed in \cref{eqn:IkU11,eqn:HSU11}:
\begin{equation}
\HS_\I^{U(1),\, \left(+1, -1\right)} (q) = \frac{1}{1-q^2}
\quad\Longleftrightarrow\quad
\text{$G$ invariants} = 1 \oplus I \oplus I^2 \oplus \cdots \equiv \big\{ I^k \big\} \,.
\label{eqn:HSU11Redo}
\end{equation}
Clearly, the Hilbert series with a single denominator factor implies that the whole set of $G$-invariant $\BB$ polynomials are generated as polynomials of a single operator, which is $I=\phi_1 \phi_1^*$ in this example.

Next, for the example in \cref{eqn:ExampleU12}, the Hilbert series was computed in \cref{eqn:HSU12}:
\begin{equation}
\HS_\I^{U(1),\, 2\times\left(+1, -1\right)} = \frac{1 +q^2}{\left( 1-q^2 \right)^3}
= \left( 1 + q^2 + q^4 + q^6 + \cdots \right)^3 \left( 1 + q^2 \right) \,,
\label{eqn:HSU12Redo}
\end{equation}
implying the following structure for the set of non-redundant $G$-invariant polynomials
\begin{equation}
\text{non-redundant $G$ invariants} = \Big\{ P_1^{k_1} \Big\} \otimes \Big\{ P_2^{k_2} \Big\}
\otimes \Big\{ P_3^{k_3} \Big\} \otimes \big( 1 \oplus S \big) \,,
\label{eqn:HironakaU12}
\end{equation}
for some operators $P_1, P_2, P_3$, and $S$, all quadratic in $\BB$. Among these operators, $P_1, P_2, P_3$ are needed to all powers; they are called \emph{primary invariants}. On the other hand, the operator $S$ is only needed up to its first power; it is called a \emph{secondary invariant}. Generally speaking, the structure of the type in \cref{eqn:HironakaU12} is called a \emph{Hironaka decomposition}. It is fairly nontrivial, implying several properties:
\begin{itemize}
\item There are no redundancies in the set of polynomials generated by the primary invariants $\big\{ P_1^{k_1} P_2^{k_2} P_3^{k_3} \big\}$,
\begin{align}
\sum c_{k_1k_2k_3}  P_1^{k_1} P_2^{k_2} P_3^{k_3} = 0
\quad\iff\quad
c_{k_1k_2k_3}=0 \quad \forall\, k_1,k_2,k_3 \,.
\label{eqn:Punique}
\end{align}
\item The truncation in the power of the secondary invariant $S$ means that its higher powers are already captured by the polynomials of primary invariants:
\begin{equation}
S^2 \subset \Big\{ P_1^{k_1} P_2^{k_2} P_3^{k_3} \Big\} \,.
\label{eqn:S2inP}
\end{equation}
\item The direct sum structure in \cref{eqn:HironakaU12} also implies no overlap between the two sets below, except the zero element:
\begin{equation}
\Big\{ P_1^{k_1} P_2^{k_2} P_3^{k_3} \Big\} \otimes S
\;\;\cap\;\; 
\Big\{ P_1^{k_1} P_2^{k_2} P_3^{k_3} \Big\}
\;\;=\;\; 0 \,.
\label{eqn:NoOverlap}
\end{equation}
\end{itemize}
Using the four basic $G$-invariant products in \cref{eqn:IsU12}, one can choose
\begin{equation}
P_1 = I_1 \,,\qquad
P_2 = I_2 \,,\qquad
P_3 = I_3 + I_4 \,,\qquad
S= I_3 - I_4 \,,
\label{eqn:PSU12}
\end{equation}
then the above properties can indeed be all satisfied. In particular, the property in \cref{eqn:S2inP} is satisfied because
\begin{equation}
I_1 I_2 = I_3 I_4
\qquad\Longrightarrow\qquad
S^2 = P_3^2 - 4 P_1 P_2 \,.
\end{equation}
So \cref{eqn:PSU12} gives an  instantiation  of the Hironaka decomposition in \cref{eqn:HironakaU12}.

In general, the Hironaka decomposition states that there are a set of primary invariants $\{P_a\}$ and second invariants $\{S_i\}$ such that each $G$-invariant polynomial of $\BB$ has a \emph{unique} decomposition as in \cref{eqn:PSdecomposition}:
\begin{equation}
\sum_{i=0}^{n_S^{}} p_i \big( P_1, \cdots, P_{n_P^{}} \big)\, S_i \,,
\label{eqn:Hironaka}
\end{equation}
with $p_i \big( P_1, \cdots, P_{n_P^{}} \big)$ certain polynomials of the primary invariants $\{P_a\}$, and $S_0\equiv 1$. This captures all the three properties mentioned above in \cref{eqn:Punique,eqn:S2inP,eqn:NoOverlap}.

If a Hironaka decomposition is found explicitly by constructing a set of primary and secondary invariants, such as in \cref{eqn:PSU12}, then one can readily compute the Hilbert series from it. However, the Hironaka decomposition is often difficult to find directly, except for extremely simple cases like the ones discussed above. Therefore, in practical cases one often first computes the Hilbert series with the Molien-Weyl formula, and then uses the result as a guidance to find the Hironaka decomposition.

Mathematically, the set of all $\BB$ polynomials in \cref{eqn:BBpolys} form a \emph{polynomial ring} (or \emph{polynomial algebra}) $\rBB$, because it is equipped with an \emph{addition}, a \emph{multiplication}, and also a \emph{scalar multiplication} with a set of coefficients. Suppose that the coefficient set is $\mathbb{C}$ (complex numbers), then a standard notation for this ring is (see \eg\ \cite{Sturmfels:inv, bruns1998cohen})
\begin{equation}
\rBB = \mathbb{C} \big[ \BB_1, \BB_2, \cdots, \BB_n \big] = \mathbb{C} \big[ \BB \big] \,.
\label{eqn:rBB}
\end{equation}
When a symmetry group $G$ is in consideration, the $G$-invariant subset of $\rBB$ gives a subring $\rI \subset \ring$ (as it is closed under the three operations mentioned above), which is usually denoted as
\begin{equation}
\rI = \mathbb{C} \big[ \BB \big]^G \subset \rBB \,.
\label{eqn:rI}
\end{equation}
The Hilbert series defined in \cref{eqn:HSqdef}, once summed, reflects the structure and properties of $\rI$. This is extensively studied in invariant theory, with many facts known. For instance, a theorem of Hochster and Roberts \cite{Hochster74} states that when $G$ is a linearly reductive group,\footnote{This covers all finite groups, all simple compact Lie groups, $U(1)$ factors, and their products, so essentially all the cases that we will ever discuss in this paper.}
the subring $\rI$ is a \emph{Cohen-Macaulay ring}, which then guarantees a Hironaka decomposition (see \eg\ \cite{Sturmfels:inv}). In such cases, one can always write the Hilbert series into a rational form
\begin{equation}
\HS_\I^{G,\, \RBB} (q) = \frac{N_\I(q)}{D(q)} \,,
\label{eqn:HSInv}
\end{equation}
with the denominator $D(q)$ encoding the primary invariants, and the numerator $N_\I(q)$ encoding the secondary invariants. They are both polynomials in $q$ with integer coefficients. The denominator has the general form
\begin{equation}
D(q) = \prod_{p=1}^{n_P^{}} \left( 1 - q^{d_p} \right) \,,
\label{eqn:DqGeneral}
\end{equation}
where $n_P^{}$ is the number of primary invariants, and $d_p$ is the degree of each of
them.\footnote{In the language of invariant theory, a set of primary invariants is often referred to as a \emph{homogeneous system of parameters}. There could be different choices for it, but all choices share the same $n_P^{}$, which is called the \emph{Krull dimension}.}
The numerator has the general form
\begin{equation}
N_\I(q) = 1 + \sum_{s=1}^{n_S^{}} q^{d_s} \,,
\label{eqn:NqGeneral}
\end{equation}
where $n_S^{}$ is the number of secondary invariants,\footnote{In the math literature, $S_0\equiv 1$ is often called a secondary invariant as well. In that language, the number of secondary invariants would be $n_S^{}+1$.}
and $d_s$ gives the degree of each of them. Another known fact is that when $G$ is a semisimple Lie group, the ring $\rI$ is not only a Cohen-Macaulay ring, but also a \emph{Gorenstein ring} (see \eg\ \cite{Gray:2008yu, bruns1998cohen}). A theorem by Stanley \cite{Stanley78} then implies that the numerator of the Hilbert series $N_\I(q)$ is a palindromic polynomial, namely that
\begin{equation}
N_\I(q) = q^h\, N_\I(q^{-1}) \,,
\end{equation}
with $h$ denoting the highest degree in $N_\I(q)$. This property can be explicitly verified with the examples in \cref{eqn:HSU12,eqn:HSU1345}, as well as the various Hilbert series results that we explored in \cref{tab:U1,tab:SU2adj,tab:SU3ffbar,tab:SU3adj} in \cref{appsec:Examples}.

\subsection{Extension: Hilbert series for a module of group covariants}
\label{subsec:Covariants}

In \cref{subsec:Invariants}, we elaborated how to use the Hilbert series to understand the structure of sets of group invariants. We explained how to compute the Hilbert series with Molien-Weyl formula. The result reflects a Hironaka decomposition (\cref{eqn:Hironaka}), with a palindromic numerator, reflecting the fact that the ring of group invariants is a Gorenstein (hence also Cohen-Macaulay) ring. What if we are interested in some set of group covariants instead? In this subsection, we extend the Hilbert series technique to such cases where we wish to study all the structures that transform under a given representation of the group, including by going beyond the invariant representation. As we will see, group covariants generally form a \emph{module} over a ring of invariants, whose structure will also be reflected by the Hilbert series.

\subsubsection{Motivation}
\label{subsubsec:Motivation}

In particle physics, it is frequently necessary to systematically list all conceivable structures with a given transformation property. Several examples come to mind:
\begin{itemize}
\item In the Operator Product Expansion, one combines quantum fields and their derivatives as building blocks of structures with definite transformation properties under Lorentz and internal symmetry groups.
\item In counting the form factors required in characterizing matrix elements of operators between (multi-)particle states one combines momenta, polarization tensors and spinors into structures that transform under space-time transformations as the matrix element does.
\item When studying the possible composite particles which arise in a confining gauge theory from a given set of fundamental fields, a systematic method to enumerate gauge-singlet flavor covariants may prove useful.
\end{itemize}

In this paper we will focus on a fourth example, namely that of a spurion analysis in the flavor sector (Minimal Flavor Violation). In this context, we treat the Yukawa coupling matrix $Y_d$ below as a spurion under the flavor symmetries
\begin{equation}
\L \supset \bar{q}\, Y_d\, d\, H + \text{h.c.} \,,
\label{eqn:toyLagYd}
\end{equation}
where the flavor and gauge indices are omitted and $q$, $d$, and $H$ are the left-handed quark doublets, the right handed down-type quarks, and the Higgs doublet, respectively. When the Yukawa $Y_d$ is promoted to a spurious field transforming as a bi-fundamental representation under the group $SU(3)_q \times SU(3)_d $, the interaction in \cref{eqn:toyLagYd} becomes invariant under this ``flavor'' group. Taking $Y_d$ as the building block, one can then ask how many $SU(3)_d$ adjoints can be built out of it, which is relevant for, \eg\ constructing the $f(Y_d)$ factor in the formally flavor invariant effective operator
\begin{equation}
\L \supset \Big( \bar{d}\, f(Y_d)\, \gamma^\mu d \Big) \left( H^\dag i\overleftrightarrow{D}_\mu H \right) \,.
\end{equation}

\subsubsection{General formulation and solution}
\label{subsubsec:Formulation}

In general, this type of questions can be formulated as follows. Given a symmetry group $G$ with a set of building blocks $\BB$ that form a representation $R_\BB$ of $G$, what are the $\BB$ polynomials that form a certain rep $R$ of $G$? Similar to our notation for invariants, we denote the Hilbert series for rep-$R$ covariants as $\HS_R^{G,\, \RBB}$. We see that the special case of $R=\I$ (the trivial rep) is what we have discussed in \cref{subsec:Invariants}. Now we would like to extend the discussion to a general irrep $R$.

The Hilbert series $\HS_R^{G,\, \RBB}$ for group covariants can be calculated by slightly modifying the Molien-Weyl formula given in \cref{eqn:Molien}. We recall that the key idea behind the Molien-Weyl formula is character orthonormality. Specifically, the group invariants are the trivial irreps in the decompositions of $R_{\BB^k}^{}$ (see \cref{eqn:RBBkInvs}), which are selected out by \cref{eqn:nkRBBk} using character orthonormality. Now we are interested in some general irrep $R$ instead:
\begin{equation}
R_{\BB^k} = n_R^{}(k)\, R\, \oplus \text{ other irreps} \,.
\label{eqn:RBBkR}
\end{equation}
They can be selected out in a similar way
\begin{equation}
n_R^{}(k) = \int \dd\mu_G(x)\, \chi_R^*(x)\, \chi_{R_{\BB^k}}^{}(x) \,.
\label{eqn:nkRMolien}
\end{equation}
The only difference is the replacement of the inserted character:
\begin{equation}
\chi_\I^*(x) = 1
\qquad\longrightarrow\qquad
\chi_R^*(x) \,.
\end{equation}
This leads us to the slightly modified Molien-Weyl formula
\vspace{3mm}
\begin{tcolorbox}[colback=light-gray]
\begin{center}
\begin{minipage}{5.5in}\vspace{-5mm}
\begin{equation}
\HS_R^{G,\, \RBB} (q) \equiv \sum_{k=0}^\infty n_R^{}(k)\, q^k
= \int \dd\mu_G(x)\, \chi_R^*(x)\, \frac{1}{\det \big[ 1 - q\, g_\RBB^{}(x) \big] } \,.
\label{eqn:MolienR}
\end{equation}
\end{minipage}
\end{center}
\end{tcolorbox}

\subsubsection{A simple example}
\label{subsubsec:SimpleExample}

Let us make use of \cref{eqn:MolienR} to compute a simple example. Consider again the example of two complex scalars with the same charge under a $G=U(1)$ group as given in \cref{eqn:ExampleU12}. We know that for the set of $U(1)$ invariants, the Hilbert series is given in \cref{eqn:HSU12}, which corresponds to a Hironaka decomposition in \cref{eqn:HironakaU12}, with an instantiation given in \cref{eqn:PSU12}. Now suppose that we are interested in the set of $\BB$ polynomials that have charge $+2$ under $G=U(1)$. The character of the charge $+2$ irrep of $U(1)$ is (see \cref{eqn:chiU1} in \cref{appsec:Weyl})
\begin{equation}
\chi_{+2}^{U(1)}(z) = z^2 \,.
\end{equation}
Using this in \cref{eqn:MolienR}, we get the Hilbert series
\begin{align}
\HS_{+2}^{U(1),\, 2\times\left(+1, -1\right)}(q)
&= \oint_{|z|=1} \frac{\dd z}{2\pi i}\, \frac{1}{z}\, z^{-2}\, \frac{1}{(1-qz)^2 (1-qz^{-1})^2} \notag\\[8pt]
&= \left[ \frac{\dd}{\dd z}\, \frac{1}{z (1-qz)^2} \right]\Bigg|_{z=q}
= \frac{3q^2 -q^4}{\left( 1 - q^2 \right)^3} \,.
\label{eqn:HS2U12}
\end{align}
Comparing this result with the Hilbert series for $G$-invariants, $\HS_\I^{U(1),\, 2\times\left(+1, -1\right)}$ in \cref{eqn:HSU12}, we see that they share the same denominator. This should not come as a surprise since one can multiply any covariant of charge $+2$ by a polynomial of the primary invariants and obtain another covariant of charge $+2$. However, a new feature is that the coefficients in the numerator are no longer all positive, indicating that this set of polynomials do not have a Hironaka decomposition. This is actually a quite common feature for group covariant polynomial sets in general, as can be seen from the various examples tabulated in \cref{tab:U1,tab:SU2adj,tab:SU3ffbar,tab:SU3adj} in \cref{appsec:Examples}. We will explain this feature below, but let us emphasize here that once fully expanded, the Hilbert series $\HS_R^{G,\, \RBB}$ must still have all of its coefficients positive, because by the construction in \cref{eqn:MolienR}, those coefficients are $n_R^{}(k)$, the number of $G$-covariant rep-$R$ polynomials of $\BB$ at each total degree $k$. This expectation is indeed satisfied by the example here, as can be seen by rewriting it as
\begin{equation}
\HS_{+2}^{U(1),\, 2\times\left(+1, -1\right)} = \frac{3q^2 -q^4}{\left( 1 - q^2 \right)^3}
= \frac{2q^2}{\left( 1 - q^2 \right)^3} + \frac{q^2}{\left( 1 - q^2 \right)^2} \,.
\end{equation}

\subsubsection{Structure of group covariants: a module over invariants}
\label{subsubsec:Module}

To better understand the general structure of group-covariant polynomials implied by the Hilbert series, such as that given in \cref{eqn:HS2U12}, especially the origin and interpretation of the negative terms in the numerator, let us look into their mathematical nature a little bit.

Given a symmetry group $G$ and a set of building blocks $\BB$, we note that a $G$-covariant rep-$R$ polynomial of $\BB$ generically comes with multiple components, like a vector.\footnote{This is in contrast with a $G$-invariant polynomial of $\BB$, which has only one component, since invariant irreps have dimension one.}
The set of these vectors, which we will denote by $\Mod_R^{G,\, \RBB}$, is not a ring in general, because it is not equipped with a multiplication. However, it is closed under linear combinations with coefficients that are drawn from a ring of $G$ invariants, such as $\rI$ defined in \cref{eqn:rI}:
\begin{equation}
\eR_i \in \rI \,,\quad
\eM_i \in \Mod_R^{G,\, \RBB}
\qquad\Longrightarrow\qquad
\sum_i \eR_i\, \eM_i \in \Mod_R^{G,\, \RBB} \,.
\label{eqn:MRrI}
\end{equation}
In this case, it is said that $\Mod_R^{G,\,\RBB}$ is a \emph{module} over the ring $\rI$, or simply an $\rI$-module. Mathematically, a \emph{module over a ring} is a notion similar to a vector space. It is a set of vectors closed under linear combinations, except that the combination coefficients are not numbers, but elements in the ring specified.\footnote{The elements of the ring need not have an inverse under multiplication. If they do have an inverse (and the multiplication is commutative) the ring would also constitute a \emph{field}, and the module would be a true vector space.}

\subsubsection*{Modules over different rings}

It is often possible to view a given set of vectors as (different) modules over different rings --- the nature of the module depends on the choice of the ring. For example, we have explained above that the set of rep-$R$ covariants $\Mod_R^{G,\, \RBB}$ can be viewed as a module over the ring of all $G$ invariants $\rI$, because the defining condition in \cref{eqn:MRrI} holds. However, the same condition remains true if one restricts the linear combination coefficients $\eR_i$ to the subring
\begin{equation}
\rPI \equiv \mathbb{C} \big[ P_1, P_2, \cdots, P_{n_P} \big] \subset \rI \,,
\label{eqn:rPI}
\end{equation}
which is formed by all the polynomials of the primary invariants (with $n_P^{}$ denoting the number of primary invariants, as in \cref{eqn:DqGeneral}). That is, we have
\begin{equation}
\eR_i \in \rPI \,,\quad
\eM_i \in \Mod_R^{G,\, \RBB}
\qquad\Longrightarrow\qquad
\sum_i \eR_i\, \eM_i \in \Mod_R^{G,\, \RBB} \,.
\label{eqn:MRrPI}
\end{equation}
This is saying that $\Mod_R^{G,\, \RBB}$ can also be viewed as a module over the ring $\rPI$. We introduce the following notations to avoid ambiguity
\begin{subequations}\label{eqn:TwoModules}
\begin{alignat}{2}
\Mod_R^{G,\, \RBB} &\;\;\text{viewed as an $\rI$-module} &&\qquad\Longrightarrow\qquad
{}_\rI\Mod_R^{G,\, \RBB} \,, \\[5pt]
\Mod_R^{G,\, \RBB} &\;\;\text{viewed as an $\rPI$-module} &&\qquad\Longrightarrow\qquad
{}_\rPI\Mod_R^{G,\, \RBB} \,.
\end{alignat}
\end{subequations}
Some common features of these two modules, as well as their differences will be discussed later.

\subsubsection*{General properties of a module}

The defining feature of a module is that it is equipped with a linear combination operation. The nature of this operation of course rests on the choice of the ring $\ring$, from which the linear combination coefficients are drawn. Once it is specified, the linear combination operation in an $\ring$-module $\Mod$ has several general definitions and properties that will be relevant for our discussions later:
\begin{itemize}
\item \textbf{Generating set:} A set of vectors in $\Mod$ is called a \emph{generating set} if every vector of $\Mod$ is a (finite) linear combination of them. If there exists a finite generating set, then $\Mod$ is said to be finitely generated.
\item \textbf{Linear (in)dependence:} A finite set of vectors in $\Mod$ are said to be linearly dependent if there exists a nontrivial linear combination of them that vanishes. Otherwise, they are said to be linearly independent.
\item \textbf{Basis and free module:} For a finitely generated module, a linearly independent generating set is called a \emph{basis}. However, this is not guaranteed to exist, which is a crucial difference with a vector space. In a vector space, if a finite generating set $\{\eM_1,\, \eM_2,\, \cdots,\, \eM_m\}$ is linearly dependent:
\begin{equation}
\eR_1\, \eM_1 + \eR_2\, \eM_2 + \cdots + \eR_m\, \eM_m = 0 \,,
\label{eqn:eReMrelation}
\end{equation}
with at least one coefficient nonzero --- say $\eR_m \ne 0$ without loss of generality, then one can divide the above relation by $\eR_m$ to write $\eM_m$ into a linear combination of the remaining set $\{\eM_1, \cdots, \eM_{m-1}\}$. It means that the remaining set works as a generating set. One can keep reducing the generating set this way, until it is linearly independent, and hence a basis. For a module, however, the coefficients are not numbers; they are elements of a ring, for which a division may not be defined. So \cref{eqn:eReMrelation} with $\eR_m \ne 0$ does not guarantee that $\eM_m$ can be written as a linear combination of the rest $\{\eM_1,\cdots, \eM_{m-1}\}$. This is to be contrasted with the algorithm in vector space described above. Therefore, a minimal generating set of a module may still be linearly dependent. On the other hand, if a module has a basis, it is called a \emph{free module}.
\item \textbf{Rank of finitely generated modules:} In this paper, we define the rank of a module $\Mod$ as the maximal number of linearly independent vectors in it. To obtain the rank constructively, it is useful to introduce a concept of \emph{filling set}. We call a finite set of vectors $\{\eM_1, \cdots, \eM_k\}$ in $\Mod$ a \emph{filling set} if for any vector $\eM\in\Mod$, there exists a linear relation in which the coefficient of $\eM$ is nonzero:
\begin{equation}
\eR\, \eM + \tilde\eR_1\, \eM_1 + \cdots + \tilde\eR_k\, \eM_k = 0 \,,\qquad
\text{with}\quad
\eR \ne 0 \,.
\label{eqn:FillingSetdef}
\end{equation}
By definition, adding any vector of $\Mod$ to a filling set will make it linearly dependent. A filling set ``fills'' the module in this sense. However, a filling set can be linearly dependent itself. If one finds a linearly independent filling set, then its cardinality gives the rank of the module.\footnote{One can show that linearly independent filling sets must all have the same cardinality.}
This can always be achieved for a finitely generated module --- one can start with a finite generating set, which by definition is a filling set, and then keep reducing it to a linearly independent one. The reduction procedure goes as follows. Suppose a filling set $\{\eM_1, \cdots, \eM_k\}$ is linearly dependent. Without loss of generality, we can assume
\begin{equation}
\eR_1\, \eM_1 + \cdots + \eR_k\, \eM_k = 0 \,,\qquad
\text{with}\quad
\eR_k \ne 0 \,.
\label{eqn:eReMzero}
\end{equation}
Then for any vector $\eM\in\Mod$, combining \cref{eqn:FillingSetdef,eqn:eReMzero} allows us to eliminate $\eM_k$ to find a linear relation among $\eM$ and the remaining set $\{\eM_1, \cdots, \eM_{k-1}\}$:
\begin{equation}
\eR_k\, \eR\, \eM + \left( \eR_k \tilde\eR_1 - \tilde\eR_k \eR_1 \right) \eM_1 + \cdots
+ \left( \eR_k \tilde\eR_{k-1} - \tilde\eR_k \eR_{k-1} \right) \eM_{k-1} = 0 \,,
\label{eqn:vkm1}
\end{equation}
in which the coefficient of $\eM$ is nonzero $\eR_k\, \eR \ne 0$. Therefore, the remaining set $\{\eM_1, \cdots, \eM_{k-1}\}$ is still a filling set. Following this procedure, one can keep reducing a filling set to obtain a linearly independent one.
\item \textbf{Cardinality of a minimal generating set:} The number of vectors in a minimal generating set is greater or equal to the rank of the module. The equality holds for free modules for which a minimal generating set is a basis.
\end{itemize}

\subsubsection*{Illustrating examples of modules}

To illustrate the above general concepts and properties of a module, let us revisit the example discussed in \cref{subsubsec:SimpleExample}. In this example, the symmetry group is $G=U(1)$, and the building blocks are two complex scalars $\{\phi_1, \phi_1^*, \phi_2, \phi_2^*\}$ (each with charge $\pm 1$). We have studied the set of group invariants $\Mod_\I^{U(1),\, 2\times(+1,-1)} = \rI$, as well as the set of charge $+2$ group covariants $\Mod_{+2}^{U(1),\, 2\times(+1,-1)}$. The corresponding Hilbert series were computed in \cref{eqn:HSU12,eqn:HS2U12} respectively:
\begin{equation}
\HS_\I^{U(1),\, 2\times\left(+1, -1\right)} = \frac{1+q^2}{\left( 1 - q^2 \right)^3} \,,\qquad
\HS_{+2}^{U(1),\, 2\times\left(+1, -1\right)} = \frac{3q^2-q^4}{\left( 1 - q^2 \right)^3} \,.
\end{equation}
Both $\Mod_\I^{U(1),\, 2\times(+1,-1)}$ and $\Mod_{+2}^{U(1),\, 2\times(+1,-1)}$ can be viewed as modules. Below, let us check their properties regarding the linear combination operations.

Let us begin with $\Mod_\I^{U(1),\, 2\times(+1,-1)}$. We recall from \cref{eqn:PSU12} that there are three primary invariants and one secondary invariant, which can be taken as
\begin{equation}
P_1 = \phi_1 \phi_1^* \,,\quad
P_2 = \phi_2 \phi_2^* \,,\quad
P_3 = \phi_1 \phi_2^* + \phi_2 \phi_1^* \,,\quad
  S = \phi_1 \phi_2^* - \phi_2 \phi_1^* \,.
\label{eqn:PSU12Redo}
\end{equation}
The denominator in the Hilbert series $\HS_\I^{U(1),\, 2\times\left(+1, -1\right)}$ encodes the ring $\rPI$ --- the set of polynomials in primary invariants:
\begin{equation}
\rPI = \mathbb{C} \big[ P_1, P_2, P_3\big]
\qquad\Longleftrightarrow\qquad
\frac{1}{D(q)} = \frac{1}{\left( 1 - q^2 \right)^3} \,.
\label{eqn:rPIU12}
\end{equation}
Then the set of all invariants $\Mod_\I^{U(1),\, 2\times(+1,-1)} = \rI$ can be viewed as an $\rPI$-module
\begin{equation}
{}_\rPI\Mod_\I^{U(1),\, 2\times\left(+1, -1\right)}:\qquad
\rI =  \rPI \otimes \left( 1 \oplus S \right) = \rPI \mqty(1\\S) \,.
\label{eqn:rPIModInvU12}
\end{equation}
As indicated here, the two vectors\footnote{For the module discussed here, the vectors are one dimensional.}
$\{1, S\}$ can be taken as a generating set of this module: every vector in $\Mod_\I^{U(1),\, 2\times(+1,-1)}$ can be obtained by a linear combination of these two vectors with coefficients in $\rPI$; this is nothing but the statement of Hironaka decomposition (\cref{eqn:Hironaka}). Since $\{1, S\}$ is a finite set, the module ${}_\rPI\Mod_\I^{U(1),\, 2\times\left(+1, -1\right)}$ is finitely generated. Furthermore, the two vectors $\{1, S\}$ are linearly independent due to the uniqueness of the Hironaka decomposition. Therefore, they also form a basis of the module, meaning that ${}_\rPI\Mod_\I^{U(1),\, 2\times\left(+1, -1\right)}$ is a free module. The rank of this module is given by the cardinality of the basis
\begin{equation}
\rank \left( {}_\rPI\Mod_\I^{U(1),\, 2\times\left(+1, -1\right)} \right) = 2 \,.
\end{equation}
The properties of this module is encoded in the numerator of the Hilbert series:
\begin{equation}
N_\I (q) = 1 + q^2 \,.
\end{equation}
Specifically, it reflects the structure of the basis, and evaluating it at $q=1$ gives the rank of the module $N_\I(q=1)=2$.

Alternatively, the set of all invariants $\Mod_\I^{U(1),\, 2\times(+1,-1)} = \rI$ can also be viewed as an $\rI$-module
\begin{equation}
{}_\rI\Mod_\I^{U(1),\, 2\times\left(+1, -1\right)}:\qquad
\rI =  \rI \otimes \mqty(1) \,.
\label{eqn:rIModInvU12}
\end{equation}
As indicated here, a generating set of this module is simply the trivial vector $\{1\}$, because any vector in $\rI$ can be obtained by a linear combination of it --- just taking the element itself as the combination coefficient. The generating set $\{1\}$ is a finite set and also linearly independent; it forms a basis. Therefore the module ${}_\rI\Mod_\I^{U(1),\, 2\times\left(+1, -1\right)}$ is finitely generated and also free. Clearly, its rank is
\begin{equation}
\rank \left( {}_\rI\Mod_\I^{U(1),\, 2\times\left(+1, -1\right)} \right) = 1 \,.
\end{equation}
It is instructive to note that in this module, the vector set $\{1, S\}$ is linearly dependent, because combination coefficients are now from the whole $\rI$ (instead of within $\rPI$):
\begin{equation}
S \cdot 1 - 1 \cdot S = 0 \,.
\end{equation}

Now let us move on to $\Mod_{+2}^{U(1),\, 2\times(+1,-1)}$, the set of vectors that have charge $+2$ under the group $G=U(1)$. Since it is closed under linear combinations with coefficients drawn from $\rPI = \mathbb{C} \big[ P_1, P_2, P_3 \big]$, it can be viewed as an $\rPI$-module:
\begin{equation}
{}_\rPI\Mod_{+2}^{U(1),\, 2\times(+1,-1)} \,.
\end{equation}
To find a generating set of it, we recall that any vector $\eM$ therein is a polynomial in $\BB=\{\phi_1, \phi_1^*, \phi_2, \phi_2^*\}$, which to have charge $+2$, must have the following decomposition
\begin{equation}
\eM = f_1(\BB)\, \phi_1^2 + f_2(\BB)\, \phi_2^2 + f_3(\BB)\, \phi_1\phi_2 \,,
\label{eqn:eMf3}
\end{equation}
with $f_1(\BB), f_2(\BB), f_3(\BB)$ some neutral (charge $0$) factors. For convenience, let us denote the three vectors in this decomposition as
\begin{equation}
\eM_1 = \phi_1 \phi_1 \,,\qquad
\eM_2 = \phi_2 \phi_2 \,,\qquad
\eM_3 = \phi_1 \phi_2 \,.
\label{eqn:eMsU12}
\end{equation}
The neutral factors in \cref{eqn:eMf3} are also polynomials in $\BB$, which can be written into polynomials in the primary and secondary invariants $\{P_1, P_2, P_3, S\}$. However, the following relations
\begin{subequations}\label{eqn:SeMtoPeM}
\begin{align}
S\, \eM_1 &=  P_3\, \eM_1 - 2 P_1\, \eM_3 \,, \\[5pt]
S\, \eM_2 &= -P_3\, \eM_2 + 2 P_2\, \eM_3 \,, \\[5pt]
S\, \eM_3 &=  P_2\, \eM_1 -   P_1\, \eM_2 \,,
\end{align}
\end{subequations}
allow us to eliminate the secondary invariant $S$ in favor of primary invariants $P_1,P_2,P_3$. Therefore, one can always reorganize the decomposition in \cref{eqn:eMf3} into a new linear combination of $\eM_1, \eM_2, \eM_3$ with coefficients drawn only from $\rPI = \mathbb{C} \big[ P_1, P_2, P_3 \big]$:
\begin{equation}
\eM = \eR_1\, \eM_1 + \eR_2\, \eM_2 + \eR_3\, \eM_3 \,,
\quad\text{with}\quad
\eR_1, \eR_2, \eR_3 \in \rPI \,.
\end{equation}
Therefore, $\{\eM_1, \eM_2, \eM_3\}$ is a generating set of the module ${}_\rPI\Mod_{+2}^{U(1),\, 2\times(+1,-1)}$. This explains the absence of a factor $(1+q^2)$ in the numerator of \cref{eqn:HS2U12}. As this is a finite set, the module is finitely generated. However, this generating set is not linearly independent (hence not a basis). There is one linear relation among them:
\begin{equation}
P_3\, \eM_3 = P_2\, \eM_1 + P_1\, \eM_2 \,.
\label{eqn:eMRelationU12}
\end{equation}
To find the rank of the module, we begin with the generating set $\{\eM_1, \eM_2, \eM_3\}$ as a filling set. Because of the linear relation above (and $P_3\ne0$), this filling set can be reduced to a smaller filling set $\{\eM_1, \eM_2\}$, which one can then check is linearly independent. This tells us the rank:
\begin{equation}
\rank \left( {}_\rPI\Mod_{+2}^{U(1),\, 2\times\left(+1, -1\right)} \right) = 2 \,.
\end{equation}
On the other hand, one can also check that the generating set $\{\eM_1, \eM_2, \eM_3\}$ is a minimal generating set of the module. It has cardinality $3$, larger than the rank. This indicates that the module is not a free module. The above properties of the module ${}_\rPI\Mod_{+2}^{U(1),\, 2\times\left(+1, -1\right)}$ are encoded in the numerator of the Hilbert series:
\begin{equation}
N_{+2} (q) = 3q^2 - q^4 \,.
\end{equation}
Specifically, the $+3q^2$ term corresponds to the minimal generating set $\{\eM_1, \eM_2, \eM_3\}$, and the $-q^4$ term reflects the one linear redundancy relation among them (\cref{eqn:eMRelationU12}). Evaluating the numerator at $q=1$ gives the rank of the module $N_{+2}(q=1)=2$.

Alternatively, $\Mod_{+2}^{U(1),\, 2\times\left(+1,-1\right)}$ can also be viewed as an $\rI$-module:
\begin{equation}
{}_\rI\Mod_{+2}^{U(1),\, 2\times\left(+1, -1\right)} \,.
\end{equation}
The decomposition in \cref{eqn:eMf3} then directly claims $\{\eM_1, \eM_2, \eM_3\}$ as a generating set for this module. As before, this is a finite set, meaning that ${}_\rI\Mod_{+2}^{U(1),\, 2\times\left(+1, -1\right)}$ is finitely generated. However, we now have more linear redundancies among these generating vectors, because both \cref{eqn:SeMtoPeM,eqn:eMRelationU12} count as linear relations when we allow the combination coefficients to be drawn from $\rI$ (instead of within $\rPI$). Working through the filling set reduction procedure, one now finds a linearly independent filling set as just $\{\eM_1\}$, meaning that
\begin{equation}
\rank \left( {}_\rI\Mod_{+2}^{U(1),\, 2\times\left(+1, -1\right)} \right) = 1 \,.
\end{equation}

\subsubsection*{Extracting information of the module from Hilbert series}

In general, a set of group covariants $\Mod_R^{G,\, \RBB}$ can be viewed as an $\rPI$-module, as well as an $\rI$-module. We emphasize again that they are the same vector set, but different modules. This is because what are allowed to be the linear combination coefficients rests on the choice of ring, and therefore so does the notion of linear dependence among vectors and the rank of the module, \etc. For a linearly reductive group $G$, it is known (\eg\ \cite{broer1994hilbert}) that both modules are finitely generated. In below, we explain the connection between the module ${}_\rPI\Mod_R^{G,\, \RBB}$ and the Hilbert series $\HS_R^{G,\, \RBB}$. In the next subsection, we will discuss how to extract information about the other module ${}_\rI\Mod_R^{G,\, \RBB}$ using (multiple) Hilbert series.

The structure of the module ${}_\rPI\Mod_R^{G,\, \RBB}$ is encoded in the numerator of the Hilbert series $\HS_R^{G,\, \RBB}$. We remind the reader that there could be different choices of $\rPI$, corresponding to different denominator options $D(q)$, and the numerator will change accordingly. For each given choice of $\rPI$, the corresponding numerator reflects the properties of the module ${}_\rPI\Mod_R^{G,\, \RBB}$.

First, this applies to the special case $R=\I$ (the invariant rep) --- the set of $G$-invariant $\BB$ polynomials $\Mod_\I^{G,\, \RBB}$ can be viewed as a module over $\rPI$, namely ${}_\rPI\Mod_\I^{G,\, \RBB}$. In fact, this precisely follows from the Hironaka decomposition in \cref{eqn:Hironaka}, with its existence implying that the secondary invariants (together with $S_0=1$) form a generating set of the module, and its uniqueness requiring them to be linearly independent, making them a basis. Therefore, ${}_\rPI\Mod_\I^{G,\, \RBB}$ is a free module. We recall from \cref{eqn:HSInv} that the corresponding Hilbert series has the general form
\begin{equation}
\HS_\I^{G,\, \RBB} (q) = \frac{N_\I(q)}{D(q)} \,.
\label{eqn:HSInvRedo}
\end{equation}
Clearly, the factor $1/D(q)$ is generating the ring $\rPI$, and the term-by-term positive numerator $N_\I(q)$ encodes the basis of the free module. Therefore, the rank of the module, \ie\ the number of vectors in the basis, can be computed by evaluating the numerator at $q=1$:
\begin{equation}
\rank \left( {}_\rPI\Mod_\I^{G,\, \RBB} \right) = N_\I (q=1) \,.
\label{eqn:rankModInvrPI}
\end{equation}
Note that the ring $\rPI$ can also be encoded by a Hilbert series
\begin{equation}
\HS_\PI^{G,\, \RBB} (q) = \frac{1}{D(q)} \,.
\label{eqn:HSrPI}
\end{equation}
With this, one can rewrite the rank formula above as
\begin{equation}
\rank \left( _{\rPI}\Mod_\I^{G,\, \RBB} \right)
= \frac{\HS_\I^{G,\, \RBB}(q)}{\HS_\PI^{G,\, \RBB}(q)} \Bigg|_{q=1} \,.
\label{eqn:rankHSRatiosInv}
\end{equation}

Next, let us move on to a general irrep $R$. It is known (\eg\ \cite{broer1994hilbert}) that the Hilbert series $\HS_R^{G,\, \RBB}(q)$ also has a rational form, similar with that in \cref{eqn:HSInvRedo}:
\begin{equation}
\HS_R^{G,\, \RBB}(q) = \frac{N_R(q)}{D(q)} \,,
\label{eqn:HSR}
\end{equation}
where the denominator $D(q)$ is the same as that of $\HS_\I^{G,\, \RBB}(q)$, and the numerator $N_R(q)$ is a polynomial with integer coefficients. However, the module $_{\rPI}\Mod_R^{G,\, \RBB}$ may not be a free module, and a minimal generating set could contain more vectors than the rank of the module. When this happens, one has to include the full generating set in $N_R(q)$, while their linear relations are redundancies that need to be subtracted at higher degrees, leading to the negative terms in $N_R(q)$. The numerator $N_R(q)$ \emph{can be written} in a way such that the positive terms indicate the degrees and number of the elements of a generating set, while the negative terms describe redundancies (\ie\ linear dependencies) among them. Evaluating the numerator at $q=1$ again gives us the rank of the module:
\begin{equation}
\rank \left( {}_\rPI\Mod_R^{G,\, \RBB} \right) = N_R (q=1) \,.
\label{eqn:rankModRrPI}
\end{equation}
The rank formula in \cref{eqn:rankModInvrPI} is the special case of taking $R=\I$ here. Again similar to \cref{eqn:rankHSRatiosInv}, this rank formula can be written as a ratio of two Hilbert series:
\begin{equation}
\rank \left( _{\rPI}\Mod_R^{G,\, \RBB} \right)
= \frac{\HS_R^{G,\, \RBB}(q)}{\HS_\PI^{G,\, \RBB}(q)} \Bigg|_{q=1} \,.
\label{eqn:rankHSRatiosR}
\end{equation}

As mentioned before, studying the other module ${}_{\rI}\Mod_R^{G,\, \RBB}$ is the subject of the next subsection. But let us quickly advance here that its rank can be computed in a very similar way, by taking a ratio of two Hilbert series:
\begin{equation}
\rank \left( _{\rI}\Mod_R^{G,\, \RBB} \right)
= \frac{\HS_R^{G,\, \RBB}(q)}{\HS_\I^{G,\, \RBB}(q)} \Bigg|_{q=1} \,.
\label{eqn:rankHSRatiosRrIadv}
\end{equation}

\subsection{Independent group covariants and rank saturation}
\label{subsec:RankSaturation}

In \cref{subsec:Covariants}, we explained how to compute the Hilbert series $\HS_R^{G,\, \RBB}$ for a general set of rep-$R$ group covariants $\Mod_R^{G,\, \RBB}$, as well as how to interpret the results. Specifically, the summed Hilbert series has a rational form $\HS_R^{G,\, \RBB}(q) = \frac{N_R(q)}{D(q)}$, with the denominator $D(q)$ encoding a set of primary invariants, and the factor $\frac{1}{D(q)}$ corresponding to their polynomial ring $\rPI$. The numerator $N_R(q)$ then depicts the structure of $\Mod_R^{G,\, \RBB}$ viewed as an $\rPI$-module, \ie\ ${}_\rPI\Mod_R^{G,\, \RBB}$, indicating a generating set (positive terms) and the redundancies or linear relations in it (negative terms, if any). Evaluating the numerator at $q=1$ gives the rank of this module. The special case of group invariants is captured by taking $R=\I$, the invariant (trivial) representation.

Alternatively, the set of rep-$R$ group covariants $\Mod_R^{G,\, \RBB}$ can also be viewed as an $\rI$-module, \ie\ ${}_\rI\Mod_R^{G,\, \RBB}$.\footnote{As explained around \cref{eqn:MRrI}, $\Mod_R^{G,\, \RBB}$ can be viewed as a module over \emph{any} ring of $G$ invariants. When the chosen ring is too small, the module may not be finitely generated.}
We explained this in \cref{subsec:Covariants}, but did not discuss how to obtain information about this module from the Hilbert series. In this section, we attack this problem.

We are interested in this module ${}_\rI\Mod_R^{G,\, \RBB}$ (in addition to ${}_\rPI\Mod_R^{G,\, \RBB}$) because it reflects a notion of ``independent'' group covariants in the following sense. Any specific rep-$R$ group covariant vector $\eM \in \Mod_R^{G,\, \RBB}$ would induce a set of them --- $\rI \otimes \{v\}$, by multiplying it with all the invariants $\rI$. Therefore, we are interested in some sort of quotient ``$\Mod_R^{G,\, \RBB} / \rI$'', which vaguely speaking, would capture a set of ``independent'' such inducing vectors. A mathematically well-defined version of such a quotient is the module ${}_\rI\Mod_R^{G,\, \RBB}$, and the ``independent'' inducing vectors are a set of linearly independent vectors in this module. The rank of the module tells us how many of them there could be.

On general ground, it is known \cite{broer1994hilbert} that the module ${}_\rI\Mod_R^{G,\, \RBB}$ is finitely generated. But a generating set of it might have complicated linear relations, especially when there are secondary invariants in $\rI$. Remarkably, one can use Hilbert series to compute the rank of this module without an explicit knowledge of those linear relations. Below we explain.

\subsubsection{Computing the rank of ${}_\rI\Mod_R^{G,\, \RBB}$ with Hilbert series}
\label{subsubsec:ComputingRank}

As already mentioned in \cref{eqn:rankHSRatiosRrIadv}, the rank of the module ${}_\rI\Mod_R^{G,\, \RBB}$ can be computed by taking a ratio of two Hilbert series:
\begin{equation}
\rank \left( _{\rI}\Mod_R^{G,\, \RBB} \right)
= \frac{\HS_R^{G,\, \RBB}(q)}{\HS_\I^{G,\, \RBB}(q)} \Bigg|_{q=1}
= \frac{N_R(q=1)}{N_\I(q=1)} \,.
\label{eqn:rankHSRatiosRrI}
\end{equation}
This is due to the following relation between the ranks of the two modules:
\begin{equation}
\rank\left( _{\rPI}\Mod_R^{G,\, \RBB} \right)
= \left( n_S^{} + 1 \right)\, \rank\left( _{\rI}\Mod_R^{G,\, \RBB} \right) \,,
\label{eqn:rankRelation}
\end{equation}
where $n_S^{}$ is the number of secondary invariants (as in \cref{eqn:Hironaka}). Note from \cref{eqn:NqGeneral} that $N_\I(q=1)=n_S^{}+1$. Combining this with the rank formula in \cref{eqn:rankModRrPI}, one can readily derive \cref{eqn:rankHSRatiosRrI} from \cref{eqn:rankRelation}. Therefore, all we need to explain is the rank relation in \cref{eqn:rankRelation}. We do so below.

First, let a set of vectors $\{\eM_1, \eM_2, \cdots, \eM_k\}$ in $\Mod_R^{G,\, \RBB}$ be a linearly independent filling set for the module $_{\rI}\Mod_R^{G,\, \RBB}$. The cardinality of this filling set then gives the rank of this module:
\begin{equation}
\rank\left( _{\rI}\Mod_R^{G,\, \RBB} \right) = k \,.
\label{eqn:rankk1}
\end{equation}
Now the key observation is that the following set of vectors in $\Mod_R^{G,\, \RBB}$
\begin{align}
&\big\{ 1, S_1, S_2, \cdots, S_{n_S^{}} \big\} \otimes \big\{ \eM_1, \eM_2, \cdots, \eM_k \big\}
\notag\\[5pt]
&\hspace{30pt}
= \Big\{ \eM_1, S_1\eM_1, S_2\eM_1, \cdots, S_{n_S^{}}\eM_1,\, \cdots\cdots,\, \eM_k, S_1\eM_k, S_2\eM_k, \cdots, S_{n_S^{}}\eM_k \Big\} \,,
\end{align}
gives a linearly independent filling set for the module $_{\rPI}\Mod_R^{G,\, \RBB}$.\footnote{One can check that this set of vectors are linearly independent and also fill the module $_{\rPI}\Mod_R^{G,\, \RBB}$. Here we skip the details of showing these.} The cardinality of this set is $\left( n_S^{} + 1 \right) k$, which means
\begin{equation}
\rank\left( _{\rPI}\Mod_R^{G,\, \RBB} \right) = \left( n_S^{} + 1 \right) k
= \left( n_S^{} + 1 \right) \rank\left( _{\rI}\Mod_R^{G,\, \RBB} \right) \,.
\label{eqn:rankk2}
\end{equation}
This is the rank relation in \cref{eqn:rankRelation}. As explained above, this in turn will prove the rank formula in \cref{eqn:rankHSRatiosRrI}.

Since we know how to compute the Hilbert series $\HS_R^{G,\, \RBB}(q)$ and $\HS_\I^{G,\, \RBB}(q)$ with Molien-Weyl formula, we can use \cref{eqn:rankHSRatiosRrI} to compute the rank of the module ${}_\rI\Mod_R^{G,\, \RBB}$. We have computed a variety of examples, tabulated in
\cref{tab:U1,tab:SU2adj,tab:SU3ffbar,tab:SU3adj} in \cref{appsec:Examples}. From these results, we observe that $\rank \big( {}_\rI\Mod_R^{G,\, \RBB} \big) \leq \dim(R)$, with the equality occurring in most cases, but not all of them.

\subsubsection{A bound on the rank of ${}_\rI\Mod_R^{G,\, \RBB}$}
\label{subsubsec:RankBound}

For connected semisimple Lie groups, there is an upper bound on the rank of the module ${}_\rI\Mod_R^{G,\, \RBB}$ --- it cannot be larger than the dimension of the representation $R$, which is the number of components that each vector of it has:
\begin{equation}
\rank \left( {}_\rI\Mod_R^{G,\, \RBB} \right) \le \dim(R) \,.
\label{eqn:rankBound}
\end{equation}
When the equal sign is realized, we say that there is a \emph{\textbf{rank saturation}} for the rep-$R$ group covariants made out of $\BB$. This implies that the building blocks $\BB$ can generate the most general rep-$R$ group covariants.

The proof of \cref{eqn:rankBound} proceeds by contradiction --- a set of vectors with elements more than $\dim(R)$ must be linearly dependent. Let us assume that we are given a set of $\dim(R)$ linearly independent vectors:
\begin{equation}
\{\eM_1, \eM_2, \cdots, \eM_k\} \,,
\qquad\text{with}\quad
k=\dim(R) \,.
\label{eqn:eMset}
\end{equation}
We will show that adding one more vector to it will make the set linearly dependent. We start by constructing a $k\times k$ matrix
\begin{equation}
A \equiv \mqty(\eM_1 & \eM_2 & \cdots & \eM_k) \,,
\label{eqn:matrixA}
\end{equation}
which is invertible, as $\det(A)\ne 0$ due to the linear independence of the set in \cref{eqn:eMset}. Then we can construct a matrix $B$ using the first minors of $A$, denoted $M(A)_{ij}$ below:
\begin{equation}
B_{ij} \equiv (-1)^{i+j} M(A)_{ji} \,.
\label{eqn:matrixB}
\end{equation}
From the standard procedure of constructing the inverse of matrix $A$, we know that the matrix $B$ satisfies
\begin{equation}
AB = BA = \det(A)\, \mathbb{1}_{k\times k}\,.
\label{eqn:ABBA}
\end{equation}
Note that the matrix elements of $A$ and $B$, as well as the determinant $\det(A)$ are all polynomials in
$\BB$. Now consider adding one more vector $\eM \in {}_\rI\Mod_R^{G,\, \RBB}$ to the set in \cref{eqn:eMset}. Using \cref{eqn:ABBA}, we get
\begin{equation}
A \left(B\, \eM\right) - \det\left( A \right) \eM = 0 \,,
\end{equation}
which is a nontrivial linear relation among $\{\eM_1, \cdots, \eM_k, \eM\}$:
\begin{equation}
\eR_1\, \eM_1 + \cdots + \eR_k\, \eM_k + \eR\, \eM = 0 \,,
\qquad\text{with}\quad
\mqty(\eR_1 \\ \vdots \\ \eR_k) = B\, \eM \,,\quad
\eR = -\det(A) \,.
\label{eqn:eMrelation}
\end{equation}
To complete the proof, we must also show that the linear combination coefficients $\eR_i$ and $\eR$ here are group invariants, \ie\ elements of the ring $\rI$. To this end, we study how the various quantities in \cref{eqn:eMrelation} transform under the group $G$. First, recall that $\eM_i$ and $\eM$ are rep-$R$ vectors of the group $G$:
\begin{subequations}
\begin{align}
\eM_i &\quad\longrightarrow\quad
g_R^{}\, \eM_i \,, \\[5pt]
\eM &\quad\longrightarrow\quad
g_R^{}\, \eM \,,
\end{align}
\end{subequations}
where $g_R^{}$ denotes the representation matrix of the group element $g\in G$ in the representation $R$ (following our notation established in \cref{eqn:gRBB}). Then by the constructions in \cref{eqn:matrixA,eqn:matrixB}, we have
\begin{subequations}
\begin{align}
A &\quad\longrightarrow\quad
g_R^{}\, A \,, \\[5pt]
\det\left(A\right) &\quad\longrightarrow\quad
\det\left(g_R^{}\right) \det\left(A\right) \,, \\[5pt]
B &\quad\longrightarrow\quad
\det\left(g_R^{}\right) B\, g_R^{-1} \,.
\end{align}
\end{subequations}
It follows that the linear combination coefficients in \cref{eqn:eMrelation} transform as
\begin{subequations}\label{eqn:eRieRTrans}
\begin{align}
\eR_i &\quad\longrightarrow\quad
\det\left(g_R^{}\right) \eR_i \,, \\[5pt]
\eR &\quad\longrightarrow\quad
\det\left(g_R^{}\right) \eR \,,
\end{align}
\end{subequations}
Now if $G$ is a connected semisimple Lie group, we have $\det\left(g_R^{}\right)=1$, meaning that $\eR_i$ and $\eR$ are indeed elements of the ring $\rI$. In this case, \cref{eqn:eMrelation} shows that the set in \cref{eqn:eMset} supplemented by any additional vector $\eM \in {}_\rI\Mod_R^{G,\, \RBB}$ will become linearly dependent, meaning that no linearly independent set can have more than $\dim(R)$ vectors. This proves the bound stated in \cref{eqn:rankBound} for connected semisimple Lie groups.

The bound in \cref{eqn:rankBound} can be violated for groups containing $U(1)$ factors, as $\det\left(g_R^{}\right)$ can be nontrivial in this case. To see an example, let us consider the group $G=U(1)$ and the building blocks $\BB = \{ \phi_1, \cdots, \phi_n\}$, with $n>1$ and each $\phi_i$ having charge $+1$. Now suppose that we are interested in the set of group covariants in the representation $R=+1$. In this case, we have $\dim(R)=1$ but $\rank\left(_\rI\Mod_R^{G,\,\RBB}\right)=n>1$, violating the bound. However, the bound would apply if the building blocks in this example were supplemented by at least one charge $-1$ component, say $\chi$. In this case, even if the combination coefficients $\eR_i$ and $\eR$ found in \cref{eqn:eMrelation} transform nontrivially, due to $\det\left(g_R^{}\right)\ne 1$ in \cref{eqn:eRieRTrans}, one can always multiply the relation in \cref{eqn:eMrelation} by proper powers of $\chi$ such that the linear combination coefficients become group invariants. One can generalize this argument to show that the rank bound \cref{eqn:rankBound} applies to the case of $U(1)$ group as long as the building blocks $\BB$ contain both positive and negative charges (with all charges restricted to integers).

As mention before, we observe that the rank bound in \cref{eqn:rankBound} holds for all the examples that we explored in \cref{tab:U1,tab:SU2adj,tab:SU3ffbar,tab:SU3adj} in \cref{appsec:Examples}. As to rank saturation, we see that it occurs for all the $G=U(1)$ cases that we studied in \cref{tab:U1}. For the case of $G=SU(2)$, we have explored the scenarios of having $m$ adjoint reps ($\mathbf{3}$) as building blocks in \cref{tab:SU2adj}. We see that rank saturation occurs for bosonic (integer spin) reps for $m\ge2$, but not at $m=1$. For the case of $G=SU(3)$ with $m$ pairs of $\left( \mathbf{3} + \mathbf{\bar{3}} \right)$ reps as building blocks that we explored in \cref{tab:SU3ffbar}, rank saturation occurs for $m\ge2$, but not at $m=1$. For $G=SU(3)$ with $m$ adjoint reps ($\mathbf{8}$) as building blocks in \cref{tab:SU3adj}, rank saturation occurs for $m\geq2$, except for those representations with nonzero triality.

\subsubsection{Condition for rank saturation}
\label{subsubsec:Condition}

In \cref{subsubsec:RankBound}, we explained that there is an upper bound on the rank of the module $_\rI\Mod_R^{G,\,\RBB}$. We explored a variety of cases in \cref{tab:U1,tab:SU2adj,tab:SU3ffbar,tab:SU3adj} in \cref{appsec:Examples}, and observed that the bound is saturated for most of them, but not all. It calls for a more systematic understanding of when the rank saturation would occur. There is a very useful theorem by Brion \cite{brion1993modules} (see also \cite{broer1994hilbert}), which states that
\begin{equation}
\rank \left( {}_\rI\Mod_R^{G,\, \RBB} \right) = \dim \left( R^H \right) \,.
\label{eqn:TheoremBrion}
\end{equation}
Here $H$ denotes the unbroken subgroup of $G$ when the building blocks $\BB$ adopt a generic vev, and $R^H$ denotes the subspace of $R$ that is invariant under $H$. In other words, the representation $R$ of $G$ also forms a representation of its subgroup $H\subset G$, but typically a reducible one. One can decompose this reducible rep into irreps of $H$ and ask how many invariant irreps there are. This is the number $\dim(R^H)$.

With the theorem above, one can compute the rank of ${}_\rI\Mod_R^{G,\, \RBB}$ without computing the Hilbert series. Specifically, for a given group $G$ and a set of building blocks $\BB$, we first figure out the unbroken subgroup $H \subset G$, imagining that $\BB$ takes a generic vev. Then for a goal
representation $R$ of $G$ specified, we decompose it into irreps of the subgroup $H$ to work out the number of $H$ invariants in $R$. According to the theorem in \cref{eqn:TheoremBrion}, this number gives the rank of ${}_\rI\Mod_R^{G,\, \RBB}$. 

The theorem in \cref{eqn:TheoremBrion} also makes the bound in \cref{eqn:rankBound} manifest --- of course we have
\begin{equation}
\dim(R^H) \le \dim(R) \,.
\label{eqn:rankCondition}
\end{equation}
Now the condition for rank saturation is simply when the equality in \cref{eqn:rankCondition} is realized. In particular, when the building blocks $\BB$ break $G$ to the trivial subgroup $H=\Ge$, then we are guaranteed to have rank saturation for any goal representation $R$. When $H$ is nontrivial, it is also possible to have
$\dim(R^H) = \dim(R)$, depending on the choice of $R$. Examples of this include the case of $G=SU(3)$ with the building blocks $\BB$ being two or three adjoint reps listed in the last two columns of \cref{tab:SU3adj}, where we have a nontrivial unbroken subgroup $H=Z_3$.

One can verify that \cref{eqn:TheoremBrion} works for all of the ranks explored in \cref{tab:U1,tab:SU2adj,tab:SU3ffbar,tab:SU3adj} in \cref{appsec:Examples}. For example, consider the case of $G=SU(3)$ with a single octet $(\mathbf{8})$ as building blocks. This is the first column in \cref{tab:SU3adj}. The vev of a single octet can be brought into diagonal form by a $G$-transformation. The resulting diagonal matrix commutes with the two generators of the Cartan subalgebra, so that the unbroken subgroup is the Cartan subgroup $H=U(1)^2$. Denoting by $(Q_1,Q_2)$ these two $U(1)$ charges, the decompositions of rep $R$ under the subgroup $H$ follow from the familiar ``branching rules'':
\begin{subequations}
\begin{align}
\mathbf{3}  &\quad\longrightarrow\quad
(1,0) \oplus (-1,1) \oplus (0,-1) \,, \\[15pt]
\mathbf{6}  &\quad\longrightarrow\quad
(2,0) \oplus (-2,2) \oplus (0,-2) \oplus (-1,0) \oplus(1,-1) \oplus(0,1) \,, \\[15pt]
\mathbf{8}  &\quad\longrightarrow\quad
(0,0) \oplus (0,0) \oplus (1,-1) \oplus (-1,1) \oplus(2,-1) \oplus(-2,1)
\notag\\[5pt]
&\hspace{50pt}
\oplus(1,-2) \oplus(-1,2)
\,, \\[15pt]
\mathbf{10} &\quad\longrightarrow\quad
(0,0) \oplus (3,0) \oplus (1,1) \oplus (-3,3) \oplus(2,-1) \oplus(-1,1)
\notag\\[5pt]
&\hspace{50pt}
\oplus(-2,1) \oplus(-1,2)  \oplus(0,-3) \oplus(1,-2) \,.
\end{align}
\end{subequations}
We read off that there are $0, 0, 2, 1$ charge-less components in the cases $R = \mathbf{3}, \mathbf{6}, \mathbf{8}, \mathbf{10}$ respectively, in accord with the rank of the first column of \cref{tab:SU3adj}.\footnote{The rep $R=\mathbf{27}$ can be decomposed in a similar fashion, and it has three charge-less components. One can also see this from $\mathbf{6} \otimes \mathbf{\bar{6}} = \mathbf{27} \oplus \mathbf{8} \oplus \mathbf{1}$.}

\section{Most general Minimal Flavor Violation}
\label{sec:MFV}

In this section, we will apply the previously outlined techniques to count the number of distinct flavor structures that can arise within the framework of (quark) MFV. First, we will review the MFV hypothesis and systematically obtain all the different flavor representations arising in the quark-MFV SMEFT at dimension six. Next, we will compute the Hilbert series for these flavor covariants. The simplest ones among these will be cross checked with conventional methods by applying the Cayley-Hamilton theorem, where we also provide explicit expressions for (a choice of) the generating set of the corresponding modules.

The main conclusion of this section is that the sole assumption of \emph{MFV does not restrict the EFT} or in other words, one can obtain the most general SMEFT while being compatible with the MFV hypothesis:
\vspace{2mm}
\begin{tcolorbox}[colback=light-gray]
\begin{center}
\begin{minipage}{5.5in}\vspace{-5pt}
\begin{equation}
\text{MFV SMEFT} \equiv \text{SMEFT} \,.
\label{eqn:main}
\end{equation}
\end{minipage}
\end{center}
\end{tcolorbox}
The reason is that for the case of MFV, we do reach the \emph{rank saturation} explained in \cref{subsec:RankSaturation}, and thus there are as many linearly independent flavor structures as the dimension of the MFV representation. The physics then lies in the additional assumptions on top of MFV, \eg\ one may assume that the phenomenology is dominated by operators built only with $Y_u$, or by operators in the lowest power of $Y_u$ and $Y_d$ (treating Yukawas as counting orders in a perturbative expansion) \etc\

\subsection{MFV: The key hypothesis}
\label{subsec:KeyHypothesis}

The idea of Minimal Flavor Violation was introduced to alleviate the severe problems that models of physics beyond the SM face when confronted with data on neutral Kaon  mixing. In the SM these processes, mediated by so-called flavor changing neutral currents, are automatically suppressed by the GIM mechanism~\cite{Glashow:1970gm}. The mechanism can be understood as a consequence of the approximate $U(45)$ global symmetry\footnote{The SM contains 18 left-handed quarks, 18 right-handed quarks, 6 left-handed leptons and 3 right-handed leptons amounting to 45 Weyl fermions which present a $U(45)$ symmetry in the absence of interactions.} of the free SM \cite{Dugan:1984qf}, or, more pointedly, the subgroup of $U(45)$ that is still a symmetry even in the presence of gauge fields and their interactions~\cite{Chivukula:1987py}.  

The quark flavor group is defined as the classical global symmetry of the renormalizable Lagrangian of the SM quark sector in the limit of vanishing Yukawa couplings. In other words, it corresponds to the three unitary field redefinitions that leave the SM fermionic kinetic terms invariant
\begin{equation}
G_f = U(3)_q \times U(3)_u \times U(3)_d \,,
\label{eqn:GFdef}
\end{equation}
where each factor acts on the flavor indices of the associated fermion fields
\begin{subequations}
\label{eqn:FlavorTrans}
\begin{align}
q &\longrightarrow U_q\, q \,, \\[8pt]
u &\longrightarrow U_u\, u \,, \\[8pt]
d &\longrightarrow U_d\, d \,.
\end{align}
\end{subequations}
This flavor symmetry is partially broken in the SM by the Yukawa couplings
\begin{equation}
\L_{\text{Yukawa}} = -\bar{q}\, Y_u\, u \,\widetilde{H} - \bar{q}\, Y_d\, d\, H + \text{h.c.} \,,
\end{equation}
but one can assign spurious transformation properties to the Yukawas to make the SM Lagrangian \emph{formally} invariant under the flavor group. The Yukawa spurions must transform as
\begin{subequations}
\begin{align}
Y_u &\longrightarrow U_q\, Y_u\, U_u^{\dagger} \,, \\[8pt]
Y_d &\longrightarrow U_q\, Y_d\, U_d^{\dagger} \,,
\end{align}
\end{subequations}
which corresponds to bi-fundamental representations of non-abelian subgroup of $G_f$
\begin{subequations}
\begin{align}
Y_u &\sim \mathbf{(3,\overline 3,1)} \,, \\[8pt]
Y_d &\sim \mathbf{(3,1,\overline 3)} \,.
\end{align}
\end{subequations}
Regarding the $U(1)$ transformations, each fundamental/anti-fundamental representation has a charge $\pm 1$ under the associated abelian group $U(3)_X = S U(3)_X \times U(1)_X$ for $X=q, u, d$.

A given effective theory is defined to satisfy the criterion of \emph{Minimal Flavor Violation}~\cite{DAmbrosio:2002ex,Cirigliano:2005ck,Davidson:2006bd,Gavela:2009cd,Alonso:2011jd} if all higher dimensional operators built from SM fields and the Yukawa spurions are formally invariant under the flavor group. This hypothesis thus assumes that at low energies the Yukawa couplings are the only source of flavor violation both in the SM and beyond. One should stress that, at this point, any formally flavor invariant non-renormalizable operator satisfies the MFV hypothesis and hence should be considered on an equal footing.

\subsection{MFV covariants in dim-6 SMEFT}
\label{subsec:MFVCovariants}

\begin{table}[t]
\begin{center}
\fontsize{8.5pt}{11pt}\selectfont

\hspace{-3cm}
\begin{minipage}[t]{2.7cm}
\renewcommand{\arraystretch}{1.5}
\begin{tabular}[t]{c|c|c}
\multicolumn{2}{c}{$5: \psi^2H^3 + \hbox{h.c.}$}  & $SU(3)_{q, u, d}$ \\
\hline
$Q_{eH}$           & $(H^\dag H)(\bar l_p e_r H)$ & $(\mathbf{1},\mathbf{1},\mathbf{1})$ \\
$Q_{uH}$          & $(H^\dag H)(\bar q_p u_r \widetilde H )$ & $( \mathbf{3},\mathbf{\bar 3},\mathbf{1})$ \\
$Q_{dH}$           & $(H^\dag H)(\bar q_p d_r H)$ & $( \mathbf{3},\mathbf{1},\mathbf{\bar 3})$\\
\end{tabular}
\end{minipage}

\vspace{0.2cm}
\hspace{-2.5cm}
\begin{minipage}[t]{5.2cm}
\renewcommand{\arraystretch}{1.5}
\begin{tabular}[t]{c|c|c}
\multicolumn{2}{c}{$6:\psi^2 XH+\hbox{h.c.}$} & $SU(3)_{q, u, d}$ \\
\hline
$Q_{eW}$      & $(\bar l_p \sigma^{\mu\nu} e_r) \tau^I H W_{\mu\nu}^I$ & $(\mathbf{1},\mathbf{1},\mathbf{1})$ \\
$Q_{eB}$        & $(\bar l_p \sigma^{\mu\nu} e_r) H B_{\mu\nu}$ & $(\mathbf{1},\mathbf{1},\mathbf{1})$ \\
$Q_{uG}$        & $(\bar q_p \sigma^{\mu\nu} T^A u_r) \widetilde H \, G_{\mu\nu}^A$ & $( \mathbf{3},\mathbf{\bar3},\mathbf{1})$ \\
$Q_{uW}$        & $(\bar q_p \sigma^{\mu\nu} u_r) \tau^I \widetilde H \, W_{\mu\nu}^I$ & $( \mathbf{3},\mathbf{\bar3},\mathbf{1})$ \\
$Q_{uB}$        & $(\bar q_p \sigma^{\mu\nu} u_r) \widetilde H \, B_{\mu\nu}$ & $( \mathbf{3},\mathbf{\bar3},\mathbf{1})$ \\
$Q_{dG}$        & $(\bar q_p \sigma^{\mu\nu} T^A d_r) H\, G_{\mu\nu}^A$ & $( \mathbf{3},\mathbf{1},\mathbf{\bar 3})$\\
$Q_{dW}$         & $(\bar q_p \sigma^{\mu\nu} d_r) \tau^I H\, W_{\mu\nu}^I$ & $( \mathbf{3},\mathbf{1},\mathbf{\bar 3})$\\
$Q_{dB}$        & $(\bar q_p \sigma^{\mu\nu} d_r) H\, B_{\mu\nu}$& $( \mathbf{3},\mathbf{1},\mathbf{\bar 3})$ \\
\end{tabular}
\end{minipage}
\hspace{1.5cm}
\begin{minipage}[t]{5.4cm}
\renewcommand{\arraystretch}{1.5}
\begin{tabular}[t]{c|c|c}
\multicolumn{2}{c}{$7:\psi^2H^2 D$} & $SU(3)_{q, u, d}$ \\
\hline
$Q_{H l}^{(1)}$      & $(H^\dag i\overleftrightarrow{D}_\mu H)(\bar l_p \gamma^\mu l_r)$ & $(\mathbf{1},\mathbf{1},\mathbf{1})$ \\
$Q_{H l}^{(3)}$      & $(H^\dag i\overleftrightarrow{D}^I_\mu H)(\bar l_p \tau^I \gamma^\mu l_r)$ & $(\mathbf{1},\mathbf{1},\mathbf{1})$ \\
$Q_{H e}$            & $(H^\dag i\overleftrightarrow{D}_\mu H)(\bar e_p \gamma^\mu e_r)$ & $(\mathbf{1},\mathbf{1},\mathbf{1})$ \\
$Q_{H q}^{(1)}$      & $(H^\dag i\overleftrightarrow{D}_\mu H)(\bar q_p \gamma^\mu q_r)$& $(\mathbf{1}\oplus\mathbf{8},\mathbf{1},\mathbf{1})$ \\
$Q_{H q}^{(3)}$      & $(H^\dag i\overleftrightarrow{D}^I_\mu H)(\bar q_p \tau^I \gamma^\mu q_r)$& $(\mathbf{1}\oplus\mathbf{8},\mathbf{1},\mathbf{1})$\\
$Q_{H u}$            & $(H^\dag i\overleftrightarrow{D}_\mu H)(\bar u_p \gamma^\mu u_r)$&  $(\mathbf{1},\mathbf{1}\oplus\mathbf{8},\mathbf{1})$\\
$Q_{H d}$            & $(H^\dag i\overleftrightarrow{D}_\mu H)(\bar d_p \gamma^\mu d_r)$& $(\mathbf{1},\mathbf{1},\mathbf{1}\oplus\mathbf{8})$ \\
$Q_{H u d}$ + h.c.   & $i(\widetilde H ^\dag D_\mu H)(\bar u_p \gamma^\mu d_r)$& $(\mathbf{1},\mathbf{3},\mathbf{\bar3})$\\
\end{tabular}
\end{minipage}

\vspace{0.2cm}
\hspace{-2.5cm}
\begin{minipage}[t]{4.75cm}
\renewcommand{\arraystretch}{1.5}
\begin{tabular}[t]{@{}c|c|c}
\multicolumn{2}{c}{$8:(\bar LL)(\bar LL)$} & $SU(3)_{q, u, d}$ \\
\hline
$Q_{ll}$        & $(\bar l_p \gamma_\mu l_r)(\bar l_s \gamma^\mu l_t)$ & $(\mathbf{1},\mathbf{1},\mathbf{1})$\\
$Q_{qq}^{(1)}$  & $(\bar q_p \gamma_\mu q_r)(\bar q_s \gamma^\mu q_t)$ & $(\mathbf{1}\oplus\mathbf{8}\oplus\mathbf{10}\oplus\mathbf{\overline{10}}\oplus\mathbf{27},\mathbf{1},\mathbf{1})$\\
$Q_{qq}^{(3)}$  & $(\bar q_p \gamma_\mu \tau^I q_r)(\bar q_s \gamma^\mu \tau^I q_t)$ & $(\mathbf{1}\oplus\mathbf{8}\oplus\mathbf{10}\oplus\mathbf{\overline{10}}\oplus\mathbf{27},\mathbf{1},\mathbf{1})$\\
$Q_{lq}^{(1)}$                & $(\bar l_p \gamma_\mu l_r)(\bar q_s \gamma^\mu q_t)$ & $(\mathbf{1}\oplus\mathbf{8},\mathbf{1},\mathbf{1})$\\
$Q_{lq}^{(3)}$                & $(\bar l_p \gamma_\mu \tau^I l_r)(\bar q_s \gamma^\mu \tau^I q_t)$ & $(\mathbf{1}\oplus\mathbf{8},\mathbf{1},\mathbf{1})$
\end{tabular}
\end{minipage}
\hspace{3.65cm}
\begin{minipage}[t]{4.75cm}
\renewcommand{\arraystretch}{1.5}
\begin{tabular}[t]{c|c|c}
\multicolumn{2}{c}{$8:(\bar LL)(\bar RR)$} & $SU(3)_{q, u, d}$ \\
\hline
$Q_{le}$               & $(\bar l_p \gamma_\mu l_r)(\bar e_s \gamma^\mu e_t)$ &  $(\mathbf{1},\mathbf{1},\mathbf{1})$\\
$Q_{lu}$               & $(\bar l_p \gamma_\mu l_r)(\bar u_s \gamma^\mu u_t)$ & $(\mathbf{1},\mathbf{1}\oplus\mathbf{8},\mathbf{1})$ \\
$Q_{ld}$               & $(\bar l_p \gamma_\mu l_r)(\bar d_s \gamma^\mu d_t)$ & $(\mathbf{1},\mathbf{1},\mathbf{1}\oplus\mathbf{8})$ \\
$Q_{qe}$               & $(\bar q_p \gamma_\mu q_r)(\bar e_s \gamma^\mu e_t)$ &  $(\mathbf{1}\oplus\mathbf{8},\mathbf{1},\mathbf{1})$\\
$Q_{qu}^{(1)}$         & $(\bar q_p \gamma_\mu q_r)(\bar u_s \gamma^\mu u_t)$ &  $(\mathbf{1}\oplus\mathbf{8},\mathbf{1}\oplus\mathbf{8},\mathbf{1})$\\
$Q_{qu}^{(8)}$         & $(\bar q_p \gamma_\mu T^A q_r)(\bar u_s \gamma^\mu T^A u_t)$ &  $(\mathbf{1}\oplus\mathbf{8},\mathbf{1}\oplus\mathbf{8},\mathbf{1})$\\
$Q_{qd}^{(1)}$ & $(\bar q_p \gamma_\mu q_r)(\bar d_s \gamma^\mu d_t)$ &  $(\mathbf{1}\oplus\mathbf{8},\mathbf{1},\mathbf{1}\oplus\mathbf{8})$\\
$Q_{qd}^{(8)}$ & $(\bar q_p \gamma_\mu T^A q_r)(\bar d_s \gamma^\mu T^A d_t)$ & $(\mathbf{1}\oplus\mathbf{8},\mathbf{1},\mathbf{1}\oplus\mathbf{8})$
\end{tabular}
\end{minipage}

\vspace{0.2cm}
\hspace{-1.2cm}
\begin{minipage}[t]{5.25cm}
\renewcommand{\arraystretch}{1.5}
\begin{tabular}[t]{c|c|c}
\multicolumn{2}{c}{$8:(\bar RR)(\bar RR)$} & $SU(3)_{q, u, d}$ \\
\hline
$Q_{ee}$               & $(\bar e_p \gamma_\mu e_r)(\bar e_s \gamma^\mu e_t)$ & $(\mathbf{1},\mathbf{1},\mathbf{1})$ \\
$Q_{uu}$        & $(\bar u_p \gamma_\mu u_r)(\bar u_s \gamma^\mu u_t)$ & $(\mathbf{1},\mathbf{1}\oplus\mathbf{8}\oplus\mathbf{10}\oplus\mathbf{\overline{10}}\oplus\mathbf{27},\mathbf{1})$ \\
$Q_{dd}$        & $(\bar d_p \gamma_\mu d_r)(\bar d_s \gamma^\mu d_t)$ & $(\mathbf{1},\mathbf{1},\mathbf{1}\oplus\mathbf{8}\oplus\mathbf{10}\oplus\mathbf{\overline{10}}\oplus\mathbf{27})$ \\
$Q_{eu}$                      & $(\bar e_p \gamma_\mu e_r)(\bar u_s \gamma^\mu u_t)$ & $(\mathbf{1},\mathbf{1}\oplus\mathbf{8},\mathbf{1})$ \\
$Q_{ed}$                      & $(\bar e_p \gamma_\mu e_r)(\bar d_s\gamma^\mu d_t)$ & $(\mathbf{1},\mathbf{1},\mathbf{1}\oplus\mathbf{8})$ \\
$Q_{ud}^{(1)}$                & $(\bar u_p \gamma_\mu u_r)(\bar d_s \gamma^\mu d_t)$ & $(\mathbf{1},\mathbf{1}\oplus\mathbf{8},\mathbf{1}\oplus\mathbf{8})$ \\
$Q_{ud}^{(8)}$                & $(\bar u_p \gamma_\mu T^A u_r)(\bar d_s \gamma^\mu T^A d_t)$ & $(\mathbf{1},\mathbf{1}\oplus\mathbf{8},\mathbf{1}\oplus\mathbf{8})$ \\
\end{tabular}
\end{minipage}
\hspace{3.7cm}
\begin{minipage}[t]{5.5cm}
\renewcommand{\arraystretch}{1.5}
\begin{tabular}[t]{c|c|c}
\multicolumn{2}{c}{$8:(\bar LR)(\bar L R)+\hbox{h.c.}$} & $SU(3)_{q, u, d}$ \\
\hline
$Q_{quqd}^{(1)}$ & $(\bar q_p^j u_r) \epsilon_{jk} (\bar q_s^k d_t)$  & $(\mathbf{\bar 3}\oplus \mathbf{6},\mathbf{\bar 3},\mathbf{\bar 3})$\\
$Q_{quqd}^{(8)}$ & $(\bar q_p^j T^A u_r) \epsilon_{jk} (\bar q_s^k T^A d_t)$ &  $(\mathbf{\bar 3}\oplus \mathbf{6},\mathbf{\bar 3},\mathbf{\bar 3})$\\
$Q_{lequ}^{(1)}$ & $(\bar l_p^j e_r) \epsilon_{jk} (\bar q_s^k u_t)$  & $( \mathbf{3},\mathbf{\bar 3},\mathbf{1})$\\
$Q_{lequ}^{(3)}$ & $(\bar l_p^j \sigma_{\mu\nu} e_r) \epsilon_{jk} (\bar q_s^k \sigma^{\mu\nu} u_t)$ & $( \mathbf{3},\mathbf{\bar 3},\mathbf{1})$\\
\end{tabular}

\vspace{0.5cm}
\begin{tabular}[t]{c|c|c}
\multicolumn{2}{c}{$8:(\bar LR)(\bar RL)+\hbox{h.c.}$} & $SU(3)_{q, u, d}$ \\
\hline
$Q_{ledq}$ & $(\bar l_p^j e_r)(\bar d_s q_{tj})$ & $( \mathbf{\bar 3},\mathbf{ 1},\mathbf{3})$
\end{tabular}
\end{minipage}
\end{center}
\caption{Baryon number preserving fermionic operators in dim-6 SMEFT \cite{Grzadkowski:2010es} and the corresponding spurious transformations of the Wilson coefficients under the non-abelian part of the MFV quark flavor group $SU(3)_q \times S U(3)_u \times SU(3)_d \subset G_f$. Transformations under the $U(1)$ factors can be inferred from \cref{eqn:FlavorTrans}. Operators are organized into eight classes as in Ref.~\cite{Alonso:2013hga}, with classes 1-4 collecting bosonic operators not shown here. The subscripts $p, r, s, t$ denote flavor indices.}
\label{tab:MFVdim6SMEFT}
\end{table}

In the absence of new light degrees of freedom, the SMEFT~\cite{Buchmuller:1985jz,Grzadkowski:2010es} provides a powerful tool to parameterize any deviation from the renormalizable SM Lagrangian in a model independent manner. The SMEFT accounts for all possible higher dimensional operators built out of the SM matter content and respecting the SM symmetries. At dimension 5, there are only two allowed hermitian operators, namely the lepton number violating Weinberg operator~\cite{Weinberg:1979sa} plus/minus its hermitian conjugate. The leading non-renormalizable operators preserving baryon and lepton numbers arise at dimension 6, and add up to 76 hermitian operators for a single generation and to 2499 operators for three generations~\cite{Alonso:2013hga}:
\begin{align}
\L_\text{dim-6 SMEFT} = \L_\text{SM} + \sum_{i=1}^{2499} \frac{C_i}{\Lambda^2}\, \O_i \,.
\end{align}
Since it is often impractical to work and compute observables with 2499 operators, extra assumptions are often invoked on the flavor structure of the SMEFT with the objective of reducing the number of free parameters. One of the most popular hypothesis is the MFV reviewed in \cref{subsec:KeyHypothesis}. In \cref{tab:MFVdim6SMEFT}, we list the baryon number preserving fermionic operators of the dim-6 SMEFT in the \emph{Warsaw basis} \cite{Grzadkowski:2010es}, together with the required transformation properties\footnote{In the context of an extended flavored scalar sector, a subset of these representations were already derived in Ref.~\cite{Arnold:2009ay}, in particular those operators involving a single quark bilinear.}
of the corresponding Wilson coefficient under the flavor group in order for the operator to satisfy the MFV criterion. The baryon number violating operators in dim-6 SMEFT are listed in \cref{tab:BNVMFVdim6SMEFT}. One can accommodate these operators in the MFV by relaxing one of the $U(1)$ factors in the flavor group $G_f$ in \cref{eqn:GFdef}, such as relaxing $U(1)_q$:
\begin{equation}
G'_f = SU(3)_q \times SU(3)_u \times SU(3)_d \times U(1)_u \times U(1)_d \,.
\label{eqn:GFRelaxed}
\end{equation}

\begin{table}[t]
\begin{center}
\fontsize{8.5pt}{11pt}\selectfont
\renewcommand{\arraystretch}{1.5}
\begin{tabular}[t]{c|c|c|c}
\multicolumn{2}{c}{$B$-violating} & \multicolumn{1}{c}{$SU(3)_{q, u, d}$} & $U(1)_u \times U(1)_d$ \\
\hline
$Q_{duq}$ & $\epsilon_{\alpha\beta\gamma} \epsilon_{jk} \big[ (d_p^\alpha)^T C u_r^\beta \big] \big[ (q_s^{\gamma j})^T C l_t^k \big]$ & $\mathbf{(\bar{3}, \bar{3}, \bar{3})}$ & $(-1,-1)$ \\
$Q_{qqu}$ & $\epsilon_{\alpha\beta\gamma} \epsilon_{jk} \big[ (q_p^{\alpha j})^T C q_r^{\beta k} \big] \big[ (u_s^\gamma)^T C e_t \big]$ & $\mathbf{(\bar{6}, \bar{3}, 1)}$ & $(-1,0)$ \\
$Q_{qqq}$ & $\epsilon_{\alpha\beta\gamma} \epsilon_{jn} \epsilon_{km} \big[ (q_p^{\alpha j})^T C q_r^{\beta k} \big] \big[ (q_s^{\gamma m})^T C l_t^n \big]$ & $\mathbf{(10 \oplus 8 \oplus 1, 1, 1)}$ & $(0,0)$ \\
$Q_{duu}$ & $\epsilon_{\alpha\beta\gamma} \big[ (d_p^\alpha)^T C u_r^\beta \big] \big[ (u_s^\gamma)^T C e_t \big]$ & $\mathbf{(1, \bar{6} \oplus 3, \bar{3})}$ & $(-2,-1)$ \\
\end{tabular}
\end{center}
\caption{Baryon number violating operators in dim-6 SMEFT \cite{Grzadkowski:2010es, Alonso:2014zka} and the corresponding spurious transformations of the Wilson coefficients under the relaxed version of the MFV quark flavor group $G'_f = SU(3)_q \times S U(3)_u \times SU(3)_d \times U(1)_u \times U(1)_d$. $U(1)_q$ and therefore the baryon number symmetry $U(1)_\text{BN}$ is not imposed. Here $C$ is the charge conjugation matrix, with $C\equiv i\gamma^0\gamma^2$ in the Dirac representation. The subscripts $p, r, s, t$ denote flavor indices.}
\label{tab:BNVMFVdim6SMEFT}
\end{table}

\subsection{Hilbert series for MFV covariants in dim-6 SMEFT}
\label{subsec:HSMFV}

Our goal is to count and obtain the different flavor structures that can contribute to MFV SMEFT operators. To illustrate, let us examine the following operator
\begin{align}
Q_{Hq}^{(1)} = \frac{(C_{H q}^{(1)})_{pr}}{\Lambda^2}\,
\big( H^\dag i\overleftrightarrow{D}_\mu H \big) \big( \bar{q}_p \gamma^\mu q_r \big) \,.
\end{align}
As shown in \cref{tab:MFVdim6SMEFT}, for this operator to satisfy the MFV criterion, the Wilson coefficient $C_{Hq}^{(1)}$ needs to transform either as a singlet or an octet of $SU(3)_q$. To make $U(3)_u \times U(3)_d$ invariants, it is useful to define the following combinations of Yukawas
\begin{equation}
h_u \equiv Y_u Y_u^\dagger \longrightarrow U_q\, h_u \, U_q^{\dagger} \,,\qquad
h_d \equiv Y_d Y_d^\dagger \longrightarrow U_q\, h_d \, U_q^{\dagger} \,,
\label{eqn:huhddef}
\end{equation}
which indeed transform as $h_{u,d}\sim  \mathbf{({1}\oplus{8},{1},{1})}$. Any matrix polynomial of $h_u, h_d$ also has these transformation properties. Consequently, one might be tempted to claim that an infinite number of terms with arbitrary powers (\eg\ $h_u^n$) can contribute to the operator $C_{Hq}^{(1)}$ within MFV:
\begin{align}
C_{Hq}^{(1)} = c_0\, \mathbb{1} + c_1\, h_u + c_2\, h_d
+ c_3\, h_u^2 + c_4\, h_u h_d + c_5\, h_d h_u + c_6\, h_d^2 + \dots
\label{eqn:MFVInfiniteSum}
\end{align}
However, this is not the case since there exist only a finite number of independent covariants that can be constructed out of the Yukawa spurions. As we learned in the preceding section, one can employ the tools of invariant theory to tackle this problem. The set of covariants transforming as $\mathbf{({1}\oplus{8},{1},{1})}$ form a module over the ring of flavor invariants; see \cref{eqn:MRrI}. Importantly, this module is finitely generated, which implies that there are a finite number of covariants in the generating set~\cite{Popov1994}. It should be kept in mind that the coefficients are $G_f$ invariant polynomials of $h_u$ and $h_d$.

It is worth noting that in previous references~\cite{Colangelo:2008qp,Mercolli:2009ns} it was already demonstrated that the sum in \cref{eqn:MFVInfiniteSum} can be reduced to a finite number of terms. Through the application of the Cayley-Hamilton theorem, this was shown for a few specific representations: $\mathbf{({8},{1},{1})}$, $\mathbf{(3,\bar 3,1)}$, and $\mathbf{(3,1,\bar 3)}$. In this work, we go further by extending this result to encompass all flavor representations. This is made possible after recognizing that the underlying structure of the problem corresponds to a finitely generated module.

The pertinent questions are then: how many independent covariants are there and do they lead to specific correlations among flavor observables that allow us to distinguish MFV SMEFT from a general SMEFT? In order to answer these questions, we have computed the Hilbert series for all the flavor covariants in \cref{tab:MFVdim6SMEFT} that arise in baryon preserving MFV dim-6 SMEFT. The general background of the computation has been explained in \cref{sec:Hilbert} and the computational details of some representative examples of the flavor covariants are given in the second part of \cref{appsec:Examples}. The Hilbert series of these MFV covariants read
\begin{subequations}\label{eqn:HSMFV}
\begin{align}
\HS_\mathbf{(1,1,1)} &= \frac{1}{D(q)}
\mqty( 1 + q^{12} ) \,, \label{eqn:HS111} \\[8pt]
\HS_\mathbf{(8,1,1)} &= \frac{1}{D(q)}\, 2
\mqty( q^2 + 2q^4 + 2q^6 + 2q^8 + q^{10} ) \,, \label{eqn:HS811} \\[8pt]
\HS_\mathbf{(1,8,1)} &= \frac{1}{D(q)}\, q^2
\mqty( 1 + 2q^2 + 3q^4 + 4q^6 + 4q^8 + 2q^{10} + q^{12} - q^{16} ) \,, \label{eqn:HS181} \\[8pt]
\HS_\mathbf{(10,1,1)} &= \frac{1}{D(q)}\, q^4
\mqty( 1 + 6q^2 + 7q^4 + 8q^6 + 4q^8 - 3q^{12} - 2q^{14} - q^{16} ) \,, \label{eqn:HS1011} \\[8pt]
\HS_\mathbf{(1,10,1)} &= \frac{1}{D(q)}\, q^6
\mqty( 2 + 3q^2 + 6q^4 + 7q^6 + 6q^8 + 2q^{10} - 3q^{14} - 2q^{16} - q^{18} ) \,, \label{eqn:HS1101} \\[8pt]
\HS_\mathbf{(27,1,1)} &= \frac{1}{D(q)}\, q^4\,
\mqty( 3 + 8q^2 + 17q^4 + 20q^6 + 19q^8 + 8q^{10} \\[4pt]
-q^{12} -8 q^{14} - 7q^{16} - 4q^{18} - q^{20} ) \,, \label{eqn:HS2711} \\[8pt]
\HS_\mathbf{(1,27,1)} &= \frac{1}{D(q)}\, q^4\,
\mqty( 1 + 2q^2 + 6q^4 + 10q^6 + 17q^8 + 18q^{10} + 16q^{12} \\[4pt]
+ 6q^{14} - 2q^{16} - 8q^{18} - 7q^{20} - 4q^{22} -q^{24} ) \,, \label{eqn:HS1271} \\[8pt]
\HS_\mathbf{(3,\overline 3,1)} &= \frac{1}{D(q)}\, q
\mqty( 1 + 2q^2 + 4q^4 + 4q^6 + 4q^8 + 2q^{10} + q^{12} ) \,, \label{eqn:HS33bar1} \\[8pt] 
\HS_\mathbf{(1,3,\overline 3)} &= \frac{1}{D(q)}\, q^2
\mqty( 1 + 2q^2 + 4q^4 + 4q^6 + 4q^8 + 2q^{10} + q^{12} ) \,, \label{eqn:HS133bar} \\[8pt]
\HS_\mathbf{(8,8,1)} &= \frac{1}{D(q)}\, q^2\,
\mqty( 1 + 6q^2 + 17q^4 + 30q^6 + 39q^8 + 38q^{10} + 24q^{12} \\[4pt]
+ 6q^{14} - 7q^{16} - 12q^{18} - 9q^{20} - 4q^{22} - q^{24} ) \,, \label{eqn:HS881} \\[8pt]
\HS_\mathbf{(1,8,8)} &= \frac{1}{D(q)}\, q^4\,
\mqty( 2 + 8q^2 + 19q^4 + 32q^6 + 40q^8 + 36q^{10} + 21q^{12} \\[4pt]
+ 4q^{14} - 9q^{16} - 12q^{18} - 8q^{20} - 4q^{22} - q^{24} ) \,, \label{eqn:HS188} \\[8pt]
\HS_\mathbf{(\overline{3},\overline{3},\overline{3})} &= \frac{1}{D(q)}\, q^2\,
\mqty( 1 + 4q^2 + 9q^4 + 14q^6 + 15q^8 + 12q^{10} \\[4pt]
+ 5q^{12} - 3q^{16} - 2q^{18} - q^{20} ) \,, \label{eqn:HS3bar3bar3bar} \\[8pt]
\HS_\mathbf{(6,\overline{3},\overline{3})} &= \frac{1}{D(q)}\, q^2\,
\mqty( 1 + 4q^2 + 12q^4 + 22q^6 + 32q^8 + 32q^{10} + 24q^{12} \\[4pt]
+ 8q^{14} - 4q^{16} - 10q^{18} - 8q^{20} - 4q^{22} - q^{24} ) \,, \label{eqn:HS63bar3bar}
\end{align}
\end{subequations}
with the denominator factor given by
\begin{equation}
\frac{1}{D(q)} \equiv
\frac{1}{\left(1-q^2\right)^2 \left(1-q^4\right)^3 \left(1-q^6\right)^4 \left(1-q^8\right)} \,.
\label{eqn:MFVDenominator}
\end{equation}
The Hilbert series for all the other MFV covariants listed in \cref{tab:MFVdim6SMEFT} coincide with one of the expressions above:
\begin{subequations}\label{eqn:HSMFVsame}
\begin{align}
\HS_\mathbf{(3,\overline{3},1)} &= \HS_\mathbf{(3,1,\overline{3})} \,, \\[5pt]
\HS_\mathbf{(8,8,1)} &= \HS_\mathbf{(8,1,8)} \,, \\[5pt]
\HS_\mathbf{(10,1,1)} &= \HS_\mathbf{(\overline{10},1,1)} \,, \\[5pt]
\HS_\mathbf{(1,10,1)} &= \HS_\mathbf{(1,\overline{10},1)}
= \HS_\mathbf{(1,1,10)} = \HS_\mathbf{(1,1,\overline{10})} \,, \\[5pt]
\HS_\mathbf{(1,8,1)} &= \HS_\mathbf{(1,1,8)} \,, \\[5pt]
\HS_\mathbf{(1,27,1)} &= \HS_\mathbf{(1,1,27)} \,.
\end{align}
\end{subequations}

For the choice of MFV symmetry group in \cref{eqn:GFdef}, none of the baryon number violating covariants listed in \cref{tab:BNVMFVdim6SMEFT} can be generated, so their Hilbert series vanish. The reason is clear, as the Yukawas do not carry baryon number charge, it is impossible to construct baryon number singlets with the baryon number violating operators and the Yukawa spurions. Nevertheless, the outcome changes if one instead imposes the relaxed version of the flavor group, $G'_f$ in \cref{eqn:GFRelaxed}, which lacks one of the $U(1)$ factors, and therefore does not impose the baryon number $U(1)_\text{BN}$. The corresponding Hilbert series for the baryon number-violating covariants can be computed using the same methodology detailed in \cref{sec:Hilbert,appsec:Examples} yielding
\begin{subequations}\label{eqn:HSMFVBNV}
\begin{align}
\H^{G'_f}_{\mathbf{(\overline{6},\overline{3},1)}\otimes (-1,0)} &= \frac{1}{D(q)}\, q^3\,
\mqty(2 + 5q^2 + 10q^4 + 11q^6 + 10q^8  \\[4pt]
+ 4q^{10} - 3q^{14} - 2q^{16} - q^{18}) \,, \\[8pt]
\HS^{G'_f}_{\mathbf{(1,\overline{6},\overline{3})}\otimes (-2,-1)} &= \frac{1}{D(q)}\, q^5\,
\mqty(2 + 5q^2 + 10q^4 + 11q^6 + 10q^8  \\[4pt]
+ 4q^{10} - 3q^{14} - 2q^{16} - q^{18}) \,, \\[8pt]
\HS^{G'_f}_{\mathbf{(1,3,\overline{3})}\otimes (-2,-1)} &= \frac{1}{D(q)}\, q^3\,
\mqty(1 + 2q^2 + 4q^4 + 4q^6 + 4q^8 + 2q^{10} + q^{12}) \,,
\end{align}
\end{subequations}
with the same denominator in \cref{eqn:MFVDenominator}. Hilbert series for the other baryon number violating covariants in \cref{tab:BNVMFVdim6SMEFT} coincide with one of the Hilbert series in \cref{eqn:HSMFV}:
\begin{subequations}\label{eqn:HSMFVBNVsame}
\begin{align}
\HS^{G'_f}_{\mathbf{(\bar{3},\bar{3},\bar{3})}\otimes(-1,-1)} &= \HS_\mathbf{(\bar{3},\bar{3},\bar{3})} \,, \\[5pt]
\HS^{G'_f}_{\mathbf{(10, 1, 1)}\otimes(0,0)} &= \HS_\mathbf{(10, 1, 1)} \,, \\[5pt]
\HS^{G'_f}_{\mathbf{(8, 1, 1)}\otimes(0,0)} &= \HS_\mathbf{(8, 1, 1)} \,, \\[5pt]
\HS^{G'_f}_{\mathbf{(1, 1, 1)}\otimes(0,0)} &= \HS_\mathbf{(1, 1, 1)} \,.
\end{align}
\end{subequations}

\subsection{Understanding the MFV Hilbert series}
\label{subsec:UnderstandingHS}

Based on the (summed) Hilbert series obtained in \cref{eqn:HSMFV}, and the underlying module structure behind the rep-$R$ covariants explained in \cref{subsubsec:Module}, several general properties of the MFV covariants become apparent:
\begin{enumerate}
\item The numerators have a finite number of terms, implying that the associated modules are finitely generated.  This is consistent with the theorem that for a linearly reductive group all the rep-$R$ modules are finitely generated (Theorem 3.24 in Ref.~\cite{Popov1994}).
\item Every Hilbert series shares a common denominator (\cref{eqn:MFVDenominator}) that corresponds to the primary flavor invariants (see \cref{eqn:MFVPrimaryInv} below for an explicit list). This was expected since one can multiply any rep-$R$ covariant by a polynomial of the primary invariants and obtain another rep-$R$ covariant.
\item The Hilbert series for invariants has a factor $(1+q^{12})$ in the numerator, where the $q^{12}$ term reflects the secondary invariant, \ie\ the Jarlskog determinant $J$ (see \cref{eqn:Jarlskog} below for an explicit expression). One may have expected the same factor to appear in the numerators of the other Hilbert series in \cref{eqn:HSMFV}, because multiplying a given rep-$R$ covariant by the secondary invariant $J$ gives another rep-$R$ covariant. 
Notably, the factor is not present in the numerators of the Hilbert series for covariants. The lack of such a factorization indicates that the new covariant obtained by multiplying $J$ is not independent, which can be expressed in terms of the other elements of the generating set with only the primary invariants as coefficients. This will be explicitly proved in \cref{eqn:JhuJhd}.
\item Unlike the case for invariants, the rep-$R$ covariants do not necessarily form a free module. The numerators may include negative terms, signaling redundancies within the generating set. Consequently, the generating set does not form a basis for the module.
\item The numerators, unlike those of flavor invariants, are not generally  palindromic polynomials. While the flavor invariant ring enjoys the \emph{Gorenstein property}~\cite{Gray:2008yu, bruns1998cohen} ensuring that the numerator of the Hilbert series is palindromic \cite{Stanley78}, the same is not true for general rep-$R$ covariants.
\item Interestingly, the numerator of the Hilbert series for $\mathbf{(8,1,1)}$ happens to present exclusively positive coefficients. Indeed, as we will see below, the $\mathbf{(8,1,1)}$ covariant module over the ring generated by the primary flavor invariants, ${}_\rPI\Mod_{\mathbf{(8,1,1)}}$, is a free module. Consequently, the minimal generating set serves as a basis in this case. However, when viewed as a module over the ring of all invariants, \ie\ ${}_\rI\Mod_{\mathbf{(8,1,1)}}$, the module is not free as it contains relations that involve the secondary invariant $J$. We will elaborate this in subsequent sections.
\end{enumerate}
For the rest of this subsection, we look into the structure of a few simplest flavor covariants by conventional methods, in order to better understand the above properties. We will explicitly work with the Yukawa matrices $Y_u, Y_d$ and apply the Cayley-Hamilton theorem detailed in \cref{appsec:CayleyHamilton}. This will cross check some of the MFV Hilbert series in \cref{eqn:HSMFV}.

As explained in \cref{subsubsec:Module}, a given set of covariants can be viewed both as a module over the ring of all invariants or over the ring generated by the primary invariants. It is thus important to first understand the properties of the flavor invariant ring. These were already discussed in detail by E. Jenkins and A. Manohar in Ref.~\cite{Jenkins:2009dy}.\footnote{In fact, the Hilbert series $\HS_\mathbf{(1,1,1)}$, $\HS_\mathbf{(8,1,1)}$, and $\HS_\mathbf{(10,1,1)}$ in \cref{eqn:HSMFV} were also computed in the math literature \cite{teranishi_1986, broer1994hilbert}, without making the connection to flavor physics.}
In \cref{subsubsec:HS111}, we give a brief review on this. In \cref{subsubsec:MatrixPolynomials}, we move on to studying the matrix polynomials in $h_u, h_d$ defined in \cref{eqn:huhddef}. These form the $\mathbf{(1\oplus8,1,1)}$ flavor representation, which will provide insights for quite a few other representations. Subsequently, we discuss the $\mathbf{(8,1,1)}$ flavor covariants in \cref{subsubsec:HS811}, and then the $\mathbf{(3,\overline{3},1)}$, $\mathbf{(3,1,\overline{3})}$, and $\mathbf{(1,3,\overline{3})}$ covariants in \cref{subsubsec:HS33bar1}. Finally, we discuss the more involved cases of $\mathbf{(1,8,1)}$ and $\mathbf{(1,1,8)}$ covariants in \cref{subsubsec:HS181}.

\subsubsection{Flavor invariants $\mathbf{(1,1,1)}$}
\label{subsubsec:HS111}

We have explained in \cref{subsubsec:HironakaDecomposition} that a ring of invariants admits a Hironaka decomposition (see \cref{eqn:Hironaka}), and one can extract the number and degrees of the primary and secondary invariants from the Hilbert series. For the MFV case at hand, this is the Hilbert series $\HS_\mathbf{(1,1,1)}$ in \cref{eqn:HS111}:
\begin{equation}
\HS_\mathbf{(1,1,1)} = \frac{1+q^{12}}{\left(1-q^2\right)^2 \left(1-q^4\right)^3 \left(1-q^6\right)^4 \left(1-q^8\right)} \,.
\label{eqn:HS111Redo}
\end{equation}

The denominator in \cref{eqn:HS111Redo} indicates that the quark flavor sector has 10 primary invariants: two invariants of degree two, three of degree four, four of degree six, and one of degree eight. An explicit choice of them can be \cite{Jenkins:2009dy}
\begin{subequations}\label{eqn:MFVPrimaryInv}
\begin{alignat}{2}
P_{2,0} &= \tr \big( h_u \big) \,,\qquad &
P_{0,2} &= \tr \big( h_d \big) \,, \\[8pt]
P_{4,0} &= \tr \big( h_u^2 \big) \,,\qquad &
P_{0,4} &= \tr \big( h_d^2 \big) \,, \\[8pt]
P_{2,2} &= \tr \big( h_u h_d \big) \,, && \\[8pt]
P_{6,0} &= \tr \big( h_u^3 \big) \,,\qquad &
P_{0,6} &= \tr \big( h_d^3 \big) \,, \\[8pt]
P_{4,2} &= \tr \big( h_u^2 h_d \big) \,,\qquad &
P_{2,4} &= \tr \big( h_u h_d^2 \big) \,, \\[8pt]
P_{4,4} &= \tr \big( h_u^2 h_d^2 \big) \,. && 
\end{alignat}
\end{subequations}
As expected, the total number of primary invariants, $n_P^{}=10$, coincides with the number of parameters or degrees of freedom in the quark sector, consisting of six quark masses, three mixing angles and one phase $\delta_\text{KM}$.\footnote{Note that the $\bar\theta$-parameter of QCD does not arise in this analysis because when computing the Hilbert series we are imposing the axial $U(1)_A$ (see \cref{appsec:Examples}), despite that it is anomalous.} We note that all the primary invariants are real (traces of hermitian matrices).

The numerator in \cref{eqn:HS111Redo} implies that there is one secondary invariant of order $\O(q^{12})$, which corresponds to the Jarlskog determinant \cite{Jarlskog:1985cw, Jarlskog:1985ht}:
\begin{align}
J &= \det \big( \comm{h_u}{h_d} \big)
= \tfrac13 \tr \Big( \comm{h_u}{h_d}^3 \Big)
= \tr \left( h_u h_d h_u^2 h_d^2 \right) - (u \leftrightarrow d)
= 2i \Im \tr \left( h_u h_d h_u^2 h_d^2 \right)
\notag\\[12pt]
&= 2i \left( m_t^2 - m_c^2 \right) \left( m_t^2 - m_u^2 \right) \left( m_c^2 - m_u^2 \right)
\left( m_b^2 - m_s^2 \right) \left( m_s^2 - m_d^2 \right) \left( m_b^2 - m_d^2 \right)
\notag\\[5pt]
&\hspace{100pt}
\times c_{12}\, c_{23}\, c_{13}^2\, s_{12}\, s_{23}\, s_{13} \sin\delta_\text{KM} \,.
\label{eqn:Jarlskog}
\end{align}
We will interchangeably refer to $J$ as either the secondary or Jarlskog invariant within this context. We see that unlike the primary invariants, the Jarlskog is anti-hermitian, or purely imaginary. While $J$ is an independent invariant, its square, $J^2$, can be expressed as a polynomial of the primary invariants, indicating the existence of a ``syzygy'' of order $\O(q^{24})$. It may come as a surprise that the 10 primary invariants in \cref{eqn:MFVPrimaryInv}, which are CP-even, are sufficient to determine whether the SM exhibits CP violation (and for quantifying the absolute measure of CP violation $|J|$). The reason is simple: one can obtain the value of $\cos\delta_{\rm KM}$, which is CP-even, from the 10 CP-even primary invariants. Consequently, the only information that the Jarlskog determinant adds is the overall sign through $\sin\delta_\text{KM}$. This can also be understood from the well-known fact that the unitarity triangle can be determined by measuring the length of its three sides (which are CP conserving). This completely determines the area of the triangle which is related with the absolute value of the Jarlskog determinant.

To sum up, the Hilbert series in \cref{eqn:HS111Redo} reflects a Hironaka decomposition of the flavor invariants. Any flavor invariant can be uniquely expressed as a polynomial of the 10 primary invariants in \cref{eqn:MFVPrimaryInv} and the secondary invariant $J$ in \cref{eqn:Jarlskog}, where $J$ arises linearly at most:
\begin{equation}
p_0 \big( P_1, \cdots, P_{10} \big) + p_1 \big(P_1, \cdots, P_{10} \big)\, J \,.
\label{eqn:FlavorHironaka}
\end{equation}
Here $p_0 \big( P_1, \cdots, P_{10} \big)$ and $p_1 \big( P_1, \cdots, P_{10} \big)$ are polynomials in the primary invariants. We emphasize that the choice of the primary and secondary invariants in \cref{eqn:MFVPrimaryInv,eqn:Jarlskog} is not unique. However, after selecting a set of primary and secondary invariants, the Hironaka decomposition in \cref{eqn:FlavorHironaka} for each flavor invariant is unique.

\subsubsection{Matrix polynomials in $h_u, h_d$ --- the $\mathbf{(1\oplus8,1,1)}$ covariants}
\label{subsubsec:MatrixPolynomials}

Going beyond the flavor invariants, let us now consider the matrix polynomials constructed out of the matrices $h_u, h_d$ defined in \cref{eqn:huhddef}:
\begin{equation}
p(h_u, h_d) \equiv \sum_{k,i} c_{2k,i}(h_u, h_d)\; \mm_i^{(2k)}(h_u, h_d) \,.
\label{eqn:MPdef}
\end{equation}
Here $\mm_i^{(2k)}(h_u, h_d)$ denotes a matrix monomial in $h_u, h_d$ of order $\O(q^{2k})$, \ie\ a product of $k$ powers of $h_u, h_d$, as $h_u, h_d \sim \O(q^2)$. The coefficients $c_{2k,i}(h_u, h_d)$ are flavor invariants, also made out of $h_u, h_d$; they are proportional to the identity matrix. We are interested in the matrix polynomials in \cref{eqn:MPdef} because they constitute the type of Wilson coefficients in \cref{eqn:MFVInfiniteSum}. They form the $\mathbf{(1\oplus8,1,1)}$ flavor representation, and the Hilbert series can be obtained from \cref{eqn:HSMFV} as
\begin{equation}
\HS_\mathbf{(1\oplus8,1,1)} = \HS_\mathbf{(1,1,1)} + \HS_\mathbf{(8,1,1)}
= \frac{1 + 2q^2 + 4q^4 + 4q^6 + 4q^8 + 2q^{10} + q^{12}}
{\left(1-q^2\right)^2 \left(1-q^4\right)^3 \left(1-q^6\right)^4 \left(1-q^8\right)} \,.
\label{eqn:HS1811}
\end{equation}
It implies that there are only a finite number of independent matrix polynomials (encoded by the numerator), while all the other matrix polynomials can be generated by multiplying them with the primary invariants (encoded by the denominator). This is due to the Cayley-Hamilton theorem detailed in \cref{appsec:CayleyHamilton}. To see how it works, it is informative to also compute the multi-graded Hilbert series, in which we keep track of the powers of $Y_u$ and $Y_d$ with the labels $q_u$ and $q_d$, respectively. With such a finer grading option, the numerator and the denominator of the Hilbert series become
\begin{subequations}\label{eqn:HS1811ud}
\begin{align}
N_\mathbf{(1\oplus8,1,1)}(q_u, q_d) &= 1 + q_u^2 + q_d^2 + q_u^4 + q_d^4 + 2\, q_u^2\, q_d^2
+ 2\, q_u^4\, q_d^2 + 2\, q_u^2\, q_d^4
\notag\\[5pt]
&\quad + q_u^6\, q_d^2 + q_u^2\, q_d^6 + 2\, q_u^4\, q_d^4 + q_u^6\, q_d^4 + q_u^4\, q_d^6 + q_u^6\, q_d^6 \,, \label{eqn:N1811ud} \\[10pt]
D(q_u, q_d) &= \left(1-q_u^2\right) \left(1-q_d^2\right) \left(1-q_u^4\right) \left(1-q_d^4\right) \left(1-q_u^2\, q_d^2\right)
\notag\\[5pt]
&\quad
\times \left(1-q_u^6\right) \left(1-q_d^6\right) \left(1-q_u^4\, q_d^2\right) \left(1-q_u^2\, q_d^4\right) \left(1-q_u^4\, q_d^4\right) \,.
\end{align}
\end{subequations}
The single variable Hilbert series in \cref{eqn:HS1811} is recovered from this by setting $q_u = q_d = q$.\footnote{From \cref{eqn:HS1811ud}, one can also easily reproduce the Hilbert series result for the case of a single building block only, say $h_u$, by setting $q_d=0$.}

Let us begin with understanding the number of independent monomials of lowest orders. For  $\mm^{(2)}(h_u, h_d)$ the distinct matrix monomials are simply $h_u,\, h_d$, while for $\mm^{(4)}(h_u, h_d)$ we have four independent monomials $h_u^2,\, h_d^2,\, h_u h_d,\, h_d h_u$. We see that these agree with the lowest order terms in \cref{eqn:N1811ud}: $q_u^2 + q_d^2 + q_u^4 + q_d^4 + 2\, q_u^2\, q_d^2$.

In general, there are $2^k$ matrix monomials in $\mm^{(2k)}(h_u, h_d)$. However, starting from  $\mm^{(6)}(h_u, h_d)$, the Cayley-Hamilton theorem can be employed to obtain linear relations among them. As detailed in \cref{appsec:CayleyHamilton}, this theorem allows to express the third power of any $3\times 3$ matrix $X$ in terms of its lower powers (\cref{eqn:CH3app}):
\begin{equation}
X^3 = \left(\tr_X\right) X^2 + \frac12 \left( \tr_{X^2} - \tr_X^2 \right) X
+ \frac16 \left( 2\tr_{X^3} - 3\tr_{X^2} \tr_X + \tr_X^3 \right) I_3 \,,
\label{eqn:CH3}
\end{equation}
where $I_3$ is the $3\times3$ identity matrix. Taking $X=xA+yB+zC$ and focusing on the term proportional to $xyz$, one can rewrite this identity into a form that applies to any three $3\times3$ matrices $A,B,C$ (\cref{eqn:CH3ABCapp}):
\begin{align}
&\hspace{-10pt}
ABC + ACB + BAC + BCA + CAB + CBA
\notag\\[8pt]
&= \tr_C \Big( AB + BA \Big) + \tr_B \Big( AC + CA \Big) + \tr_A \Big( BC + CB \Big)
\notag\\[5pt]
&\quad
+ \Big( \tr_{BC} - \tr_B \tr_C \Big)\, A
+ \Big( \tr_{AC} - \tr_A \tr_C \Big)\, B
+ \Big( \tr_{AB} - \tr_A \tr_B \Big)\, C
\notag\\[5pt]
&\quad
+ \Big( \tr_{ABC} + \tr_{ACB} \Big)
- \Big(  \tr_{AB} \tr_C + \tr_{AC} \tr_B + \tr_{BC} \tr_A \Big)
+ \tr_A \tr_B \tr_C \,.
\label{eqn:CH3ABC}
\end{align}
The last line is understood to come with an $I_3$ factor that we will suppress onward. 

Applying \cref{eqn:CH3ABC} to $\mm^{(6)}(h_u, h_d)$, one can find redundancy relations, such as those in \cref{eqn:hu3Reduce,eqn:huhdhdReduce}. The total number of redundancy relations are constructed by distinct combinations of $(A, B, C)$, each of which can be either $h_u$ or $h_d$. So this is counting the combinations of choosing $k=3$ repeatable factors of the $n=2$ objects $h_u$ and $h_d$, which corresponds to the multiset coefficient
\begin{align}
\mqty(\kern-.45em\mqty(n\\k)\kern-.45em) = \mqty(n+k-1\\k) = \mqty(4\\3) = 4 \,.
\end{align}
This leads us to $2^3-4 = 4$ independent matrix monomials in $\mm^{(6)}(h_u, h_d)$, agreeing with \cref{eqn:N1811ud}. One can choose four independent monomials $\mm_i^{(6)}$ with $i=1,2,3,4$, and any given $\O(q^6)$ monomial $\mm^{(6)}$ can be written as linear combinations of them together with lower order ones $\mm^{(4)}$, $\mm^{(2)}$ (and $\mm^{(0)}=I_3$ that we will suppress):
\begin{align}
\mm^{(6)} = \sum_{i=1}^4 c_i^{(0)}\, \mm_i^{(6)} + \sum_{i=1}^4 c_i^{(2)}\, \mm_i^{(4)}
+ \sum_{i=1}^2 c_i^{(4)}\, \mm_i^{(2)} + c^{(6)} \,.
\end{align}
Here $c_i^{(2k)}$ with a superscript ``$(2k)$'' denotes a flavor invariant of order $q^{2k}$ (\cf\ the order agnostic notation $c_{2k,i}$ in \cref{eqn:MPdef}). Note that $c_i^{(0)}$ has zero power in $h_u, h_d$ and therefore is just a number.

At order $\mm^{(8)}(h_u, h_d)$, we start from a total of $2^4 = 16$ matrix monomials. Out of these, one can find redundancy relations such as in \cref{eqn:hu2hdhdReduce,eqn:hu2huhdReduce}. The total number of relations are again given by distinct combinations of $(A, B, C)$ in \cref{eqn:CH3ABC}. To make a total power of $q^8$, we can take without loss of generality that
\begin{equation}
A \sim \mm^{(4)} \,,\quad
B \sim \mm^{(2)} \,,\quad
C \sim \mm^{(2)} \,,
\end{equation}
which leads us to the following number of distinct combinations of them
\begin{equation}
4\; \mqty(\kern-.45em\mqty(2\\2)\kern-.45em) = 4\; \mqty(2+2-1\\2) = 4 \times 3 = 12 \,.
\end{equation}
Therefore, we have $2^4 - 12 = 4$ independent matrix monomials at order $\mm^{(8)}(h_u, h_d)$, agreeing with \cref{eqn:N1811ud}. As a result, one can choose four independent monomials $\mm_i^{(8)}$ with $i=1,2,3,4$, and any given $\O(q^8)$ monomial $\mm^{(8)}$ can be written as linear combinations of them together with lower order monomials:
\begin{align}
\mm^{(8)} = \sum_{i=1}^4 c_i^{(0)}\, \mm_i^{(8)} + \sum_{i=1}^4 c_i^{(2)}\, \mm_i^{(6)}
+ \sum_{i=1}^4 c_i^{(4)}\, \mm_i^{(4)} + \sum_{i=1}^2 c_i^{(6)}\, \mm_i^{(2)} + c^{(8)} \,.
\end{align}

Moving on to order $\mm^{(10)}(h_u, h_d)$, one can follow a similar procedure to count the number of redundancies made out of \cref{eqn:CH3ABC}. There are two possible divisions of the total power $q^{10}$ into the factors $A, B, C$. Without loss of generality, they are
\begin{subequations}
\begin{align}
A &\sim \mm^{(6)} \,,\quad
B \sim \mm^{(2)} \,,\quad
C \sim \mm^{(2)} \,, \\[5pt]
A &\sim \mm^{(4)} \,,\quad
B \sim \mm^{(4)} \,,\quad
C \sim \mm^{(2)} \,.
\end{align}
\end{subequations}%
Adding them together, the total number of distinct $(A,B,C)$ combinations is
\begin{equation}
8\; \mqty(\kern-.45em\mqty(2\\2)\kern-.45em) + 2\; \mqty(\kern-.45em\mqty(4\\2)\kern-.45em) = 8\; \mqty(3\\2) + 2\; \mqty(5\\2) = 24 + 20 = 44 \,.
\end{equation}
However, not all these relations are linearly independent. An explicit computation shows that the rank is $30$, leading us to $2^5 - 30 = 2$ independent matrix monomials at order $\mm^{(10)}(h_u, h_d)$, again agreeing with \cref{eqn:N1811ud}. Consequently, one can choose two independent monomials $\mm_i^{(10)}$ with $i=1,2$, and any given $\O(q^{10})$ monomial $\mm^{(10)}$ can be written as linear combinations of them together with lower order ones:
\begin{align}
\mm^{(10)} &= \sum_{i=1}^2 c_i^{(0)}\, \mm_i^{(10)} + \sum_{i=1}^4 c_i^{(2)}\, \mm_i^{(8)}
+ \sum_{i=1}^4 c_i^{(4)}\, \mm_i^{(6)} + \sum_{i=1}^4 c_i^{(6)}\, \mm_i^{(4)}
\notag\\[5pt]
&\qquad
+ \sum_{i=1}^2 c_i^{(8)}\, \mm_i^{(2)} + c^{(10)} \,.
\end{align}

Finally, at order $\mm^{(12)}(h_u, h_d)$ we start with a total of $2^6 = 64$ matrix monomials. One can check, by computing them explicitly, that the number of linearly independent redundancy relations built out of \cref{eqn:CH3ABC} is equal to this number. So no independent matrix monomials survive at this order; all can be expressed as linear combinations of lower order monomials:
\begin{align}
\mm^{(12)} &= \sum_{i=1}^2 c_i^{(2)}\, \mm_i^{(10)} + \sum_{i=1}^4 c_i^{(4)}\, \mm_i^{(8)}
+ \sum_{i=1}^4 c_i^{(6)}\, \mm_i^{(6)} + \sum_{i=1}^4 c_i^{(8)}\, \mm_i^{(4)}
\notag\\[5pt]
&\qquad
+ \sum_{i=1}^2 c_i^{(10)}\, \mm_i^{(2)} + c^{(12)} \,.
\label{eqn:M12cReduction}
\end{align}
This again agrees with \cref{eqn:N1811ud}.\footnote{The $q_u^6 q_d^6$ term in \cref{eqn:N1811ud} represents the Jarlskog invariant $J$ in \cref{eqn:Jarlskog}, proportional to the identity matrix and hence belonging to the $c^{(12)}$ term above.}
Among all these invariant coefficients, only $c^{(12)}$ has enough power of $h_u, h_d$ to involve the Jarlskog $J$; the others must be polynomials of the primary invariants in \cref{eqn:MFVPrimaryInv}. Therefore, we can refine \cref{eqn:M12cReduction} as
\begin{align}
\mm^{(12)} &= \sum_{i=1}^2 \eR_i^{(2)}\, \mm_i^{(10)} + \sum_{i=1}^4 \eR_i^{(4)}\, \mm_i^{(8)}
+ \sum_{i=1}^4 \eR_i^{(6)}\, \mm_i^{(6)} + \sum_{i=1}^4 \eR_i^{(8)}\, \mm_i^{(4)}
\notag\\[5pt]
&\qquad
+ \sum_{i=1}^2 \eR_i^{(10)}\, \mm_i^{(2)} + \Big[ \eR^{(12)} + \eR^{(0)}\, J \Big] \,,
\qquad\text{with}\qquad
\eR \in \rPI \,.
\label{eqn:M12Reduction}
\end{align}
Note that we use ``$\eR$'' to denote elements in the polynomial ring of the primary invariants, $\rPI$, while ``$c$'' is used to denote elements in the ring of all invariants, $\rI$. \cref{eqn:M12Reduction} says that any monomial $\mm^{(12)}(h_u, h_d)$ can be written as a linear combination, with elements in $\rPI$ as coefficients, of $1$, $J$, and the
$2$ independent $\mm_i^{(10)}$,
$4$ independent $\mm_i^{(8)}$,
$4$ independent $\mm_i^{(6)}$,
$4$ independent $\mm_i^{(4)}$, and
$2$ independent $\mm_i^{(2)}$ that we have discussed above. An explicit choice of them are provided in \cref{tab:GeneratingSet1811}, which are organized into hermitian and anti-hermitian combinations.

The reduction in \cref{eqn:M12Reduction} has profound implications. We can use it to eventually show that all the matrix polynomials defined in \cref{eqn:MPdef} share the same reduction. To get there, however, we first need another important ingredient: the same reduction in \cref{eqn:M12Reduction} holds also for another combination --- the Jarlskog $J$ multiplied with an order $\mm^{(2)}$ matrix monomial:
\begin{align}
J \mm^{(2)} &= \sum_{i=1}^2 \eR_i^{(4)}\, \mm_i^{(10)} + \sum_{i=1}^4 \eR_i^{(6)}\, \mm_i^{(8)}
+ \sum_{i=1}^4 \eR_i^{(8)}\, \mm_i^{(6)} + \sum_{i=1}^4 \eR_i^{(10)}\, \mm_i^{(4)}
\notag\\[5pt]
&\qquad
+ \sum_{i=1}^2 \eR_i^{(12)}\, \mm_i^{(2)} + \Big[ \eR^{(14)} + \eR^{(2)}\, J \Big] \,,
\qquad\text{with}\qquad
\eR \in \rPI \,.
\label{eqn:JM2Reduction}
\end{align}
To show this, it is convenient to work with the traceless components of $h_u$ and $h_d$.  Denoting by an overline the traceless component of a matrix
\begin{equation}
\trless{A} \equiv A - \frac13 \tr(A)\, I_3 \,,
\label{eqn:trlessdef}
\end{equation}
we define 
\begin{equation}
\tu \equiv \trless{h_u} \,,\qquad
\td \equiv \trless{h_d} \,.
\label{eqn:tutd}
\end{equation}
Making use of the definition in \cref{eqn:Jarlskog} and the Cayley-Hamilton identity in \cref{eqn:CH3} for the traceless matrix $\comm{\tu}{\td}$, one can rewrite the Jarlskog as
\begin{equation}
J = \tfrac13 \tr \Big( \comm{h_u}{h_d}^3 \Big)
= \tfrac13 \tr \big( \comm{\tu}{\td}^3 \big)
= \comm{\tu}{\td}^3 - \tfrac12 \big( \tr_{\comm{\tu}{\td}^2} \big) \comm{\tu}{\td} \,.
\label{eqn:Jarlskogtutd}
\end{equation}
Here and below, $\tr_X=\tr(X)$ for clarity. With this, one can rewrite $J \tu$ and $J \td$ as matrix polynomials of $u, d$, and then make use of the identity in \cref{eqn:CH3ABC} to reduce its power. In the end, we find the following relations
\begin{subequations}\label{eqn:JhuJhd}
\begin{align}
J \tu &= \big( \tr_{\tu \td} \big)\, \tu^2 \td^2 \tu - \big( \tr_{\tu^2} \big)\, \td^2 \tu^2 \td
- \big( \tr_{\tu \td^2} \big)\, \tu^2 \td \tu
+ \big( \tr_{\tu^3} \big)\, \td^2 \tu \td
+ 2 \big( \tr_{\tu^2 \td} \big)\, \tu^2 \td^2
\notag\\[5pt]
&\hspace{20pt}
- \big( 2 \tr_{\tu^2 \td^2} - \tr_{\tu^2} \tr_{\td^2} \big)\, \tu^2 \td
- \tfrac13\, \big( \tr_{\tu^3} \tr_{\td^2} \big)\, \tu \td - \hc
\,, \\[10pt]
J \td &= \big( \tr_{\td^2} \big)\, \tu^2 \td^2 \tu - \big( \tr_{\tu \td} \big)\, \td^2 \tu^2 \td
- \big( \tr_{\td^3} \big)\, \tu^2 \td \tu
+ \big( \tr_{\tu^2 \td} \big)\, \td^2 \tu \td
+ 2 \big( \tr_{\tu \td^2} \big)\, \tu^2 \td^2
\notag\\[5pt]
&\hspace{20pt}
- \big( 2 \tr_{\tu^2 \td^2} - \tr_{\tu^2} \tr_{\td^2} \big)\, \tu \td^2
- \tfrac13\, \big( \tr_{\td^3} \tr_{\tu^2} \big)\, \tu \td - \hc \,.
\end{align}
\end{subequations}
Once plugging in \cref{eqn:tutd}, this clearly proves \cref{eqn:JM2Reduction}. Note that compared to the general expectation in \cref{eqn:JM2Reduction}, only anti-hermitian matrix polynomials in \cref{tab:GeneratingSet1811} show up in the results. This is because $J \tu$ and $J \td$ are anti-hermitian, whereas primary invariants are real.

Now with the reduction in both \cref{eqn:M12Reduction,eqn:JM2Reduction} proven, let us make use of them to show that a general matrix polynomial defined in \cref{eqn:MPdef} share the same reduction. To this end, we follow the steps below:
\begin{enumerate}
\item The same reduction holds for a matrix monomial of any order $\mm^{(2k)}$:
\begin{equation}
\mm^{(2k)} = \eR_i^{(2k-10)}\, \mm_i^{(10)} + \cdots + \eR_i^{(2k-2)}\, \mm_i^{(2)} + \Big[ \eR^{(2k)} + \eR^{(2k-12)}\, J \Big] \,.
\label{eqn:M2kReduction}
\end{equation}
Here we have suppressed the explicit sums over each individual dummy index $i$; the sums have the same form as in \cref{eqn:M12Reduction,eqn:JM2Reduction}. We can prove \cref{eqn:M2kReduction} by induction. Assuming it holds at $k=n$, then using $\mm^{(2n+2)} = (h_u \text{ or } h_d)\, \mm^{(2n)}$ together with \cref{eqn:M12Reduction,eqn:JM2Reduction}, it implies that \cref{eqn:M2kReduction} holds at $k=n+1$. Given that we have successfully demonstrated \cref{eqn:M2kReduction} up to $k=6$ in \cref{eqn:M12Reduction}, we conclude that \cref{eqn:M2kReduction} holds for all non-negative integer $k$.
\item The same reduction holds for $J \mm^{(2k)}$:
\begin{equation}
J \mm^{(2k)} = \eR_i^{(2k+2)}\, \mm_i^{(10)} + \cdots + \eR_i^{(2k+10)}\, \mm_i^{(2)} + \Big[ \eR^{(2k+12)} + \eR^{(2k)}\, J \Big] \,.
\label{eqn:JM2kReduction}
\end{equation}
We can prove this again by induction. Assuming \cref{eqn:JM2kReduction} holds at $k=n$, then using $J \mm^{(2n+2)} = (h_u \text{ or } h_d)\, J \mm^{(2n)}$ together with \cref{eqn:M12Reduction,eqn:JM2Reduction}, it implies that \cref{eqn:JM2kReduction} holds at $k=n+1$. Given that we have successfully demonstrated \cref{eqn:JM2kReduction} up to $k=1$ in \cref{eqn:JM2Reduction}, we conclude that \cref{eqn:JM2kReduction} holds for all non-negative integer $k$.
\item As each invariant has a Hironaka decomposition, \cref{eqn:M2kReduction,eqn:JM2kReduction} together means that any matrix polynomial defined in \cref{eqn:MPdef} shares same reduction:
\begin{align}
p(h_u, h_d) &\equiv \sum_{k,i} c_{2k,i}(h_u, h_d)\; \mm_i^{(2k)}(h_u, h_d) \notag\\[5pt]
&= \eR_{10,i}\, \mm_i^{(10)} + \cdots + \eR_{2,i}\, \mm_i^{(2)} + \big( \eR_0 + \eR_J\, J \big) \,,
\label{eqn:pReduction}
\end{align}
where the 18 terms in \cref{tab:GeneratingSet1811} serve as a generating set, and the linear combination coefficients are in the polynomial ring of the primary invariants, $\rPI$. Here, similar to $c_{2k,i}$, we have used the order agnostic notation $\eR_{2k,i}$.
\end{enumerate}

\paragraph{$\mathbf{(1\oplus8,1,1)}$ as an $\rPI$-module:} With the above understanding of the coefficients in \cref{eqn:pReduction}, we see that the matrix polynomials in \cref{tab:GeneratingSet1811} form a generating set of the module ${}_\rPI\Mod_\mathbf{(1\oplus8,1,1)}$. Furthermore, they precisely agree with the numerators in the Hilbert series \cref{eqn:HS1811,eqn:HS1811ud}, meaning that the decomposition in \cref{eqn:pReduction} is actually unique --- there is no redundancy relations among this generating set, so they give a basis for the module ${}_\rPI\Mod_\mathbf{(1\oplus8,1,1)}$ and the module is a free module.

\paragraph{$\mathbf{(1\oplus8,1,1)}$ as an $\rI$-module:} Once we allow the Jarlskog invariant $J$ to be part of the linear combination coefficients, the matrix polynomials in \cref{tab:GeneratingSet1811} are no longer linearly independent, due to the relation in \cref{eqn:JM2kReduction}. Therefore, the matrix polynomials in \cref{tab:GeneratingSet1811} form a generating set for the module ${}_\rI\Mod_\mathbf{(1\oplus8,1,1)}$, but not a basis for it. In fact, the lack of the factor $(1+q^{12})$ in the numerator of the Hilbert series in \cref{eqn:HS1811} implies that the module ${}_\rI\Mod_\mathbf{(1\oplus8,1,1)}$ is not a free module.

\begin{table}[t]
\renewcommand{\arraystretch}{1.2}
\setlength{\arrayrulewidth}{.2mm}
\setlength{\tabcolsep}{1em}
\centering
\begin{tabular}{cc}
\toprule
\(q^0\) & \( 1 \) \\
\midrule
\(q^2\) & \( h_u \quad,\quad   h_d \) \\
\midrule
\(q^4\) & \( h_u^2 \quad,\quad   h_d^2 \quad,\quad   h_u h_d \pm \hc \) \\
\midrule
\(q^6\) & \( h_u^2 h_d \pm \hc \quad,\quad   h_u h_d^2 \pm \hc \) \\
\midrule
\(q^8\) & \( h_u^2 h_d^2 \pm \hc \quad,\quad   h_u^2 h_d h_u - \hc \quad,\quad   h_d^2 h_u h_d - \hc \) \\
\midrule
\(q^{10}\) & \( h_u^2 h_d^2 h_u - \hc \quad,\quad   h_d^2 h_u^2 h_d - \hc \) \\
\midrule
\(q^{12}\) & \( J \) \\
\bottomrule
\end{tabular}
\caption{A generating set and basis of the module ${}_\rPI\Mod_\mathbf{(1\oplus8,1,1)}$. We have organized terms into hermitian and anti-hermitian combinations for convenience so as to understand the absence of certain terms in some results in the text. For certain polynomials, taking a hermitian conjugate is the same as taking $u\leftrightarrow d$ exchange. Note also that the anti-hermitian ones are traceless.}
\label{tab:GeneratingSet1811}
\end{table}

\subsubsection{The $\mathbf{(8,1,1)}$ covariants}
\label{subsubsec:HS811}

With all the analyses in \cref{subsubsec:MatrixPolynomials}, it is now straightforward to understand the Hilbert series for the $SU(3)_q$ octets $\mathbf{(8,1,1)}$ in \cref{eqn:HS811}:
\begin{equation}
\HS_\mathbf{(8,1,1)}(q) = \frac{2\left( q^2 + 2q^4 + 2q^6 + 2q^8 + q^{10} \right)}{\left(1-q^2\right)^2 \left(1-q^4\right)^3 \left(1-q^6\right)^4 \left(1-q^8\right)} \,.
\label{eqn:HS811Redo}
\end{equation}
Any $\mathbf{(8,1,1)}$ covariant can be constructed as the traceless component of a matrix polynomial in $h_u, h_d$ defined in \cref{eqn:MPdef}, namely that
\begin{equation}
V_\mathbf{(1\oplus8,1,1)} \sim p(h_u, h_d) \,,\qquad
V_\mathbf{(8,1,1)} \sim \trless{p(h_u, h_d)} \,.
\end{equation}
As before (see \cref{eqn:trlessdef}), we use an over line to denote the traceless component of a matrix 
\begin{equation}
\trless{A} \equiv A - \frac13 \tr(A)\, I_3 \,.
\end{equation}
Taking the traceless component of \cref{eqn:pReduction}, we find a similar reduction structure for the traceless polynomials:
\begin{equation}
\trless{p(h_u, h_d)} = \eR_{10,i}\, \trless{M_i^{(10)}(h_u, h_d)} + \eR_{8,i}\, \trless{M_i^{(8)}(h_u, h_d)}
+ \cdots + \eR_{2,i}\, \trless{M_i^{(2)}(h_u, h_d)} \,,
\label{eqn:trlessphuhdReduction}
\end{equation}
where the coefficients $\eR_{10,i}, \cdots, \eR_{2,i}$ are all made out of primary invariants. Therefore, a generating set of the module ${}_\rPI\Mod_\mathbf{(8,1,1)}$ is given by the traceless components of the 16 matrix terms in the middle sections of \cref{tab:GeneratingSet1811}, which we summarize in \cref{tab:GeneratingSet811}. These coincide with the numerator of the Hilbert series in \cref{eqn:HS811Redo}, telling us that they form a basis and ${}_\rPI\Mod_\mathbf{(8,1,1)}$ is a free module; its rank is 16.

\begin{table}[t]
\renewcommand{\arraystretch}{1.5}
\setlength{\arrayrulewidth}{.2mm}
\setlength{\tabcolsep}{1em}
\centering
\begin{tabular}{cc}
\toprule
\(q^2\) & \( \trless{h_u}\quad,\quad   \trless{h_d} \) \\
\midrule
\(q^4\) & \( \trless{h_u^2}\quad,\quad   \trless{h_d^2}\quad,\quad   \trless{h_u h_d \pm \hc } \) \\
\midrule
\(q^6\) & \( \trless{h_u^2 h_d \pm \hc}\quad,\quad   \trless{h_u h_d^2 \pm \hc} \) \\
\midrule
\(q^8\) & \( \trless{h_u^2 h_d^2 \pm \hc}\quad,\quad   \trless{h_u^2 h_d h_u - \hc}\quad,\quad   \trless{h_d^2 h_u h_d - \hc} \) \\
\midrule
\(q^{10}\) & \( \trless{h_u^2 h_d^2 h_u - \hc}\quad,\quad   \trless{h_d^2 h_u^2 h_d - \hc} \) \\
\bottomrule
\end{tabular}
\caption{A generating set and basis of the module ${}_\rPI\Mod_\mathbf{(8,1,1)}$. Our notation for the traceless component reads $\trless{A} \equiv A - \tfrac13 \tr(A)\, I_3$. }
\label{tab:GeneratingSet811}
\end{table}

The 16 terms in \cref{tab:GeneratingSet811} also form a generating set of the module ${}_\rI\Mod_\mathbf{(8,1,1)}$. But they are no longer linearly independent when the Jarlskog invariant $J$ is allowed to be involved in the linear combination coefficients. Therefore, they do not form a basis for ${}_\rI\Mod_\mathbf{(8,1,1)}$. The rank of this module is 8. In fact, ${}_\rI\Mod_\mathbf{(8,1,1)}$ is not a free module, as indicated by the lack of a factor $(1+q^{12})$ in the numerator of the Hilbert series in \cref{eqn:HS811Redo}.

\subsubsection{The $\mathbf{(3,\overline{3},1)}$, $\mathbf{(3,1,\overline{3})}$, and $\mathbf{(1,3,\overline{3})}$ covariants}
\label{subsubsec:HS33bar1}

The Hilbert series for the $\mathbf{(3,\overline{3},1)}$, $\mathbf{(3,1,\overline{3})}$, and $\mathbf{(1,3,\overline{3})}$ covariants are given in \cref{eqn:HS33bar1,eqn:HS133bar} (together with \cref{eqn:HSMFVsame}). These are also straightforward to understand in terms of the Hilbert series of the invariants and $SU(3)_q$ octets $\HS_\mathbf{(1\oplus8,1,1)}$, because the corresponding covariants can be constructed as
\begin{subequations}\label{eqn:331and133from1811}
\begin{align}
V_\mathbf{(3,\overline 3,1)} &\sim V_\mathbf{(1\oplus8,1,1)}\, Y_u \,, \\[5pt] 
V_\mathbf{(3,1,\overline 3)} &\sim V_\mathbf{(1\oplus8,1,1)}\, Y_d \,, \\[5pt]
V_\mathbf{(1,3,\overline 3)} &\sim Y_u^\dagger\, V_\mathbf{(1\oplus8,1,1)}\, Y_d \,.
\end{align}
\end{subequations}
As a consequence, the Hilbert series satisfy the following relations
\begin{subequations}
\begin{align}
\HS_\mathbf{(3,\overline 3,1)} = \HS_\mathbf{(3,1,\overline 3)}
&= q\, \HS_\mathbf{(1\oplus8,1,1)} \,, \\[5pt]
\HS_\mathbf{(1,3,\overline 3)}
&= q^2\, \HS_\mathbf{(1\oplus8,1,1)} \,.
\end{align}
\end{subequations}
This simple correspondence allows us to translate the properties of the $\mathbf{(1\oplus8,1,1)}$ module to those of $\mathbf{(3,\overline{3},1)}$, $\mathbf{(3,1,\overline{3})}$, and $\mathbf{(1,3,\overline{3})}$. In particular, they are also free modules when viewed over the ring $\rPI$ and the generating set in \cref{tab:GeneratingSet1811} can be easily translated to these modules using \cref{eqn:331and133from1811}, which we tabulate in \cref{tab:GeneratingSet33bar1,tab:GeneratingSet313bar,tab:GeneratingSet133bar}, respectively.

\begin{table}[t]
\renewcommand{\arraystretch}{1.2}
\setlength{\arrayrulewidth}{.2mm}
\setlength{\tabcolsep}{1em}
\centering
\begin{tabular}{cc}
\toprule
\(q\) & \(  Y_u \) \\
\midrule
\(q^3\) & \( h_u\, Y_u\quad,\quad   h_d\, Y_u \) \\
\midrule
\(q^5\) & \( h_u^2\, Y_u\quad,\quad   h_d^2\, Y_u\quad,\quad   (h_u h_d \pm \hc)\, Y_u \) \\
\midrule
\(q^7\) & \( (h_u^2 h_d \pm \hc)\, Y_u\quad,\quad   (h_u h_d^2 \pm \hc)\, Y_u \) \\
\midrule
\(q^9\) & \( (h_u^2 h_d^2 \pm \hc)\, Y_u\quad,\quad   (h_u^2 h_d h_u - \hc)\, Y_u\quad,\quad   (h_d^2 h_u h_d - \hc)\, Y_u \) \\
\midrule
\(q^{11}\) & \( (h_u^2 h_d^2 h_u - \hc)\, Y_u\quad,\quad   (h_d^2 h_u^2 h_d - \hc)\, Y_u \) \\
\midrule
\(q^{13}\) & \( J\, Y_u \) \\
\bottomrule
\end{tabular}
\caption{A generating set and basis of the module ${}_\rPI\Mod_\mathbf{(3,\overline 3,1)}$.}
\label{tab:GeneratingSet33bar1}
\end{table}

\begin{table}[t]
\renewcommand{\arraystretch}{1.2}
\setlength{\arrayrulewidth}{.2mm}
\setlength{\tabcolsep}{1em}
\centering
\begin{tabular}{cc}
\toprule
\(q\) & \(  Y_d \) \\
\midrule
\(q^3\) & \( h_u\, Y_d\quad,\quad   h_d\, Y_d \) \\
\midrule
\(q^5\) & \( h_u^2\, Y_d\quad,\quad   h_d^2\, Y_d\quad,\quad   (h_u h_d \pm \hc)\, Y_d \) \\
\midrule
\(q^7\) & \( (h_u^2 h_d \pm \hc)\, Y_d\quad,\quad   (h_u h_d^2 \pm \hc)\, Y_d \) \\
\midrule
\(q^9\) & \( (h_u^2 h_d^2 \pm \hc)\, Y_d\quad,\quad   (h_u^2 h_d h_u - \hc)\, Y_d\quad,\quad   (h_d^2 h_u h_d - \hc)\, Y_d \) \\
\midrule
\(q^{11}\) & \( (h_u^2 h_d^2 h_u - \hc)\, Y_d\quad,\quad   (h_d^2 h_u^2 h_d - \hc)\, Y_d \) \\
\midrule
\(q^{13}\) & \( J\, Y_d \) \\
\bottomrule
\end{tabular}
\caption{A generating set and basis of the module ${}_\rPI\Mod_\mathbf{(3,1,\overline 3)}$.}
\label{tab:GeneratingSet313bar}
\end{table}

\begin{table}[t]
\renewcommand{\arraystretch}{1.2}
\setlength{\arrayrulewidth}{.2mm}
\setlength{\tabcolsep}{1em}
\centering
\begin{tabular}{cc}
\toprule
\(q^2\) & \(  Y_u^\dagger\, Y_d \) \\
\midrule
\(q^4\) & \( Y_u^\dagger\, h_u\, Y_d\quad,\quad   Y_u^\dagger\, h_d\, Y_d \) \\
\midrule
\(q^6\) & \( Y_u^\dagger\, h_u^2\, Y_d\quad,\quad   Y_u^\dagger\, h_d^2\, Y_d\quad,\quad   Y_u^\dagger\, (h_u h_d \pm \hc)\, Y_d \) \\
\midrule
\(q^8\) & \( Y_u^\dagger\, (h_u^2 h_d \pm \hc)\, Y_d\quad,\quad   Y_u^\dagger\, (h_u h_d^2 \pm \hc)\, Y_d \) \\
\midrule
\(q^{10}\) & \( Y_u^\dagger\, (h_u^2 h_d^2 \pm \hc)\, Y_d\quad,\quad   Y_u^\dagger\, (h_u^2 h_d h_u - \hc)\, Y_d\quad,\quad   Y_u^\dagger\, (h_d^2 h_u h_d - \hc)\, Y_d \) \\
\midrule
\(q^{12}\) & \( Y_u^\dagger\, (h_u^2 h_d^2 h_u - \hc)\, Y_d\quad,\quad   Y_u^\dagger\, (h_d^2 h_u^2 h_d - \hc)\, Y_d \) \\
\midrule
\(q^{14}\) & \( Y_u^\dagger\, J\, Y_d \) \\
\bottomrule
\end{tabular}
\caption{A generating set and basis of the module ${}_\rPI\Mod_\mathbf{(1,3,\overline{3})}$.}
\label{tab:GeneratingSet133bar}
\end{table}

\subsubsection{The $\mathbf{(1,8,1)}$ covariants}
\label{subsubsec:HS181}

Let us move on to the $\mathbf{(1,8,1)}$ covariants. The Hilbert series is given in \cref{eqn:HS181}:
\begin{equation}
\HS_\mathbf{(1,8,1)} (q) = \frac{ q^2 \left(1+2 q^2+3 q^4+4 q^6+4 q^8+2 q^{10}+q^{12}-q^{16}\right)}
{\left(1-q^2\right)^2\left(1-q^4\right)^3\left(1-q^6\right)^4\left(1-q^8\right)} \,.
\label{eqn:HS181Redo}
\end{equation}
The first lesson we derive from this is that the module ${}_\rPI\Mod_\mathbf{(1,8,1)}$ is not a free module since there is a negative sign in the numerator signaling at least a linear relation among the elements of the generating set.

We can understand this Hilbert series by noting that one can construct all the $\mathbf{(1,8,1)}$ covariants as
\begin{equation}
V_\mathbf{(1,8,1)} \sim \trless{Y_u^\dagger\, V_\mathbf{(1\oplus8,1,1)}\, Y_u} \,.
\label{eqn:181from811}
\end{equation}
Then one might naively expect the Hilbert series for $\mathbf{(1,8,1)}$ covariants to be
\begin{equation}
\HS_\mathbf{(1,8,1)} \Big|_\text{naive} = q^2\, \HS_\mathbf{(1\oplus8,1,1)}
= \frac{q^2\left(1+2 q^2+4 q^4+4 q^6+4 q^8+2 q^{10}+q^{12}\right)}{\left(1-q^2\right)^2\left(1-q^4\right)^3\left(1-q^6\right)^4\left(1-q^8\right)} \,.
\label{eqn:HS181Naive}
\end{equation}
This is almost the correct result in \cref{eqn:HS181Redo}, which only differs in the coefficients in the numerator of orders $\O(q^6)$ and $\O(q^{18})$. The mismatch stems from the overcounting in \cref{eqn:181from811}. The reason is that there are covariants transforming as $\mathbf{(8,1,1)}$ that when inserted as $Y_u^\dagger\, V_{(8,1,1)}\, Y_u$ can be reduced to lower order covariants using the Cayley-Hamilton identities. For example, at the order $\O(q^6)$ we have
\begin{equation}
\trless{Y_u^\dagger\, h_u^2\, Y_u} = \trless{\left( Y_u^\dagger Y_u \right)^3}
= \big( \tr_{h_u} \big)\, \trless{Y_u^\dagger h_u Y_u}
+ \frac12\, \big( \tr_{h_u^2} - \tr_{h_u}^2 \big)\, \trless{Y_u^\dagger Y_u} \,.
\label{eqn:V181Relationq6}
\end{equation}
This explains why we need to subtract a term $q^6$ in the numerator of \cref{eqn:HS181Naive} (not to count it twice) to correctly reproduce \cref{eqn:HS181Redo}.

\begin{table}[t]
\renewcommand{\arraystretch}{1.8}
\setlength{\arrayrulewidth}{.2mm}
\setlength{\tabcolsep}{1em}
\centering
\begin{tabular}{cc}
\toprule
\(q^2\) & \( \trless{Y_u^\dagger Y_u} \) \\
\midrule
\(q^4\) & \( \trless{Y_u^\dagger h_u Y_u}\quad,\quad   \trless{Y_u^\dagger h_d Y_u} \) \\
\midrule
\(q^6\) & \( \trless{Y_u^\dagger h_d^2 Y_u}\quad,\quad   \trless{Y_u^\dagger h_u h_d Y_u} \pm \hc \) \\
\midrule
\(q^8\) & \( \trless{Y_u^\dagger h_u^2 h_d Y_u} \pm \hc\quad,\quad   \trless{Y_u^\dagger h_u h_d^2 Y_u} \pm \hc \) \\
\midrule
\(q^{10}\) & \( \trless{Y_u^\dagger h_u^2 h_d^2 Y_u} \pm \hc\quad,\quad   \trless{Y_u^\dagger h_u^2 h_d h_u Y_u} - \hc\quad,\quad   \trless{Y_u^\dagger h_d^2 h_u h_d Y_u} - \hc \) \\
\midrule
\(q^{12}\) & \( \trless{Y_u^\dagger h_u^2 h_d^2 h_u Y_u} - \hc \quad,\quad   \trless{Y_u^\dagger h_d^2 h_u^2 h_d Y_u} - \hc \) \\
\midrule
\(q^{14}\) & \( J\, \trless{Y_u^\dagger Y_u} \) \\
\bottomrule
\end{tabular}
\caption{A generating set of the module ${}_\rPI\Mod_\mathbf{(1,8,1)}$, which is not a basis.}
\label{tab:GeneratingSet181}
\end{table}

\begin{table}[t]
\renewcommand{\arraystretch}{1.8}
\setlength{\arrayrulewidth}{.2mm}
\setlength{\tabcolsep}{1em}
\centering
\begin{tabular}{cc}
\toprule
\(q^2\) & \( \trless{Y_d^\dagger Y_d} \) \\
\midrule
\(q^4\) & \( \trless{Y_d^\dagger h_u Y_d}\quad,\quad   \trless{Y_d^\dagger h_d Y_d} \) \\
\midrule
\(q^6\) & \( \trless{Y_d^\dagger h_u^2 Y_d}\quad,\quad   \trless{Y_d^\dagger h_u h_d Y_d} \pm \hc \) \\
\midrule
\(q^8\) & \( \trless{Y_d^\dagger h_u^2 h_d Y_d} \pm \hc\quad,\quad   \trless{Y_d^\dagger h_u h_d^2 Y_d} \pm \hc \) \\
\midrule
\(q^{10}\) & \( \trless{Y_d^\dagger h_u^2 h_d^2 Y_d} \pm \hc\quad,\quad   \trless{Y_d^\dagger h_u^2 h_d h_u Y_d} - \hc\quad,\quad   \trless{Y_d^\dagger h_d^2 h_u h_d Y_d} - \hc \) \\
\midrule
\(q^{12}\) & \( \trless{Y_d^\dagger h_u^2 h_d^2 h_u Y_d} - \hc\quad,\quad   \trless{Y_d^\dagger h_d^2 h_u^2 h_d Y_d} - \hc \) \\
\midrule
\(q^{14}\) & \( J\, \trless{Y_d^\dagger Y_d} \) \\
\bottomrule
\end{tabular}
\caption{A generating set of the module ${}_\rPI\Mod_\mathbf{(1,1,8)}$, which is not a basis.}
\label{tab:GeneratingSet118}
\end{table}

Using the relation in \cref{eqn:181from811}, one can obtain a generating set of the module ${}_\rPI\Mod_\mathbf{(1,8,1)}$ from that of the module ${}_\rPI\Mod_\mathbf{(1\oplus8,1,1)}$ listed in \cref{tab:GeneratingSet1811}. The $\mathbf{(1,8,1)}$ covariant in \cref{eqn:V181Relationq6} can be removed from this set, as it gets generated by the others already. This leads us to the generating set of 17 covariants listed in \cref{tab:GeneratingSet181}. However, this is not a basis for the module, because there is one linear relation among them, and the rank of the module is 16. To find the linear redundancy, we first note that \cref{eqn:V181Relationq6} can be rewritten as
\begin{equation}
\trless{Y_u^\dagger \Big[ h_u^2 - \big(\tr_{h_u}\big) h_u \Big] Y_u}
= \frac12\, \big( \tr_{h_u^2} - \tr_{h_u}^2 \big)\, \trless{Y_u^\dagger Y_u} \,.
\label{eqn:V181Relationq61}
\end{equation}
Let us temporary call the matrix in the squared brackets $B$, then we have
\begin{equation}
B \equiv h_u^2 - \big(\tr_{h_u}\big) h_u
\qquad\Longrightarrow\qquad
\tr_B = \tr_{h_u^2} - \tr_{h_u}^2 \,,
\end{equation}
and therefore \cref{eqn:V181Relationq61} reads
\begin{equation}
\trless{Y_u^\dagger\, B\, Y_u} = \frac12\, \big( \tr_B \big)\, \trless{Y_u^\dagger Y_u}
\qquad\Longrightarrow\qquad
\trless{Y_u^\dagger\, \trless{B}\, Y_u} = \frac16\, \big( \tr_B \big)\, \trless{Y_u^\dagger Y_u} \,.
\label{eqn:V181Relationq62}
\end{equation}
This relation is fairly nontrivial. It says that a traceless matrix $\trless{B}$ and a number $\frac16\,\tr_B$ will give the same result once inserted into $\trless{Y_u^\dagger (\cdots) Y_u}$. This can be used to find a linear redundancy relation among the generating set in \cref{tab:GeneratingSet181}. Concretely, we multiply both sides of \cref{eqn:V181Relationq62} with the Jarlskog invariant $J$ to obtain
\begin{equation}
\frac16\, \big( \tr_B \big)\, J\, \trless{Y_u^\dagger Y_u} = \trless{Y_u^\dagger\, J\,\trless{B}\, Y_u} \,.
\label{eqn:V181Relationq18}
\end{equation}
The left-hand side is the generating vector $J\, \trless{Y_u^\dagger Y_u}$ (the last row in \cref{tab:GeneratingSet181}) multiplied by an element in $\rPI$. On the right-hand side, the matrix $J\,\trless{B}$ is an $\mathbf{(8,1,1)}$ covariant, which according to \cref{subsubsec:MatrixPolynomials,subsubsec:HS811} can be written as a linear combination of the 16 matrices in \cref{tab:GeneratingSet811}, with again elements in $\rPI$ as coefficients; much as in  \cref{eqn:JhuJhd}, this is given in terms of the anti-hermitian polynomials in \cref{tab:GeneratingSet811} only: 
\begin{align}
J\, \trless{B} &= \eR_{h_u h_d}\, \trless{h_u h_d}
+ \eR_{h_u^2 h_d}\, \trless{h_u^2 h_d}
+ \eR_{h_u h_d^2}\, \trless{h_u h_d^2}
+ \eR_{h_u^2 h_d h_u}\, \trless{h_u^2 h_d h_u}
+ \eR_{h_d^2 h_u h_d}\, \trless{h_d^2 h_u h_d}
\notag\\[5pt]
&\quad
+ \eR_{h_u^2 h_d^2}\, \trless{h_u^2 h_d^2}
+ \eR_{h_u^2 h_d^2 h_u}\, \trless{h_u^2 h_d^2 h_u}
+ \eR_{h_d^2 h_u^2 h_d}\, \trless{h_d^2 h_u^2 h_d} - \hc \,.
\end{align}
Using this, together with the general property that holds for any matrix $A$
\begin{equation}
\trless{Y_u^\dagger\, \trless{A}\, Y_u}
= \trless{Y_u^\dagger\, A\, Y_u} - \frac13\, \big(\tr_A\big)\, \trless{Y_u^\dagger Y_u} \,,
\end{equation}
we can write $\trless{Y_u^\dagger\, J\,\trless{B}\, Y_u}$ into a linear combination of the top 16 generating vectors in \cref{tab:GeneratingSet181}, not involving the one in the last row. Therefore, \cref{eqn:V181Relationq18} is a linear redundancy relation among the 17 generating vectors in \cref{tab:GeneratingSet181}, with nontrivial elements in $\rPI$ as coefficients. This redundancy relation is of order $\O(q^{18})$, corresponding to the $-q^{18}$ term in the numerator of the Hilbert series in \cref{eqn:HS181Redo}.

The set of $\mathbf{(1,1,8)}$ covariants share the same properties as those of $\mathbf{(1,8,1)}$ covariants, upon $u\leftrightarrow d$ exchange. Their Hilbert series is the same. A generating set of the module ${}_\rPI\Mod_\mathbf{(1,1,8)}$ is listed in \cref{tab:GeneratingSet118}. Again, there is one linear relation among them and therefore they do not form a basis for the module.

\subsection{Saturation for MFV SMEFT operators}
\label{subsec:MFVSaturation}

One of the goals of this work is to study the number of independent MFV covariants that arise for each operator of the SMEFT, which is a finite number, and explore whether they lead to specific correlations among flavor observables that allow us to disentangle MFV (without extra assumptions) from a general EFT. As we explained in \cref{subsec:RankSaturation}, the number of independent rep-$R$ covariants is given by the rank of the module $_{\rI}\Mod_R^{G,\,\RBB}$, which can be computed from the Hilbert series of the module and that of the ring of invariants applying the formula in \cref{eqn:rankHSRatiosRrI}:
\begin{equation}
\rank \left( _{\rI}\Mod_R^{G,\, \RBB} \right)
= \frac{\HS_R^{G,\, \RBB}(q)}{\HS_\I^{G,\, \RBB}(q)} \Bigg|_{q=1} \,.
\label{eqn:rankHSRatiosRrIadvRedo}
\end{equation}
Given the Hilbert series computed in \cref{eqn:HSMFV}, it is straightforward to apply this formula to compute the rank of all the modules arising in MFV dim-6 SMEFT. The result is that rank saturation is achieved for all the modules, \ie\ the rank of each module equals the dimension of the corresponding representation $R_\text{MFV}$:
\begin{equation}
\rank \left( {}_\rI\Mod_{R_\text{MFV}} \right) = \dim \left( R_\text{MFV} \right) \,.
\label{eqn:SaturationFlav}
\end{equation}
Moreover, utilizing the theorem in \cref{eqn:TheoremBrion}, this result can be extended to \emph{any} flavor representation. Consequently, the saturation is not confined solely to MFV dim-6 SMEFT but is applicable to SMEFT at any mass dimension, as well as to any other EFTs sharing the same flavor group. In order to apply the theorem, we need to identify the subgroup $H$ of the flavor group $G_f$ in \cref{eqn:GFdef} that remains unbroken when the Yukawas $Y_u, Y_d$ take a generic vev; we find
\begin{align}
G_f = U(3)_q \times U(3)_u \times U(3)_d
\quad\xrightarrow{\langle Y_u\rangle ,\, \langle Y_d\rangle}\quad
H = U(1)_{BN} \,.
\end{align}
Now consider an arbitrary representation $R_\text{MFV}$ of the flavor group $G_f$. As long as it is a neutral rep under the $U(1)_\text{BN}\subset G_f$ (irrespective of its transformation under the non-abelian component), we have the rank saturation
\begin{equation}
\rank \left( {}_\rI\Mod_{R_\text{MFV}} \right) = \dim \left( R_\text{MFV}^H \right) = \dim \left( R_\text{MFV} \right) \,,
\label{eqn:TheoremBrionToFlav}
\end{equation}
meaning that the most general version of this rep $R_\text{MFV}$ with vanishing baryon number can be generated by our building blocks $Y_u, Y_d$. This proves that saturation is guaranteed, \ie\ MFV places no restriction, for any flavor representation emerging at any mass dimension in the baryon number preserving SMEFT.

This remark will also hold when taking into account the baryon number violating operators in the dim-6 SMEFT in \cref{tab:BNVMFVdim6SMEFT}, as long as the relaxed MFV flavor group in \cref{eqn:GFRelaxed} is considered. Indeed, by applying \cref{eqn:rankHSRatiosRrIadvRedo} to the Hilbert series in \cref{eqn:HSMFVBNV}, one finds that \cref{eqn:SaturationFlav} also holds for this case and saturation takes place for the baryon number violating operators. Once again this result can be understood from the theorem by Brion \cite{brion1993modules} in \cref{eqn:TheoremBrion}. The unbroken subgroup is now $H'=Z_3$, the center of the non-chiral group $SU(3)_V\equiv SU(3)_{q=d=u}$:
\begin{align}
G'_f = SU(3)_q \times U(3)_u \times U(3)_d
\quad\xrightarrow{\langle Y_u\rangle, \, \langle Y_d\rangle}\quad
Z_3 \,.
\end{align}
It corresponds to $U(1)_\text{BN}$ rotations of angles $0, 2\pi/3, 4\pi/3$. Since each baryon number violating operator in \cref{tab:BNVMFVdim6SMEFT} contains exactly 3 quarks, all the corresponding $G'_f$ flavor representations are invariant under $H'$. Moreover, the group $H'$ actually coincides with the center of the QCD gauge group $SU(3)_\text{QCD}$, so any gauge invariant operator will also be invariant under the $Z_3$. Consequently, all the corresponding MFV Wilson coefficients also remain uncharged by $H'$, \ie\ $R_\text{MFV}^{H'=Z_3} = R_\text{MFV}$, and thus \cref{eqn:TheoremBrionToFlav} holds for all non-renormalizable operators (regardless of mass dimension). 

As a consequence, any SMEFT operator can be expressed as a linear combination of MFV covariants and thus the MFV symmetry principle does not restrict the EFT, \ie\ MFV SMEFT $\equiv$ SMEFT, the main conclusion quoted in \cref{eqn:main}.

It is important to emphasize that the saturation we found for MFV is nontrivial and not inherently guaranteed. This becomes evident when examining various examples illustrated in  \cref{tab:U1,tab:SU2adj,tab:SU3ffbar,tab:SU3adj}, where saturation does not occur in some cases with a low number of building blocks. 
In the context of flavor, a simple example lacking rank saturation corresponds to considering one single Yukawa coupling, such as $Y_u$ alone as the building block (without $Y_d$). In this scenario, it is evident that, for example, in the case of $\mathbf{(8,1,1)}$-covariants, only two independent ones --$h_u$ and $h_u^2$-- are available, falling short of the required eight for saturation. Indeed, this is compatible with the fact that with a single Yukawa, the unbroken subgroup is $H = U(1)_{V}^3 \times U(3)_d$, where $U(1)_{V}^3$ corresponds to the three flavor numbers for up-type quarks, \ie\ vectorial $U(1)$ phase rotations of the up-, charm- and top-quarks, respectively.

\subsection{Possible additional assumptions on top of MFV}
\label{subsec:Assumptions}

We have established that the MFV symmetry principle does not restrict the EFT and thus \emph{the physics lies in the additional assumptions on top of MFV}. Nevertheless, MFV can still be a useful organizing principle of the BSM flavor violating operators, especially when trying to unravel the origin of flavor. But it requires additional hypothesis which are distinct from the MFV symmetry principle.

One of the most common assumptions consists on treating the Yukawas as order parameters and consequently selecting covariants with the lowest number of certain Yukawa insertions. This implicitly assumes that all the flavor invariant coefficients accompanying the covariants have similar order of magnitude. In the example considered in \cref{eqn:MFVInfiniteSum}, this would correspond to assuming
\begin{equation}
c_0 \sim c_1 \sim c_2 \sim \cdots \,. 
\end{equation}
One then often keeps the leading order term in $Y_u$ for the phenomenology, while neglecting the covariants with $Y_d$ insertions, as the down-type Yukawas have smaller values. Alternatively, in the context of Two Higgs Doublet Models (2HDMs), at large $\tan\beta$ one may choose to treat both $Y_u$ and $Y_d$ with equal importance, given that they may be of comparable magnitude ($Y_u\sim Y_d$)~\cite{Isidori:2012ts}. Another alternative assumption in this category, explored in a recent study \cite{Bartocci:2023nvp}, involves considering only BSM operators that are automatically flavor invariant without Yukawa insertions. This scenario, dubbed \emph{minimal} MFV, corresponds to the zeroth order expansion in Yukawas and thus only includes the operators in \cref{tab:MFVdim6SMEFT} that include a singlet $\mathbf{(1,1,1)}$ as an allowed flavor representation for the Wilson coefficient.

We do not intend to survey the literature for all possible assumptions that have been made to turn MFV into a  predictive setting. It is not difficult to propose novel assumptions that may be based on scenarios in the UV completion  that have interesting phenomenological consequences, so we give one more example and then move on.  Consider,  for example, models in the UV that single out the charge $+2/3$ quark sector to participate in some additional very short distance very strong interactions, much like topcolor assisted technicolor \cite{Hill:1994hp}, but incorporating MFV.\footnote{The seminal work on MFV in Ref.~\cite{Chivukula:1987py} had in mind curing flavor changing neutral problems of technicolor models by assuming the enhanced approximate symmetry that leads to MFV.}  This scenario could be studied in a SMEFT  by assuming that the dependence on $Y_u$ is completely general, entering the SMEFT coefficients through fairly arbitrary functions $f(h_u)$, while $Y_d$ would appear perturbatively as a result of Feynman diagrams with explicit Higgs-quarks couplings.  

For each of these assumptions, the Hilbert series computed in \cref{eqn:HSMFV} provide useful information of the number of covariants that need to be considered at each order. In particular, if the assumption treats $Y_u$ and $Y_d$ differently, it may also be useful to use the techniques explained in this work to compute the multi-graded Hilbert series (see, \eg\ \cref{eqn:HS1811ud}).

Nevertheless, it is worth emphasizing the limitations of these assumptions or the classes of theories for which they may not hold. First, the top Yukawa is actually large, $y_t\sim\O (1)$, so neglecting higher powers of $Y_u$ is not always justified (and same for $Y_d$ in large $\tan \beta$ 2HDMs). Second, the invariant coefficients in front of each covariant are not always of the same order. Gauged flavor models~\cite{Grinstein:2010ve, Feldmann:2010yp, Guadagnoli:2011id, Buras:2011wi, Alonso:2016onw} give explicit examples of UV theories whose low-energy EFT fulfills the MFV symmetry principle, but higher-order covariants play a relevant role. For example, one of the operators that can be generated by integrating out the flavored gauge bosons reads
\begin{align}
\frac{1}{\Lambda^2} \frac{1}{ \Tr \big( Y_u^\dagger Y_u \big) }\, \big( \overline{u}_R\, \gamma_\mu\, Y_u^\dagger Y_u\, u_R \big)\, \big( \overline{u}_R\, \gamma^\mu\, Y_u^\dagger Y_u\, u_R \big) \,,
\label{eq:GaugedFlavor}
\end{align}
while the operator that one would naively expect under the above assumptions is
\begin{align}
\frac{1}{\Lambda^2}\, \big(\overline{u}_R\, \gamma_\mu\, Y_u^\dagger Y_u\, u_R \big)\, \big( \overline{u}_R\, \gamma^\mu\, u_R \big) \,.
\label{eq:noGaugedFlavor}
\end{align}
This example clearly showcases that treating the Yukawas as small order parameters is a distinct additional assumption (which does not always hold) and does not follow from the MFV symmetry principle. 

These observations suggest another possible assumption: one can remain agnostic on the dominant covariant but consider one covariant at a time. In the phenomenological analysis outlined \cref{sec:Phenomenology}, we will adopt this approach, in which the consideration of only one covariant generates correlations among different flavor observables. These correlations may, in principle, enable us to distinguish among different covariants in the generating set. It is essential to note that this assumption also has its limitations; several covariants may contribute with equal importance, and the choice of covariants in the generating set is not unique, so a different selection would yield different correlations.

Finally, the findings of this study can be valuable even in the absence of additional assumptions, as they position MFV as an organizing principle for BSM flavor-violating operators. Indeed one can parameterize any BSM flavor operators in terms of the described generating sets, shedding light on whether BSM flavor violation and the SM one genuinely share a common origin.

\section{Sample applications to phenomenology}
\label{sec:Phenomenology}

This section illustrates the impact of different MFV covariants quantitatively, under the \emph{additional assumption} of ``considering a single covariant at a time''. We emphasize that our analysis below does not cover all possible realizations of the MFV principle. Quite the opposite, it is merely a sample of extreme scenarios of MFV. In each of these scenarios, we will assume that only one specific flavor covariant dominates the phenomenology, while all the others are negligible. Realistic UV theories rarely give rise to these extreme scenarios. It is not surprising that once equipped with such extreme additional assumptions MFV is very predictive. By giving several Wilson coefficients in terms of a single BSM parameter, it relates the deviations from SM predictions of several observables. Our focus is on how  different additional assumptions lead to different predictions. Hopefully, our survey of these extreme scenarios help provide a taste of a variety of possibilities, and shed light on how to navigate towards specific realizations of the MFV as flavor measurements improve. This is not intended to be a complete phenomenological study but rather as  a sparse collection of illustrative examples: we will be considering a specific SMEFT operator and restricting our analysis to only a few experimental observables.

The specific SMEFT operator that we focus on is the $Q_{\ell q}^{(1)}$ in \cref{tab:MFVdim6SMEFT}:
\be
\L_\text{SMEFT} \supset \frac{c}{\Lambda^2}\, Q_{\ell q}^{(1)} \equiv \frac{c_{prst}}{\Lambda^2}\, (\ov{\ell}_p \gamma_\mu \ell_{r})(\ov{q}_s\gamma^\mu q_{t})
= \frac{\ov{c}_X^{}\, \delta_{pr} X_{st}}{\Lambda^2}\, (\ov{\ell}_p \gamma_\mu \ell_{r})(\ov{q}_s\gamma^\mu q_{t}) \,.
\label{eqn:Qlq1}
\ee
The coefficient $c_{prst}$ encodes the flavor information for both quarks and leptons. Following the scope of this paper, we focus on the quark part, and have expressed $c_{prst}$ as a product of three factors in the above. $\ov{c}_X^{}$ is a flavor blind global coefficient; $\delta_{pr}$ is a Kronecker-delta  reflecting our assumption  that the new physics is  lepton flavor blind. The matrix $X_{st}$ stands for a covariant of the module ${}_\rPI\Mod_\mathbf{(1\oplus 8,1,1)}$, whose indices are contracted with the quark bilinear. Specifically, we will consider the list of $X_{st}$ given in \cref{tab:PhenoCovariants}, one at a time; together they represent an equivalent basis to that in \cref{tab:GeneratingSet1811}.\footnote{The way in which a specific UV theory manifests itself at low energies can be described with any basis, but the presence of certain cancellations or suppressions can be more evident in one basis than another. Take, for example, the covariant $X=h_u^2h_dh_u$ of \cref{tab:PhenoCovariants} that is substituted for the anti-hermitian combination $X=h_u^2h_dh_u-\hc$ of \cref{tab:GeneratingSet1811}: when written in terms of quark Yukawas and CKM entries, the $X_{23}$ entry of the latter reads $X_{23} = -|V_{tb}|^2V_{cb}V_{cs}^* y_t^4y_b^2y_c^2$, which is more suppressed relative to that of the former, that reads $X_{23} = |V_{tb}|^2V_{tb}V_{ts}^* y_t^6y_b^2$ (see the third column of \cref{tab:ExplicitVariants} below).}
Notice that if $\ov{c}_X^{}$ is taken to be real, then the only source of CP violation is contained in the Yukawa matrices; on the other hand, a complex $\ov{c}_X^{}$ would represent a new source of CP violation, which is strongly constrained in generic analyses~\cite{Paradisi:2009ey}. In what follows, we will restrict to the case of a real $\ov{c}_X^{}$ for simplicity, as it is assumed in common MFV implementations, although relaxing this assumption would only have minor impact on our phenomenological analysis below.

\begin{table}[t!]
\renewcommand{\arraystretch}{1.5}
\setlength{\arrayrulewidth}{.2mm}
\setlength{\tabcolsep}{1em}
\centering
\begin{tabular}{cc}
\toprule
\(q^2\) & \(h_u\quad,\quad   h_d \) \\
\midrule
\(q^4\) & \({h_u^2}\quad,\quad   {h_d^2}\quad,\quad   h_u h_d \quad,\quad   h_d h_u \) \\
\midrule
\(q^6\) & \(h_u^2 h_d \quad,\quad h_u h_d^2 \quad,\quad h_d^2 h_u \quad,\quad h_d h_u^2 \) \\
\midrule
\(q^8\) & \( h_u^2 h_d^2 \quad,\quad h_d^2 h_u^2\quad,\quad h_u^2 h_d h_u\quad,\quad   h_d^2 h_u h_d \) \\
\midrule
\(q^{10}\) & \( h_u^2 h_d^2 h_u \quad,\quad   h_d^2 h_u^2 h_d \) \\
\bottomrule
\end{tabular}
\caption{List of the covariants $X \in {}_\rI\Mod_\mathbf{(1\oplus 8,1,1)}$ considered one at a time in the phenomenological analysis.}
\label{tab:PhenoCovariants}
\end{table}

To study example impacts of the operator in \cref{eqn:Qlq1} on flavor physics, we consider the following decays of the $B$ and $B_s$ mesons:
\begin{subequations}\label{eqn:BRs}
\begin{alignat}{3}
\text{purely leptonic:}\qquad&&
B^0 &\to l^+_il^-_i \,,\qquad&
B^0_s &\to l^+_il^-_i \,, \\[5pt]
\text{semileptonic:}\qquad&&
B^+ &\to \pi^+l^+_il^-_i \,,\qquad&
B^+ &\to K^+l^+_il^-_i \,, \\[5pt]
\text{di-neutrino:}\qquad&&
B^+ &\to \pi^+\ov{\nu}\nu \,,\qquad&
B^+ &\to K^+\ov{\nu}\nu \,,
\end{alignat}
\end{subequations}
where $l_i=e,\,\mu,\,\tau$. As the largest uncertainties are typically associated with the form factors entering the calculation of the various branching ratios, a common procedure to reduce the error is to use the ratios of these branching ratios. In particular, we will work with
\begin{subequations}\label{eqn:cRs}
\begin{alignat}{2}
\cR_{Bl_i} &\equiv \dfrac{\cB(B^0\to l^+_il^-_i)}{\cB(B^0\to l^+_il^-_i)^\text{SM}} \,,\qquad&
\cR_{B_sl_i} &\equiv \dfrac{\cB(B^0_s\to l^+_il^-_i)}{\cB(B^0_s\to l^+_il^-_i)^\text{SM}} \,,\\[8pt]
\cR_{B\pi l_i} &\equiv \dfrac{\cB(B^+\to \pi^+l^+_il^-_i)}{\cB(B^+\to \pi^+l^+_il^-_i)^\text{SM}} \,,\qquad&
\cR_{BKl_i} &\equiv \dfrac{\cB(B^+\to K^+l^+_il^-_i)}{\cB(B^+\to K^+l^+_il^-_i)^\text{SM}} \,, \\[8pt]
\cR_{BK\nu} &\equiv \dfrac{\cB(B^+\to K^+\ov{\nu}\nu)}{\cB(B^+\to K^+\ov{\nu}\nu)^\text{SM}} \,. &&
\end{alignat}
\end{subequations}
For 3-body decays, whenever possible the ratios are taken to be of specific bins of invariant mass of the lepton pair.

The strategy of our analysis below is very simple. Comparing the corresponding theoretical predictions with the experimental measurements of the above observables, we extract the constraints on $\frac{\ov{c}_X^{}}{\Lambda^2}$ in the SMEFT Lagrangian in \cref{eqn:Qlq1} for a choice of $X_{st}$. We also study the observable correlations predicted by this choice.  Below, we present these results for the extreme scenario consisting of choosing $X_{st}$ equal to  one covariant in \cref{tab:PhenoCovariants} at a time.

\subsection{Theoretical predictions}
\label{subsec:Model}

After the electroweak symmetry breaking, the SMEFT Lagrangian in \cref{eqn:Qlq1} contributes to the $\cO_9$ and $\cO_{10}$ operators of the effective Hamiltonian for $\Delta B=1$ transitions \cite{Bobeth:2007dw}
\begin{equation}
\Heff^{\Delta B=1} = -\frac{4G_F}{\sqrt2}\, V_{tb}V_{tq}^*
\left[ C^{(q)}_{9,\, pr}(\mu)\, \cO^{(q)}_{9,\, pr}(\mu) + C^{(q)}_{10,\, pr}(\mu)\, \cO^{(q)}_{10,\, pr}(\mu) \right] + \hc \,,
\label{eqn:HeffDeltaB1}
\end{equation}
where a sum over $q=d,s$ is implict, $V$ denotes the CKM mixing matrix, and the two operators are defined in the mass basis of the quarks:
\begin{subequations}
\begin{align}
\cO^{(q)}_{9,\, pr}(\mu) &= \dfrac{\alpha_\text{em}}{4\pi}\left(\ov{q}\gamma_\mu P_L b\right)\left(\ov{l}_p \gamma^\mu l_r\right) \,, \\[8pt]
\cO^{(q)}_{10,\, pr}(\mu) &= \dfrac{\alpha_\text{em}}{4\pi}\left(\ov{q}\gamma_\mu P_L b\right)\left(\ov{l}_p \gamma^\mu\gamma_5 l_r\right) \,.
\end{align}
\end{subequations}
Similarly, it also contributes to a low-energy operator involving neutrinos:
\begin{equation}
\cH_\text{eff}^{\ov{\nu}\nu} = -\dfrac{4G_F}{\sqrt2}\, C^{(q)}_{\ov{\nu}\nu,\, pr}(\mu)\, V_{tb}V_{tq}^*\, \dfrac{\alpha_\text{em}}{4\pi} \left(\ov{q}\gamma_\mu P_L b\right)\left(\ov{\nu}_p \gamma^\mu P_L\nu_r\right) + \hc \,.
\label{eqn:HeffDeltaBnu}
\end{equation}

The Wilson coefficients in the effective Hamiltonians above contain both the SM contribution and SMEFT corrections
\begin{equation}
C = C^\text{SM} + \delta C \,.
\end{equation}
Matching them with the SMEFT Lagrangian in \cref{eqn:Qlq1} at tree-level,\footnote{Matching is done at a renormalization scale $\mu\approx M_W$. Running has very minor effects. $\delta C$ has very weak $\mu$-dependence since  in QCD its anomalous dimension vanishes. Neglecting electromagnetic interactions, $C_{10}$ and $C_{\nu\nu}$ are scale independent, and the running of $C_9$ is exclusively due to mixing from 4-quark operators~\cite{Grinstein:1988me}.} we obtain the SMEFT correction part
\begin{subequations}
\begin{align}
\delta C_{9,\,pr}^{(q)} &= \delta C_9^{(q)}\, \delta_{pr} \,, \\[5pt]
\delta C^{(q)}_{10,\, pr} &= -\delta C^{(q)}_{9,\, pr} = -\delta C_9^{(q)}\, \delta_{pr} \,, \\[5pt]
\delta C^{(q)}_{\ov\nu\nu,\,pr} &= \delta C^{(q)}_{\ov\nu\nu}\, \delta_{pr} \,,
\end{align}
\end{subequations}
with
\begin{equation}
\delta C_9^{(q)}
= \frac12\, \delta C^{(q)}_{\ov\nu\nu}
= \left( \frac{2G_F\, \alpha_\text{em}}{\sqrt2\pi} \right)^{-1} \dfrac{\ov{c}_X^{}}{\Lambda^2}\, \Big[ X_{i_q3}/(V_{tb}V_{tq}^*) \Big] \,,
\label{eqn:deltaCMatching}
\end{equation}
where $i_q=1,2$ respectively for $q=d,s$.

The observables in \cref{eqn:cRs} can be computed including the effects of $\delta C_9^{(q)}$ and $\delta C^{(q)}_{\ov\nu\nu}$. Making use of the results in Ref.~\cite{Bobeth:2007dw} for the purely leptonic $B$ and $B_s$ decays, in Refs.~\cite{Bobeth:2007dw,Ali:2013zfa} for the semileptonic decays, and in Ref.~\cite{Bause:2021cna} for the di-neutrino decays, we obtain the following expressions:
\begin{subequations}\label{eqn:cRsFormulae}
\begin{align}
\cR_{Bl_i} &= \left| 1 - \frac{1}{C^\text{SM}_{10}}\, \delta C^{(d)}_9 \right|^2 \,, \\[15pt]
\cR_{B_sl_i} &= \left| 1 - \frac{1}{C^\text{SM}_{10}}\, \delta C^{(s)}_9 \right|^2 \,, \\[15pt]
\cR_{B\pi l_i} &= 1 + \frac{C^\text{SM,eff}_9 + \tfrac{2m_b}{M_B+M_\pi}\, C_7^\text{SM,eff}-C^\text{SM}_{10}}{\left|C^\text{SM,eff}_9+\tfrac{2m_b}{M_B+M_\pi}\, C_7^\text{SM,eff}\right|^2+\left|C^\text{SM}_{10}\right|^2}\; 2\Re\left(\delta C_9^{(d)}\right) \,, \\[15pt]
\cR_{BKl_i} &= 1 + \frac{C^\text{SM,eff}_9+\frac{2m_b}{M_B+M_K}\, C_7^\text{SM,eff}-C^\text{SM}_{10}}{\left|C^\text{SM,eff}_9+\frac{2m_b}{M_B+M_K}\, C_7^\text{SM,eff}\right|^2+\left|C^\text{SM}_{10}\right|^2}\; 2 \Re\left(\delta C_9^{(s)}\right) \,, \\[15pt]
\cR_{BK\nu} &= \left| 1 + \frac{1}{C^\text{SM}_{\ov{\nu}\nu}}\, \delta C^{(s)}_{\ov{\nu}\nu} \right|^2 \,,
\end{align}
\end{subequations}
for any lepton in the final state $l_i=e,\,\mu,\,\tau$. Note that we have used the approximation $\delta C_9^{(q)}\ll C_9^\text{SM,eff}$ for the expressions of $\cR_{B\pi l_i}$ and $\cR_{BKl_i}$ above. Note also that  we have   treated $C^\text{SM,eff}_9$ and $C^\text{SM,eff}_7$ as approximately constants, that is, as independent of the invariant mass of the lepton pair, $q^2$. This is empirically justified  by direct examination of the results of their calculation in the relevant $q^2$ bins, $(1,6)~\text{GeV}^2$ and $(15,22)~\text{GeV}^2$  (see Figs.~37 and 38 in Ref.~\cite{Parrott:2022zte}). Finally, we  approximated the form factor $f_T(q^2)$ to be equal to $f_+(q^2)$ (see Fig.~2 in Ref.~\cite{Parrott:2022zte}).

\subsection{Experimental measurements}
\label{subsec:Measurements}

\begin{table}[t]
\centering
\resizebox{\textwidth}{!}{ 
\begin{tabular}{l|c|c|}
\hline
\hline
&&\\[-3mm]
Branching Ratios & SM prediction & Experimental data\\[1mm]
\hline
&&\\[-3mm]
$B^0\to e^+e^-$
& $(2.41\pm0.13)\times 10^{-15}$~\cite{Beneke:2019slt}
& $<3.0\times 10^{-9}$~\cite{LHCb:2020pcv}
\\[1mm]
$B^0_s\to e^+e^-$ 
& $(8.60\pm0.36)\times 10^{-14}$~\cite{LHCb:2020pcv} 
& $<11.2$~\cite{LHCb:2020pcv}
\\[1mm]
$B^0\to\mu^+\mu^-$
& $(1.01\pm0.07)\times 10^{-10}$~\cite{Bause:2022rrs}
& $(1.20\pm0.84)\times 10^{-10}$~\cite{LHCb:2021awg}
\\[1mm]
$B^0_s\to\mu^+\mu^-$
& $(3.66\pm0.04)\times 10^{-9}$~\cite{Beneke:2019slt}
& $(3.09^{+0.46+0.15}_{-0.43-0.11})\times 10^{-9}$~\cite{LHCb:2021vsc}
\\[1mm]
$B^0\to \tau^+\tau^-$
& $(2.14\pm0.12)\times 10^{-8}$~\cite{Fleischer:2017ltw}
& $<2.1\times10^{-3}$~\cite{ParticleDataGroup:2022pth}
\\[1mm]
$B^0_s\to \tau^+\tau^-$
& $(7.56\pm0.35)\times10^{-7}$~\cite{Fleischer:2017ltw}
& $<6.8\times 10^{-3}$~\cite{ParticleDataGroup:2022pth}
\\[1mm]
\hline
&&\\[-3mm]
$B^+\to\pi^+\mu^+\mu^-$ & $(17.9\pm1.9\pm1.1\pm1.5)\times 10^{-9}$~\cite{Bause:2022rrs} & $(18.3\pm2.4\pm0.5)\times 10^{-9}$~\cite{LHCb:2015hsa}\\[1mm]
$B^+\to K^+e^+e^-|_{(1,6)}$&
$(1.94\pm0.16\pm0.097)\times10^{-7}$~\cite{Parrott:2022zte}&
$(1.66^{+0.32}_{-0.29}\pm0.04)\times10^{-7}$~\cite{BELLE:2019xld}\\[1mm]
$B^+\to K^+e^+e^-|_{(1.1,6)}$&
$(1.91\pm0.16\pm0.095)\times10^{-7}$~\cite{Parrott:2022zte}&
$(1.401^{+0.074}_{-0.069}\pm0.064)\times10^{-7}$~\cite{LHCb:2021trn}\\[1mm]
$B^0\to K^0\mu^+\mu^-|_{(1.1,6)}$&
$(1.60\pm0.14\pm0.032)\times10^{-7}$~\cite{Parrott:2022zte}&
$(0.92^{+0.17}_{-0.15}\pm0.044)\times10^{-7}$~\cite{LHCb:2014cxe}\\[1mm]
$B^0\to K^0\mu^+\mu^-|_{(15,22)}$&
$(1.070\pm0.095\pm0.021)\times10^{-7}$~\cite{Parrott:2022zte}&
$(0.67\pm0.11\pm0.035)\times10^{-7}$~\cite{LHCb:2014cxe}\\[1mm]
$B^+\to K^+\mu^+\mu^-|_{(1,6)}$&
$(1.95\pm0.16\pm0.039)\times10^{-7}$~\cite{Parrott:2022zte}&
$(2.30^{+0.41}_{-0.38}\pm0.05)\times10^{-7}$~\cite{BELLE:2019xld}\\[1mm]
$B^+\to K^+\mu^+\mu^-|_{(1.1,6)}$&
$(1.91\pm0.16\pm0.038)\times10^{-7}$~\cite{Parrott:2022zte}&
$(1.186\pm0.034\pm0.059)\times10^{-7}$~\cite{LHCb:2014cxe}\\[1mm]
$B^+\to K^+\mu^+\mu^-|_{(15,22)}$&
$(1.16\pm0.10\pm0.023)\times10^{-7}$~\cite{Parrott:2022zte}&
$(0.847\pm0.028\pm0.042)\times10^{-7}$~\cite{LHCb:2014cxe}\\[1mm]
\hline
&&\\[-3mm]
$B^+\to K^+\ov{\nu}\nu$ & $(5.58\pm0.37)\times10^{-6}$~\cite{Parrott:2022zte}&$(2.3\pm0.5^{+0.5}_{-0.4})\times 10^{-5}$~\cite{Belle-II:2023esi}
\\[1mm]
\hline
\hline
\end{tabular}}
\caption{\emph{SM predictions and experimental results on the branching ratios of the different $B$ and $B_s$ decays. The upper bounds are given at $95\%$ C.L.. For the semileptonic $B\to K$ decays into charged leptons the bin  in the invariant lepton pair squared-mass   is indicated in} $\text{GeV}^2$.}
\label{tab:Data}
\end{table}

The SM predictions and the experimental measurements for the different branching ratios listed in \cref{eqn:BRs} are summarized in \cref{tab:Data}. Note that for the purely leptonic decays, the $B$ and $B_s$ meson oscillations to the corresponding CP conjugate state is already taken into consideration, and therefore the reported SM predictions, as well as the experimental result, refer to the time-integrated or untagged decays. Using only the decays that have been observed, \cref{tab:Data} leads to the following experimental determinations for the ratios of the branching ratios $\cR$ listed in \cref{eqn:cRs}:
\begin{subequations}\label{eqn:cRsData}
\begin{align}
\cR_{B\mu} &=1.19\pm0.84 \,, \\[8pt]
\cR_{B_s\mu} &=0.84\pm0.13 \,, \label{eqn:cRBsmuData}
\\[8pt]
\cR_{B\pi\mu} &=1.0\pm0.2 \,, \\[10pt]
\cR_{BKe} &=
\begin{cases}
0.86\pm0.19&\text{ from } B^+\to K^+e^+e^-|_{(1,6)}\\
0.733\pm0.088&\text{ from } B^+\to K^+e^+e^-|_{(1.1,6)}
\end{cases} \,, \\[10pt]
\cR_{BK\mu} &=
\begin{cases}
0.58\pm0.12&\text{ from } B^0\to K^0\mu^+\mu^-|_{(1.1,6)}\\
0.63\pm0.12&\text{ from } B^0\to K^0\mu^+\mu^-|_{(15,22)}\\
1.18\pm0.23&\text{ from } B^+\to K^+\mu^+\mu^-|_{(1,6)}\\
0.621\pm0.064&\text{ from } B^+\to K^+\mu^+\mu^-|_{(1.1,6)}\\
0.730\pm0.078&\text{ from } B^+\to K^+\mu^+\mu^-|_{(15,22)}
\end{cases} \,, \\[10pt]
\cR_{BK\nu} &= 4.1 \pm 1.3 \,.
\end{align}
\end{subequations}

\subsection{Constraints on $\delta C_9^{(d)}$, $\delta C_9^{(s)}$, and $\delta C_{\ov{\nu}\nu}^{(s)}$}

By comparing the theoretical predictions in \cref{eqn:cRsFormulae} with the experimental measurements in \cref{eqn:cRsData} we can extract  constraints on $\delta C_9^{(d)}$, $\delta C_9^{(s)}$, and $\delta C_{\ov{\nu} \nu}^{(s)}$. To this end, we also need the experimental measurements of the various input parameters in the theoretical predictions in \cref{eqn:cRsFormulae}, for which we take the following
\begin{subequations}
\begin{alignat}{2}
M_{\pi^\pm } &= 139.57039\pm0.00018\MeV~\text{\cite{ParticleDataGroup:2022pth}}
\,,\quad&
m_b^{\ov{\text{MS}}}(\mu_b) &= 4.209(21)\GeV~\text{\cite{Hatton:2021syc}}
\,, \notag\\[5pt]
M_{K^0} &= 497.611\pm0.013\MeV~\text{\cite{ParticleDataGroup:2022pth}}
\,,\quad&
C^\text{SM}_9(\mu_b) &= 4.114(14)~\text{\cite{Blake:2016olu}}
\,, \notag\\[5pt]
M_{K^\pm} &= 493.677\pm0.016\MeV~\text{\cite{ParticleDataGroup:2022pth}}
\,,\quad&
C^\text{SM}_{10}(\mu_b) &= -4.193(33)~\text{\cite{Blake:2016olu}}
\,, \notag\\[5pt]
M_{B^0} &= 5279.66\pm0.12\MeV~\text{\cite{ParticleDataGroup:2022pth}}
\,,\quad&
C^\text{SM}_{\ov\nu\nu} &= -12.71\pm0.15~\text{\cite{Bause:2021cna}}
\,, \notag\\[5pt]
M_{B^\pm} &= 5279.34\pm0.12\MeV~\text{\cite{ParticleDataGroup:2022pth}}
\,, \notag
\end{alignat}
\end{subequations}
where $\mu_b=4.2\GeV$, and the values for $C^\text{SM}_9(\mu_b)$ and $C^\text{SM}_{10}(\mu_b)$ are given at the NNLO accuracy. For the case of $C_9$, we need the effective Wilson coefficient $C^\text{SM,eff}_9$ that includes the contributions from current-current and QCD-penguin operators. From Fig.~38 in Ref.~\cite{Parrott:2022zte}, we estimate it to be  
\be
C_{9}^\text{SM,eff}(q^2,\mu_b)=\Big( C_{9}^\text{SM}(\mu_b) + 0.5 \Big) \pm 0.2 \,,
\ee
where the uncertainty reflects the mild $q^2$ dependence and the thickness of the error band in the figure into the bins $q^2\in(1,6)\GeV^2$ and $q^2\in(15,22)\GeV^2$. We neglect the imaginary part because it is much smaller than the  real one. Similarly, from Fig.~37 in Ref.~\cite{Parrott:2022zte}, we estimate the effective Wilson coefficient $C_{7}^\text{SM,eff}$ as
\be
C_{7}^\text{SM,eff}(q^2,\mu_b) = -0.35 \pm 0.02 \,.
\ee

With the above input parameters, we can now convert the measurements in \cref{eqn:cRsData} into constraints on $\delta C_9^{(d)}$, $\delta C_9^{(s)}$, and $\delta C_{\ov{\nu} \nu}^{(s)}$ using \cref{eqn:cRsFormulae}. We obtain
\begin{subequations}\label{eqn:deltaCBounds}
\begin{align}
\delta C^{(d)}_9 &= \left\{
\begin{array}{cl} 
\quad
+0.4^{+1.4}_{-2.1}
\;\;\;\text{or}\;\;
-8.8^{+2.1}_{-1.4}
&\quad\quad\;\;\;\text{from }
B^0\to \mu^+\mu^- \\[4pt]
+0.05\pm0.42
&\quad\quad\;\;\;\text{from }
B^+\to\pi^+\mu^+\mu^-
\end{array} \right. \;, 
\label{eqn:deltaC9dBounds} \\[10pt]
\delta C^{(s)}_9 &= \left\{
\begin{array}{cl}
-0.34^{+0.29}_{-0.32}
\;\;\;\text{or}\;\;
-8.05^{+0.32}_{-0.29}
&\quad\quad\text{from }
B_s^0\to \mu^+\mu^- \\[4pt]
-0.65\pm0.08
&\quad\quad\text{from a fit to }
B\to K\ell^+\ell^-
\end{array} \right. \;, 
\label{eqn:deltaC9sBounds} \\[10pt]
\delta C_{\ov\nu\nu}^{(s)} &=
\;\;
-13.1\pm4.1
\;\;\;\text{or}\;\;\;
39.0\pm12
\quad\quad\;\;\;\text{from }
B^+\to K^+\ov{\nu}\nu \;.
\label{eqn:deltaCnuBounds}
\end{align}
\end{subequations}
The second line of \cref{eqn:deltaC9sBounds} is the result of a global fit to  all the $B\to Ke^+e^-$ and $B\to K\mu^+\mu^-$ data listed in \cref{tab:Data}.

A few comments are in order:
\begin{itemize}
\item The purely leptonic decays, both charged leptons and neutrinos, lead to two different set of solutions for the corresponding Wilson coefficients: the first one is a relatively small deviation from the SM contribution, while the second is larger and with opposite sign. They correspond to the usual constructive and destructive interference scenarios formed between the MFV contributions and the SM ones. On the other hand, this does not occur for the constraints from the semileptonic decays, because the expressions for $\cR_{B\pi l_i}$ and $\cR_{BKl_i}$ in \cref{eqn:cRsFormulae} have been obtained under the assumption that $\delta C_9^{(q)}\ll C_9^\text{SM,eff}$.
\item From \cref{eqn:deltaC9dBounds,eqn:deltaC9sBounds} we see that the constraints on the same Wilson coefficient from the first solution of the purely leptonic decay and that from the semileptonic decays are well compatible, given the present sensitivity.
\item The SMEFT prediction in Eq.~\eqref{eqn:deltaCMatching}, namely, $\delta C_9^{(s)}=\frac12\delta C_{\ov\nu\nu}^{(s)}$, is violated at slightly less than  $3\sigma$.
\item Finally, $\delta C_9^{(s)}$ from the semileptonic decays and $\delta C_{\ov\nu\nu}^{(s)}$ reflect a  tension with the SM prediction in excess of $3\sigma$.
\end{itemize} 
The tension in the last two comments demand further scrutiny, and in fact are the focus of much ongoing work both in experiment and in theory. However, our aim here is to give a sense  of how different assumptions supplementing the MFV hypothesis may lead to distinct predictions. In this spirit, we take the results of the various fits at face value and carry on. 

\subsection{Constraints on $\ov{c}_X^{} /\Lambda^2$}

Now we would like to translate the constraints on $\delta C_9^{(d)}$, $\delta C_9^{(s)}$, and $\delta C_{\ov{\nu}\nu}^{(s)}$ obtained in \cref{eqn:deltaCBounds} into constraints on the covariant-specific Wilson coefficient $\ov{c}_X^{} /\Lambda^2$. We recall from \cref{eqn:deltaCMatching} that
\begin{equation}
\delta C_9^{(q)}
= \frac12\, \delta C^{(q)}_{\ov\nu\nu}
= \left( \frac{2G_F\, \alpha_\text{em}}{\sqrt2\pi} \right)^{-1} \dfrac{\ov{c}_X^{}}{\Lambda^2}\, \Big[ X_{i_q3}/(V_{tb}V_{tq}^*) \Big] \,.
\label{eqn:deltaCMatchingRedo}
\end{equation}
Clearly, the result depends on which covariant $X$ we choose through the factor
\begin{equation}
X_{i_q3}/(V_{tb}V_{tq}^*)
\quad\text{with}\quad
i_q=1,2
\quad\text{respectively for}\quad
q=d,s \,.
\label{eqn:Xfactor}
\end{equation}
To compute this factor, we express the building blocks of the covariant $X$ (listed in \cref{tab:PhenoCovariants}), \ie\ the $h_u$ and $h_d$ defined in \cref{eqn:huhddef}, in terms of physical observables. First, we take a quark basis such that the Yukawa matrices $Y_u, Y_d$ are written in terms of the quark masses and the CKM mixing matrix:
\begin{subequations}\label{eqn:YuYdValues}
\begin{align}
Y_u &= V^\dag\, \diag\left( y_u ,\, y_c ,\, y_t \right) \,, \\[5pt]
Y_d &= \diag\left( y_d ,\, y_s ,\, y_b \right) \,,
\end{align}
\end{subequations}
where $y_\psi=\sqrt2\, m_\psi/v$ with $v=246.22\GeV$~\cite{ParticleDataGroup:2022pth} being the Higgs vev. This choice is particularly useful as we focus on observables involving mesons with down-type quarks. It follows that $h_d$ is diagonal, and $h_u$ encodes flavor changing transitions:
\begin{subequations}
\begin{align}
h_u &= Y_u\, Y_u^\dagger = V^\dagger\, \diag\left( y_u^2 ,\, y_c^2 ,\, y_t^2 \right) V \,, \\[5pt]
h_d &= Y_d\, Y_d^\dagger = \diag\left( y_d^2 ,\, y_s^2 ,\, y_b^2 \right) \,.
\end{align}
\end{subequations}
With these, one can write out the factor in \cref{eqn:Xfactor} for each covariant $X$ in \cref{tab:PhenoCovariants}. Their leading contributions are summarized in \cref{tab:ExplicitVariants}, where we have assumed the counting  $|V_{tb}|\sim1$, $|V_{ts}|\sim\lambda^2$,  $|V_{td}|\sim\lambda^3$ ,  $y_t\sim1$, and $y_c\sim\lambda^3$, with $\lambda\approx0.2$ a small parameter,  and have neglected $y_u$ altogether. Notice that, within this setup, the two covariants $h_u$ and $h_u^2$ have entries $\cO(1)$, while all the others have much smaller entries. This feature does not need to be universal as it follows strictly from the values of the Yukawas determined by \cref{eqn:YuYdValues} and $y_\psi=\sqrt2\, m_\psi/v$. If instead, we embed MFV in a 2HDM with large $\tan\beta$, then the value of $y_b$ does not need to be much smaller then $y_t$ and therefore none of the covariant entries proportional to the down-type Yukawas would  be as suppressed as in Table~\ref{tab:ExplicitVariants}.

\begin{table}[t!]
\renewcommand{\arraystretch}{1.5}
\setlength{\arrayrulewidth}{.2mm}
\setlength{\tabcolsep}{1.5em}
\centering
\begin{tabular}{ccc}
\toprule
Covariant $X$ & $X_{13}/(V_{tb} V_{td}^*)$ & $X_{23}/(V_{tb} V_{ts}^*)$ \\
\midrule
$h_u$ & $y_t^2\sim\cO(1)$ & $y_t^2\sim\cO(1)$
\\
$h_d$ & $0$ & $0$ 
\\
$h_u^2$ &  $y_t^4\sim\cO(1)$ & $y_t^4\sim\cO(1)$
\\
$h_d^2$ &  $0$ & $0$
\\
$h_uh_d$ & $y_t^2 y_b^2\sim\cO(10^{-3})$ & $y_t^2 y_b^2\sim\cO(10^{-3})$
\\
$h_dh_u$ & $y_t^2 y_d^2\sim\cO(10^{-9})$ & $y_t^2 y_s^2\sim\cO(10^{-7})$
\\
$h_u^2h_d$ & $y_t^4 y_b^2\sim\cO(10^{-3})$ & $y_t^4 y_b^2\sim\cO(10^{-3})$
\\
$h_uh_d^2$ & $y_t^2 y_b^4\sim\cO(10^{-7})$ & $y_t^2 y_b^4\sim\cO(10^{-7})$
\\
$h_d^2h_u$ & $y_t^2 y_d^4\sim\cO(10^{-18})$ & $y_t^2 y_s^4\sim\cO(10^{-13})$
\\
$h_dh_u^2$  & $y_t^4 y_d^2\sim\cO(10^{-9})$ & $y_t^4 y_s^2\sim\cO(10^{-7})$
\\
$h_u^2h_d^2$ & $y_t^4 y_b^4\sim\cO(10^{-7})$ & $y_t^4 y_b^4\sim\cO(10^{-7})$
\\
$h_d^2h_u^2$ & $y_t^4 y_d^4\sim\cO(10^{-18})$ & $y_t^4 y_s^4\sim\cO(10^{-13})$
\\
$h_u^2h_dh_u$ & $|V_{tb}|^2 y_t^6 y_b^2\sim\cO(10^{-3})$ & $|V_{tb}|^2 y_t^6 y_b^2\sim\cO(10^{-3})$
\\
$h_d^2h_uh_d$ & $y_t^2 y_b^2 y_d^4\sim\cO(10^{-22})$ & $y_t^2 y_b^2 y_s^4\sim\cO(10^{-17})$
\\
$h_u^2h_d^2h_u$ & $|V_{tb}|^2 y_t^6 y_b^4\sim\cO(10^{-7})$ & $|V_{tb}|^2 y_t^6 y_b^4\sim\cO(10^{-7})$
\\
$h_d^2h_u^2h_d$ & $y_t^4 y_b^2 y_d^4\sim\cO(10^{-22})$ & $y_t^4 y_b^2 y_s^4\sim\cO(10^{-17})$
\\
\bottomrule
\end{tabular}
\caption{Leading contributions to $X_{13}/(V_{tb}V_{td}^*)$ and $X_{23}/(V_{tb}V_{ts}^*)$ in terms of physical observables for each covariant $X$ in \cref{tab:PhenoCovariants}.}
\label{tab:ExplicitVariants}
\end{table}

Now we are ready to translate the constraints into estimates of the scale of the SMEFT.  Specifically, the various constraints in \cref{eqn:deltaCBounds} translate into corresponding  results for the cut-off scale $\Lambda/\sqrt{\ov{c}_X}$. For their numerical evaluation we use~\cite{ParticleDataGroup:2022pth}:
\begin{subequations}
\begin{align}
G_F&=1.1663787(6)\times 10^{-5}\GeV^{-2}\,, \\[5pt]
\alpha_\text{em}(M_Z)&=1/127.951(9) \,, \\[5pt]
|V_{tb}|&=1.014\pm0.029 \,.
\end{align}
\end{subequations}
We can distinguish two classes of results. On the one, hand one may derive lower bounds on $\Lambda/\sqrt{\ov{c}_X}$ when the Wilson coefficient is compatible with zero, as is the case for the fit results $\delta C_9^{(d)} = 0.4^{+1.4}_{-2.1} ,\, 0.05 \pm 0.42$, and $\delta C_9^{(s)} = -0.34^{+0.29}_{-0.32}$. On the other hand, for fits that give $\delta C$ that deviates by more than $3\sigma$ from zero, as is the case for the rest of our fits, a specific value for $\Lambda/\sqrt{\ov{c}_X}$ is obtained. Moreover, following common lore we assume $\ov{c}_X\lesssim\cO(1)$, hence only the covariants $h_u$ and $h_u^2$ lead to sensible results, since for all the others the value of $\Lambda/\sqrt{\ov{c}_X}$ is typically smaller than the electroweak scale and therefore the effective description  breaks down. Stated differently, in the latter cases, if $\Lambda/\sqrt{\overline c_X}$ is taken above the electroweak scale, then the MFV contributions are so small that they do not lead to sizable deviations from the SM predictions for the various observables, and therefore it is not possible to improve the agreement with the data. 

Therefore, we focus only on the $h_u$ and $h_u^2$ covariants. The strongest lower bound arises from $\delta C_9^{(d)}=0.05\pm0.42$ and leads to $\Lambda/\sqrt{\ov{c}_X}\gtrsim5\TeV$ at $95\%$ C.L.. On the other hand, $\delta C_9^{(d)}=-8.8^{+2.1}_{-1.4}$ and $\delta C_9^{(s)}=-8.05^{+0.32}_{-0.29}$ imply $\Lambda/\sqrt{\ov{c}_X}\sim1.5\TeV$, the two values of $\delta C_{\ov\nu\nu}^{(s)}$ give $\Lambda/\sqrt{\ov{c}_X}\sim1\TeV,\,2\TeV$ respectively, while the most precise $\delta C_9^{(s)}=-0.65\pm0.08$ leads to $\Lambda/\sqrt{\ov{c}_X}\sim 6\TeV$. Assuming that $\ov{c}_X\sim\cO(1)$, then  these values represent the scale at which new physics should be present in order to produce the corresponding observable effects.

Before closing the section, it is interesting to ponder on the generality of the $\ov{c}_X\lesssim\cO(1)$ condition. Strictly speaking, since only the ratios $\bar c_X X/\Lambda^2$ enter as SMEFT coefficients, self-consistency, dimensional analysis, and SMEFT perturbative unitarity all suggest instead $\ov{c}_X X\lesssim\cO(1)$. Consider for definiteness  $(8,1,1)$ covariants. In some familiar UV completions $c_{h_u}h_u$ is dominant, and since $y_t\approx 1$, one has  $c_{h_u}\lesssim1$. A familiar example of this is the MSSM, in which powers of $h_u$ and $h_d$ in $X$ arise from Feynman diagrams with internal Higgs or Higgsino exchange so that each additional power of $h_u$ or $h_d$ appears with a loop factor suppression, $\sim1/16\pi^2$. The leading contribution is therefore $h_u$ or $h_d$, the former of $\mathcal{O}(1)$ and the latter may also be of this order if $\tan\beta$ is large. In this most familiar case, $\overline{c}_{h_x} h_x\sim \overline{c}_{h_x}\lesssim1$ and all other covariants have smaller coefficients. But this is not mandated. Consider the case of  gauged flavor realisations of  M(L)FV~\cite{Grinstein:2010ve, Feldmann:2010yp, Guadagnoli:2011id,Buras:2011wi, Alonso:2016onw}. Replacing $d_R$ for $u_R$ and $Y_d$ for $Y_u$ in Eq.~\eqref{eq:GaugedFlavor}, that gives a SMEFT operator that results from gauge flavor vectors exchange, we see that coefficient has a large enhancement factor, $\bar c\sim 1/\Tr(h_d)\approx1/y_b^2 \gg1$. While the covariant $1\otimes X$  in \cref{eq:noGaugedFlavor} has one less factor of $X$ than the covariant $X\otimes X$ in \cref{eq:GaugedFlavor}, their overall coefficient is clearly  of the same order. We see that even the covariants that have strongly suppressed entries may have an impact on the low-energy observables, while still preserving the effective description. For example, fixing $\Lambda=1\TeV$, the covariant $h_uh_d$ may explain the tension in the $B^0_s\to\mu^+\mu^-$ decay with $\ov{c}\sim50$.

\subsection{MFV correlations}

When the covariants in \cref{tab:PhenoCovariants} are considered one at a time, the Wilson coefficients corresponding to different quark flavors are not independent. They are correlated in a specific way, depending on the specific choice of the covariant. Making use of this, one can derive predictions on the ratios of the $\cR$'s in \cref{eqn:cRs}.

\subsubsection*{Purely leptonic and semi-leptonic decays}

From \cref{eqn:deltaCMatching}, the following relation holds:
\begin{equation}
\frac{\delta C_9^{(d)}}{\delta C_9^{(s)}}
= \frac{X_{13}/(V_{tb}V_{td}^*)}{X_{23}/(V_{tb}V_{ts}^*)} \,.
\label{eqn:C9Corr}
\end{equation}
Using \cref{tab:ExplicitVariants}, we get the specific expressions
\begin{equation}
\frac{\delta C_9^{(d)}}{\delta C_9^{(s)}} = \left\{
\begin{array}{cl}
1 & \quad\text{for }\;
h_u \,,\,
h_u^2 \,,\,
h_uh_d \,,\,
h_u^2h_d \,,\,
h_uh_d^2 \,,\,
h_u^2h_d^2 \,,\,
h_u^2h_dh_u \,,\,
h_u^2h_d^2h_u \,,
\\[10pt]
\dfrac{m_d^2}{m_s^2} & \quad\text{for }\;
h_d h_u \,,\,
h_dh_u^2 \,,
\\[15pt]
\dfrac{m_d^4}{m_s^4} & \quad\text{for }\;
h_d^2h_u \,,\,
h_d^2h_u^2 \,,\,
h_d^2h_uh_d \,,\,
h_d^2h_u^2h_d \,.
\end{array}
\right.
\label{eqn:C9Correlations}
\end{equation}
We see that covariants with the leftmost  spurion factor  $h_u$  predict equality of  the two Wilson coefficients, while those with the leftmost factor $h_d$ or $h_d^2$ lead to a  ratio $\delta C_9^{(d)}/\delta C_9^{(s)}$ suppressed by the ratio of the down-quark  to the strange-quark masses, raised to the 2nd or 4th  powers, respectively.

The correlations in \cref{eqn:C9Correlations}, combined with the formulae in \cref{eqn:cRsFormulae}, provides theoretical predictions on $\cR_{B_s\mu}/\cR_{B\mu}$ and $\cR_{BKe}/\cR_{B\pi\mu} = \cR_{BK\mu}/\cR_{B\pi\mu}$. Let us discuss the different scenarios of the covariants in \cref{eqn:C9Correlations} separately:
\begin{itemize}
\item Covariants in the first line of \cref{eqn:C9Correlations}: In these cases, we get
\be
\dfrac{\cR_{B_s\mu}}{\cR_{B\mu}}=1 \,,\qquad
\dfrac{\cR_{BK\mu}}{\cR_{B\pi\mu}}=\dfrac{\cR_{BKe}}{\cR_{B\pi\mu}} \approx 1 \,,
\label{eqn:rcRC91}
\ee
where the second relation is exact at the per mil level.
\item Covariants in the second and third lines of~\cref{eqn:C9Correlations}: In these cases, the predictions depend on the value of $\delta C_9^{(s)}$. Given the limited prevision in the determination of $\delta C_9^{(s)}$, the contribution of $\delta C_9^{(d)}$ is safely neglected. Thus we find the predicted correlations actually the same for the second and third lines: 
\be
\dfrac{\cR_{B_s\mu}}{\cR_{B\mu}} = 0.72\pm0.03\,,\qquad
\dfrac{\cR_{BK\mu}}{\cR_{B\pi\mu}}=\dfrac{\cR_{BKe}}{\cR_{B\pi\mu}}=
0.69\pm0.04 \,,
\label{eqn:rcRC923a}
\ee
for $\delta C_9^{(s)}=-0.65\pm0.08$, while
\be
\dfrac{\cR_{B_s\mu}}{\cR_{B\mu}} =
0.84\pm0.13 \,,
\label{eqn:rcRC923b}
\ee
for $\delta C_9^{(s)}=-8.05^{+0.32}_{-0.29}$. Recall that this second solution presented in \cref{eqn:rcRC923b} only exists for the purely leptonic decays. It also serves as a consistency check: In this case, we have $\cR_{B\mu}\approx 1$ (see \cref{eqn:cRsFormulae}) due to the $m_d^2/m_s^2$ suppressions of $\delta C_9^{(d)}$ in \cref{eqn:C9Correlations}, and therefore we expect $\cR_{B_s\mu}/\cR_{B\mu}\approx\cR_{B_s\mu}$. This can be verified by comparing \cref{eqn:rcRC923b} with \cref{eqn:cRBsmuData}.
\end{itemize}

Let us compare these predictions with the present experimental measurements
\be
\begin{aligned}
\dfrac{\cR_{B_s\mu}}{\cR_{B\mu}}&=0.71\pm0.51\,,\\[2pt]
\dfrac{\cR_{BKe}}{\cR_{B\pi\mu}} &=\begin{cases}
0.84\pm0.25&\text{ from } B^+\to K^+e^+e^-|_{(1,6)}\\
0.72\pm0.17&\text{ from } B^+\to K^+e^+e^-|_{(1.1,6)}
\end{cases} \,, \\[2pt]
\dfrac{\cR_{BK\mu}}{\cR_{B\pi\mu}} &=\begin{cases}  
0.56.\pm0.16&\text{ from } B^0\to K^0\mu^+\mu^-|_{(1.1,6)}\\
0.61\pm0.17&\text{ from } B^0\to K^0\mu^+\mu^-|_{(15,22)}\\
1.15\pm0.33&\text{ from } B^+\to K^+\mu^+\mu^-|_{(1,6)}\\
0.61\pm0.14&\text{ from } B^+\to K^+\mu^+\mu^-|_{(1.1,6)}\\
0.71\pm0.16&\text{ from } B^+\to K^+\mu^+\mu^-|_{(15,22)}
\end{cases} \,.
\end{aligned}
\label{eqn:rCRData}
\ee
Focusing first on the purely leptonic decays, we see that the experimental error on $\cR_{B_s\mu}/\cR_{B\mu}$ is still pretty large (mainly due to the uncertainty in the $\cB(B^0\to\mu^+\mu^-)$ measurements). Consequently, the MFV predictions are compatible with the experimental result, for any covariant and both type of solutions for $\delta C_9^{(s)}$. A very similar conclusion applies for all the semileptonic decays: the MFV prediction in \cref{eqn:rcRC91,eqn:rcRC923a} are both compatible within $3\sigma$ with all the experimental combinations, mainly due to the large errors in the $\cB(B^+\to\pi^+\mu^+\mu^-)$ measurements. 

\subsubsection*{Di-neutrino decays}

It is clear from \cref{eqn:deltaCMatching} that the exact same correlations in \cref{eqn:C9Corr,eqn:C9Correlations} hold also for the ratio $\delta C_{\ov{\nu}\nu}^{(d)}/\delta C_{\ov{\nu}\nu}^{(s)}$, because
\begin{equation}
\frac{\delta C_{\ov{\nu}\nu}^{(d)}}{\delta C_{\ov{\nu}\nu}^{(s)}}
= \frac{\delta C_9^{(d)}}{\delta C_9^{(s)}}
= \frac{X_{13}/(V_{tb}V_{td}^*)}{X_{23}/(V_{tb}V_{ts}^*)} \,.
\end{equation}
This ratio is relevant for the di-neutrino decays. However, now there is only one measured branching ratio available in \cref{tab:Data} --- the decay process involving the $K^+$ meson in the final state. Therefore, we cannot perform the same kind of analysis, which requires measuring at least two branching ratios. Instead, we can turn the argument around to make a prediction for the unmeasured branching ratio $\cB(B^+\to\pi^+\ov\nu\nu)$ in \cref{eqn:BRs}. Specifically, using the SM prediction
\begin{equation}
\cB(B^+\to\pi^+\ov{\nu}\nu)^\text{SM}=(14\pm1)\times10^{-8} \,,
\end{equation}
we can provide predictions for each of the MFV covariants in \cref{tab:ExplicitVariants}. For the covariants of the first line in \cref{eqn:C9Correlations}, we obtain 
\be
\cB(B^+\to\pi^+\ov{\nu}\nu)=(5.8\pm1.9)\times 10^{-7} \,,
\ee
while for those in the second and third lines of \cref{eqn:C9Correlations}, the MFV corrections are too small to generate any sensitive deviation within the present errors and therefore the prediction is the one of the SM. Comparing these predictions with the present experimental limit
\begin{equation}
\cB(B^+\to\pi^+\ov{\nu}\nu)^\text{exp}<14\times 10^{-6}
\qquad\text{at $90\%$ C.L.} \,,
\end{equation}
we find it not excluding any of the covariants in \cref{tab:PhenoCovariants}.

\subsubsection*{Summary}

The above type of analysis on the MFV correlations, once the errors on the different branching ratios improve, could be useful to point towards a specific realization of the MFV. For example, were improved  measurements  to result in the same central values but with uncertainties decreased by a factor of 2, the  predictions in \cref{eqn:rcRC91} arising from any of the covariants on the first line of \cref{eqn:C9Correlations},  would be in strong tension with data, and those in \cref{eqn:rcRC923a}, corresponding to the 2nd and 3rd lines in \cref{eqn:C9Correlations}, would be strongly favored. However, we emphasize again that our analysis above only surveys the 16 extreme scenarios of these correlations. This does not span all the possibilities, but could help provide a taste of various different scenarios. More practically, in order for MFV to generate distinguishable correlations, it is necessary to supplement the symmetry principle with additional assumptions, preferably motivated by UV considerations. This additional hypothesis, by construction, may exclude certain UV models that adhere to the MFV principle. Consequently, a case-by-case analysis is necessary to explore the implications of the different MFV realizations.

\section{Conclusions and outlook}
\label{sec:Conclusions}

Hilbert series is a powerful tool to systematically enumerate algebraic structures. While it has been extensively applied in particle physics for counting group invariants, relatively limited attention has been directed towards its use in enumerating  all algebraic
structures with a given transformation property (group covariants) which can be constructed out of a given set of fields or spurions.

In this work, we have explored the application of the Hilbert series method to the module of covariants, focusing on the number of independent ones, which equals  the rank of the module.  It can be computed either explicitly from  the ratio of two Hilbert series, or by applying an elegant mathematical theorem by Brion, which involves the unbroken subgroup when the building blocks take a generic vev as explained around \cref{eqn:TheoremBrion}. Often, the rank of the module coincides with the dimension of the representation of the group covariants, implying that the most general covariant can be constructed out of the building blocks. We denote this phenomena \emph{rank saturation}.

The methods explored in this work have a wide range of applications in the context of particle physics whenever all structures with a given transformation property need to be enumerated. This includes Operator Product Expansions, the construction of form factors contributing to a matrix element, and the study of amplitudes or any spurion analysis.

We have applied these techniques to the quark Minimal Flavor Violation spurion analysis. Our aim was to scrutinize the implications of applying the quark MFV symmetry principle without any additional assumptions, assesing whether it leads to specific correlations among flavor observables to distinguish MFV from a general SMEFT. To do so, we  first obtained all the different flavor representations arising in the MFV SMEFT at dimension six, summarized in \cref{tab:MFVdim6SMEFT,tab:BNVMFVdim6SMEFT}. Next, we computed the Hilbert series for these flavor covariants and the corresponding ranks, both for the full flavor group and the relaxed version that does not include baryon number transformations $U(1)_\text{BN}$. The main conclusion is that for all covariants rank saturation is reached. Consequently, the quark MFV symmetry principle does not restrict SMEFT  --- any arbitrary Wilson coefficient is compatible with MFV. Even though we have only computed the Hilbert series for the dim-6 SMEFT, the application of the theorem by Brion allowed us to extend this result to an arbitrary mass dimension, implying that, MFV SMEFT $\equiv$ SMEFT. Moreover the same applies to any MFV EFT sharing the same flavor group and building blocks.

Imposing the MFV symmetry principle by itself therefore does not fulfill the MFV promise of reducing the number of parameters of the SMEFT, nor exclusively describes the low-energy effects of a class of BSM theories with enhanced flavor protection. Nonetheless, MFV can still be a useful organizing principle of the BSM flavor violating operators on top of which one can impose additional assumptions. This could include for instance, treating the Yukawas as order parameters (any of the two or both) or considering one flavor covariant at a time. We considered examples of this approach, and studied the consequent phenomenology focusing on a few specific processes. Alternatively, without relying on additional assumptions, any BSM flavor operators can be parameterized in terms of the studied group covariants, offering useful insights into whether the flavor violation within and beyond the SM truly have a common origin.

Looking ahead, there are intriguing possibilities for the application of  the methods explored in this paper. Hilbert series for covariants can facilitate the study of MFV applied to other effective field theories such as Low-energy EFT (LEFT), Higgs EFT (HEFT), axion EFTs, \etc. Additionally, considering alternative choices of MFV flavor groups, such as discrete MFV, Grand Unified Theories MFV, top-specific scenarios with $SU(2)$, the different possibilities for lepton MFV, opens avenues for further explorations. The study of saturation under the different assumptions also presents opportunities for general model building since it allows us to disentangle the hypothesis that actually restrict the EFT.

\acknowledgments
We thank Andreas Helset and Aneesh Manohar for useful discussions.
L.M. thanks the Department of Physics of the University of California San Diego for hospitality during the development of this project.
The work of B.G., X.L., and P.Q. is supported by the U.S. Department of Energy under grant number DE-SC0009919.
L.M. acknowledges partial financial support by the Spanish Research Agency (Agencia Estatal de Investigaci\'on) through the grant IFT Centro de Excelencia Severo Ochoa No CEX2020-001007-S and by the grants PID2019-108892RB-I00 and PID2022-137127NB-I00 funded by MCIN/AEI/10.13039/501100011033, by the European Union's Horizon 2020 research and innovation programme under the Marie Sk\l odowska-Curie grant agreements No 860881-HIDDeN and 101086085-ASYMMETRY.

\appendix
\section*{Appendices}
\addcontentsline{toc}{section}{\protect\numberline{}Appendices}%

\section{Weyl characters and Haar measure}
\label{appsec:Weyl}

In this appendix, we summarize some basic properties of Weyl characters and Haar measure, and provide a list of their explicit expressions that are relevant for Hilbert series calculations in this paper. For deeper explanations, derivations, and a more thorough list of them, we refer the reader to appendices A and B in Ref.~\cite{Henning:2017fpj}.

\begin{itemize}
\item \textbf{Character:} The character of a group element $g(x) \in G$ is defined as the trace of its representation matrix:
\begin{equation}
\chi_R^{} \big( g(x) \big) \equiv \tr \big( g_R^{}(x) \big) \,.
\label{eqn:chidef}
\end{equation}
It is a function that depends on the group element $g(x)$ as well as the representation $R$. Consider the group $G=SU(2)$ for example. Any of its elements can be parameterized by three real parameters $x_a$ with $a=1, 2, 3$:
\begin{equation}
g(x) = e^{i x_a t^a} \,,
\label{eqn:gxExample}
\end{equation}
where $t^a$ are the three generators of $SU(2)$. Now suppose we are talking about the fundamental representation $R=\mathbf{2}$, the representation matrices of the three generators are (half of) the Pauli matrices
\begin{equation}
t_\textbf{2}^a = \frac{\sigma^a}{2}
\qquad\Longrightarrow\qquad
g_\mathbf{2} (x) = e^{i x_a \sigma^a/2} \,,\quad
\chi_\mathbf{2}^{} (x) = \tr \left( e^{i x_a \sigma^a/2} \right) \,.
\label{eqn:chiExample}
\end{equation}
To compute this character, we diagonalize the Hermitian matrix in the exponent:
\begin{equation}
x_a t_\mathbf{2}^a
= \frac12 \mqty( x_3 & x_1 - i x_2 \\ x_1 + i x_2 & - x_3 )
= U_x\, \frac12 \mqty( \theta(x) & 0 \\ 0 & -\theta(x) )\, U_x^\dagger \,,
\end{equation}
with the real eigenvalue
\begin{equation}
\theta(x) \equiv \sqrt{x_1^2 + x_2^2 + x_3^2} \,,
\label{eqn:thetaxExample}
\end{equation}
and some (special) unitary matrix $U_x$. Accordingly, the representation matrix can be diagonalized as
\begin{equation}
g_\mathbf{2} (x) = U_x\, \mqty(z_1(x) & 0 \\ 0 & z_2(x))\, U_x^\dagger
\qquad\text{with}\qquad
\left\{ \begin{array}{l}
z_1(x) = e^{i\theta(x)/2} \\[3pt]
z_2(x) = e^{-i\theta(x)/2} \end{array} \right. \,,
\label{eqn:gxDiagExample}
\end{equation}
leading us to the character
\begin{equation}
\chi_\mathbf{2}^{} (x) = \tr \big( g_\mathbf{2}^{}(x) \big) = z_1(x) + z_2(x) = z_1(x) + z_1^{-1}(x) \,.
\label{eqn:chizExample}
\end{equation}

A few basic properties of the character follow from the definition in \cref{eqn:chidef}. First, in the conjugate representation $\overline{R}$, the representation matrices are given by the complex conjugation $g_{\overline{R}}^{} = g_R^*$, therefore, the character of $\overline{R}$ is also given by the complex conjugation
\begin{equation}
\chi_{\overline{R}}^{} (g) = \tr\left( g_{\overline{R}}^{} \right)
= \tr \left( g_R^* \right) = \chi_R^* (g) \,.
\label{eqn:chistar}
\end{equation}
Next, in a direct sum of two representations $R = R_1 \oplus R_2$, the representation matrices can be made block diagonal $g_R^{} = g_{R_1}^{} \oplus g_{R_2}^{}$. This implies that
\begin{equation}
\chi_{R_1 \oplus R_2}^{} (g) = \chi_{R_1}^{} (g) + \chi_{R_2}^{} (g) \,.
\label{eqn:chioplus}
\end{equation}
Finally, in a direct product of two representations $R = R_1 \otimes R_2$, the representation matrices are the tensor products $g_R^{} = g_{R_1}^{} \otimes g_{R_2}^{}$, which results into
\begin{equation}
\chi_{R_1 \otimes R_2}^{} (g) = \chi_{R_1}^{} (g) \cdot \chi_{R_2}^{} (g) \,.
\label{eqn:chiotimes}
\end{equation}
\item \textbf{Weyl character:} For a connected Lie group $G$, the Weyl character is the character function restricted to its Cartan subgroup $H\subset G$. The dimension of $G$ is the number of its generators, while the dimension of its Cartan subgroup $H$ is the number of mutually commuting generators, \ie\ the rank of $G$, which is typically much smaller than the dimension of $G$. For the $G=SU(2)$ example in \cref{eqn:gxExample}, its Cartan subgroup can be chosen as the $U(1)$ subgroup generated by the third generator $t^3$:
\begin{equation}
h(\theta) = e^{i \theta t^3} \in H \subset G \,,
\label{eqn:chiExampleh}
\end{equation}
and the Weyl character in the fundamental representation of $SU(2)$ is
\begin{equation}
\chi_\mathbf{2}^{} \big( h(\theta) \big) = z + z^{-1}
\qquad\text{with}\qquad
z=e^{i\theta/2} \,.
\label{eqn:chiSU2f}
\end{equation}
It is a function of a single real parameter $\theta$. Similarly, the Weyl character in the adjoint representation of $SU(2)$ is
\begin{equation}
t_\mathbf{3}^3 = \mqty( 1 & 0 & 0 \\ 0 & 0 & 0 \\ 0 & 0 & -1 )
\qquad\implies\qquad
\chi_\mathbf{3}^{} \big( h(\theta) \big) = z^2 + 1 + z^{-2} \,,
\label{eqn:chiSU2adj}
\end{equation}
with the same $z$ defined in \cref{eqn:chiSU2f}.
\item \textbf{Haar measure:} For a compact Lie group $G$, the Haar measure $\dd\bar\mu_G (g)$ is the unique normalized measure of its volume that is invariant under both left and right translations:
\begin{equation}
\int \dd\bar\mu_G^{} (g) = 1 \,,\qquad\text{and}\qquad
\dd\bar\mu_G^{} (g) = \dd\bar\mu_G^{} \big( \tilde{g} \cdot g \big) = \dd\bar\mu_G^{} \big( g\cdot \tilde{g} \big) \,,
\end{equation}
for any fixed group element $\tilde{g}$. To obtain it, one can first construct the Cartan-Maurer form
\begin{equation}
i\, \omega (g) = i\, \omega_a(g)\, t^a \equiv g^{-1} \dd g \,,
\end{equation}
where $a=1, 2, \cdots, n$, with $n=\dim(G)$. The Haar measure is then given by
\begin{equation}
\dd\bar\mu_G^{} (g) = \omega_1(g) \wedge \omega_2(g) \wedge \cdots \wedge \omega_n(g) \,.
\end{equation}
We reiterate that this is a measure of an $n$ dimensional volume, with $n=\dim(G)$.
\item \textbf{Character orthonormality and Weyl integration formula:} Both the Haar measure and the character are functions over the whole group. An important property of them is that upon integrating over the Haar measure, the characters of two unitary irreps $R_1, R_2$ satisfy the following orthonormality relation
\begin{equation}
\int \dd\bar\mu_G^{} (g)\, \chi_{R_1}^* (g)\, \chi_{R_2}^{} (g) = \delta_{R_1 R_2}^{} \,.
\label{eqn:chiOthapp}
\end{equation}
The \emph{Weyl integration formula} (\eg\ \cite{brocker2013representations}) states that this orthonormality holds also for the Weyl characters with a ``modified Haar measure'' $\dd\mu_G^{}(h)$ over elements $h$ in the Cartan subgroup $H\subset G$, namely
\begin{equation}
\int \dd\mu_G^{} (h)\, \chi_{R_1}^* (h)\, \chi_{R_2}^{} (h) = \delta_{R_1 R_2}^{} \,.
\label{eqn:chiWeylOthapp}
\end{equation}
This is because each group element $g\in G$ can be linked with an element $h \in H \subset G$ by a group conjugation:
\begin{equation}
g = \tilde{g}\, h\, \tilde{g}^{-1} \,,
\qquad\text{with}\quad
\tilde{g} \in G/H \,,
\label{eqn:gtoh}
\end{equation}
which brings the character to the Weyl character in any representation $\chi_R^{}(g) = \chi_R^{}(h)$. An example of this conjugation is the ``diagonalization'' procedure in \cref{eqn:gxDiagExample}, corresponding to the map in \cref{eqn:thetaxExample} between a general group element $g(x)$ in \cref{eqn:gxExample} and a Cartan element $h(\theta)$ in \cref{eqn:chiExampleh}. Taking \cref{eqn:gtoh} as a parameterization for all the elements $g\in G$, one can factorize the Haar measure as
\begin{equation}
\dd\bar\mu_G^{} (g) = \dd\mu_G^{} \big( \tilde{g} \big)\, \dd\mu_G^{} (h) \,.
\label{eqn:dmuFactorization}
\end{equation}
When integrating over the characters as in \cref{eqn:chiOthapp}, the $\dd\mu_G^{} \big( \tilde{g} \big)$ part can be done trivially, because the integrand does not depend on $\tilde{g}$. This leads to the Weyl integration formula in \cref{eqn:chiWeylOthapp}. However, the remaining part $\dd\mu_G^{} (h)$ in the measure factorization \cref{eqn:dmuFactorization} is fairly nontrivial, which needs to be carefully worked out following the parameterization in \cref{eqn:gtoh}. The result is the Haar measure of the Cartan subgroup $\dd\bar\mu_H^{} (h)$ multiplied by a proper Jacobian factor (see \eg\ \cite{Henning:2017fpj} for derivation):
\begin{align}
\dd\mu_G^{} (h) &= \dd\bar\mu_H^{} (h)\, \frac{1}{|W|} \det \left( 1 - h_\text{adj} \big|_{G/H} \right)
\notag\\[5pt]
&= \left( \prod_{i=1}^r \frac{\dd z_i}{2\pi i\, z_i} \right) \frac{1}{|W|} \prod_{\alpha\,\in\text{ roots}} \Big( 1 - z_1^{\alpha_1} z_2^{\alpha_2} \cdots z_r^{\alpha_r} \Big) \,.
\label{eqn:HaarModified}
\end{align}
Here, $|W|$ is the number of elements in the Weyl group $W$ of $G$. $r$ is the rank of $G$, and $\{ z_i=e^{i\theta_i} \}$ parameterize the Cartan subgroup $H = U(1)^r$. $h_\text{adj}$ denotes the representation matrix of the group element $h\in H$ under the adjoint representation. It is a diagonal matrix as $h$ belongs to the Cartan subgroup. It has $n=\dim(G)$ diagonal entries, as we are talking about the adjoint representation. $h_\text{adj} \big|_{G/H}$ then denotes the $(n-r)\times(n-r)$ submatrix of $h_\text{adj}$ where only the $G/H$ subspace of the adjoint space are kept. The ``modified Haar measure'' $\dd\mu_G^{} (h)$ in \cref{eqn:HaarModified} is what we refer to as the Haar measure in this paper. When the integrand is invariant under the Weyl group, it can be further reduced to
\begin{equation}
\dd\mu_G^{} (h) \quad\longrightarrow\quad
\left( \prod_{i=1}^r \frac{\dd z_i}{2\pi i\, z_i} \right) \prod_{\alpha\,\in\text{ positive roots}} \Big( 1 - z_1^{\alpha_1} z_2^{\alpha_2} \cdots z_r^{\alpha_r} \Big) \,.
\end{equation}
\end{itemize}
Next, we provide a list of Weyl characters and Haar measure that are relevant for the Hilbert series calculations in this paper.

For the group $G=U(1)$, Weyl characters are the same as the characters. An irrep can be labeled by the charge $R=Q$, and the (Weyl) characters are
\begin{equation}
\chi_{R=Q}^{U(1)}(z) = z^Q \,.
\label{eqn:chiU1}
\end{equation}
The group $SU(2)$ has rank $1$; its Weyl characters are functions of a single phase variable $z$. One can label $SU(2)$ irreps by the spin $j$ or the dimension $R=\dim=2j+1$. The Weyl characters are
\begin{equation}
\chi_{R=\dim}^{SU(2)}(z) = \sum_{k=1}^{\dim} z^{\dim+1-2k}
= z^{\dim-1} + z^{\dim-3} + \cdots + z^{-(\dim-1)} \,.
\label{eqn:chiSU2}
\end{equation}
The group $SU(3)$ has rank $2$; its Weyl characters are functions of two phase variables $z_1, z_2$. We label the irreps of $SU(3)$ by their dimensions $R=\dim$, except for the invariant irrep $R=\I$. The relevant irreps for the MFV Hilbert series calculations in \cref{sec:MFV} are $R=\I, \mathbf{3}, \mathbf{6}, \mathbf{8}, \mathbf{10}, \mathbf{27}$. Their Weyl characters read
\begin{subequations}\label{eqn:chiSU3}
\begin{align}
\chi_\I^{SU(3)}(z_1, z_2) &= 1 \,, \\[5pt]
\chi_{\mathbf{3}}^{SU(3)}(z_1, z_2) &= z_1 + \frac{1}{z_2} + \frac{z_2}{z_1} \,, \\[5pt]
\chi_{\mathbf{6}}^{SU(3)}(z_1, z_2) &= z_1^2 + \frac{1}{z_2^2} + \frac{z_2^2}{z_1^2} + \frac{z_1}{z_2} 
+ z_2 + \frac{1}{z_1} \,, \\[5pt] 
\chi_{\mathbf{8}}^{SU(3)}(z_1, z_2) &= \frac{z_1^2}{z_2} + \frac{z_2^2}{z_1} + \frac{z_1}{z_2^2} + \frac{z_2}{z_1^2} 
+ z_1 z_2 + \frac{1}{z_1 z_2} + 2 \,, \\[5pt]
\chi_{\mathbf{10}}^{SU(3)}(z_1, z_2) &= z_1^3 + \frac{1}{z_2^3} + \frac{z_2^3}{z_1^3} + \frac{z_1^2}{z_2} + z_1 z_2 
+ \frac{z_1}{z_2^2} + \frac{1}{z_1 z_2} + \frac{z_2^2}{z_1} + \frac{z_2}{z_1^2} + 1 \,, \\[5pt]
\chi_{\mathbf{27}}^{SU(3)}(z_1, z_2) &= \left( 1 + \frac{z_1}{z_2^2} + \frac{z_2^2}{z_1} \right)
\left( 1 + \frac{z_2}{z_1^2} + \frac{z_1^2}{z_2} \right) \left( 1 + z_1 z_2 + \frac{1}{z_1 z_2} \right) \,.
\end{align}
\end{subequations}
Following the general property in \cref{eqn:chistar}, the Weyl characters of the conjugate representations can be obtained by replacing each variable $z_i = e^{i\theta_i}$ with its complex conjugate
\begin{align}
\chi_{\overline{R}}^{} (z_i) = \chi_R^{} (z_i) \big|_{z_i\rightarrow z_i^* = z_i^{-1}}\,.
\end{align}
The Weyl character of a direct sum or of a direct product of two representations can be obtained by the general properties in \cref{eqn:chioplus,eqn:chiotimes}.

The Haar measure (over the Cartan subgroup) of $U(1)$, $SU(2)$, and $SU(3)$ are
\begin{subequations}\label{eqn:HaarU1SU2SU3}
\begin{align}
\int \dd\mu_{U(1)}^{}   & = \oint_{|z|=1} \frac{\dd z}{2\pi i\, z} \,, \label{eqn:HaarU1} \\[8pt]
\int \dd\mu_{S U(2)}^{} & = \oint_{|z|=1} \frac{\dd z}{2\pi i\, z}\,
\frac12 \left( 1 - z^2 \right) \left( 1 - z^{-2} \right)
\to \oint_{|z|=1} \frac{\dd z}{2\pi i\, z} \left(1-z^2\right) \,, \label{eqn:HaarSU2} \\[8pt]
\int \dd\mu_{S U(3)}^{} & = \oint_{\left|z_1\right|=1} \frac{\dd z_1}{2\pi i\, z_1}
\oint_{\left|z_2\right|=1} \frac{\dd z_2}{2\pi i\, z_2}\,
\frac16 \left( 1 - z_1 z_2 \right) \left( 1 - \frac{z_1^2}{z_2} \right) \left( 1 - \frac{z_2^2}{z_1} \right)
\notag\\[5pt]
&\hspace{160pt}
\times \left( 1 - \frac{1}{z_1 z_2} \right) \left( 1 - \frac{z_1}{z_2^2} \right) \left( 1 - \frac{z_2}{z_1^2} \right)
\notag\\[8pt]
&\to \oint_{\left|z_1\right|=1} \frac{\dd z_1}{2\pi i\, z_1} \oint_{\left|z_2\right|=1} \frac{\dd z_2}{2\pi i\, z_2} \left( 1 - z_1 z_2 \right) \left( 1 - \frac{z_1^2}{z_2} \right) \left( 1 - \frac{z_2^2}{z_1} \right) \,, \label{eqn:HaarSU3}
\end{align}
\end{subequations}
where expressions following the arrows are reduced Haar measure for a symmetric integrand under the Weyl group.

\section{Examples of Hilbert series calculation}
\label{appsec:Examples}

In this appendix, we give explicit examples of applying the Molien-Weyl formula in \cref{eqn:MolienR} to computing the Hilbert series for some specific group invariants and covariants. Several simple examples for the group $G=U(1)$ were already discussed in \cref{sec:Hilbert}. Here we will focus on the case of $G=SU(3)$ and some MFV scenarios discussed in \cref{sec:MFV}.
In addition, we also tabulate a variety of Hilbert series results in \cref{tab:U1,tab:SU2adj,tab:SU3ffbar,tab:SU3adj} for $U(1)$, $SU(2)$, $SU(3)$ group invariants and covariants built out of various building blocks. From these results, one can visualize the general properties of Hilbert series discussed in \cref{sec:Hilbert}, especially regarding the rank of the modules of covariants and its saturation.

\subsection*{Example $SU(3)$ Hilbert series}

In order to compute the Hilbert series with the Molien-Weyl formula in \cref{eqn:MolienR}, one needs the Haar measure of the group $\dd\mu_G^{}$, the character of the desired covariant rep $\chi_R^G$, and the building block determinant $\det \left( 1 - q\, g_\RBB^{} \right)$. For $G=SU(3)$, the Haar measure is given in \cref{eqn:HaarSU3}; the characters are listed in \cref{eqn:chiSU3}. So all we still need are the building block determinant, which depends on the rep $\RBB$.

Let us begin with the case $\RBB = \mathbf{3} \oplus \mathbf{\bar 3}$. The representation matrices are
\begin{equation}
g_\mathbf{3}^{} = \mqty( z_1 & & \\ & z_2^{-1} & \\ & & z_1^{-1} z_2 ) \,, \qquad
g_\mathbf{\bar 3}^{} = \mqty( z_1^{-1} & & \\ & z_2 & \\ & & z_1 z_2^{-1} ) \,,\qquad
g_{\mathbf{3} \oplus \mathbf{\bar 3} }^{} = g_\mathbf{3}^{} \oplus g_\mathbf{\bar 3}^{} \,.
\label{eqn:g3SU3}
\end{equation}
From these, we get the determinants
\begin{subequations}\label{eqn:detSU3ffbar}
\begin{align}
\textstyle\det_\mathbf{3}^{} (q, z_1,z_2) &\equiv \det \left( 1 - q\, g_\mathbf{3}^{} \right)
= \left( 1 - q\, z_1 \right) \left( 1 - q\, z_2^{-1} \right) \left( 1 - q\, z_1^{-1} z_2 \right) \,, \\[5pt]
\textstyle\det_\mathbf{\bar 3}^{} (q, z_1,z_2) &\equiv \det \left( 1 - q\, g_\mathbf{\bar 3}^{} \right)
= \left( 1 - q\, z_1^{-1} \right) \left( 1 - q\, z_2 \right) \left( 1 - q\, z_1 z_2^{-1} \right) \,, \\[5pt]
\textstyle\det_{\mathbf{3} \oplus \mathbf{\bar 3}}^{} (q, z_1,z_2) &\equiv \det \left( 1 - q\, g_{\mathbf{3} \oplus \mathbf{\bar 3}}^{} \right) = \textstyle\det_\mathbf{3}^{} (q, z_1,z_2)\, \textstyle\det_\mathbf{\bar 3}^{} (q, z_1,z_2) \,.
\end{align}
\end{subequations}
Inserting these into the Molien-Weyl formula, we get the Hilbert series for invariants
\begin{small}
\begin{align}
\HS_\I^{SU(3),\, \mathbf{3} \oplus \mathbf{\bar 3}} &= \int \dd\mu_{SU(3)}^{} (z_1, z_2)\, \frac{1}{\textstyle\det_{\mathbf{3} \oplus \mathbf{\bar 3}}^{} (q, z_1,z_2)} \notag\\[8pt]
&= \oint_{\left|z_1\right|=1} \frac{\dd z_1}{2\pi i\, z_1} \oint_{\left|z_2\right|=1} \frac{\dd z_2}{2\pi i\, z_2} \left( 1 - z_1 z_2 \right) \left( 1 - \frac{z_1^2}{z_2} \right) \left( 1 - \frac{z_2^2}{z_1} \right)
\notag\\[5pt]
&\hspace{20pt}
\times \frac{1}{\left( 1 - q\, z_1 \right) \left( 1 - q\, z_2^{-1} \right) \left( 1 - q\, z_1^{-1} z_2 \right)}
\frac{1}{\left( 1 - q\, z_1^{-1} \right) \left( 1 - q\, z_2 \right) \left( 1 - q\, z_1 z_2^{-1} \right)}
\notag\\[8pt]
&= \frac{1}{1-q^2} \,,
\end{align}
\end{small}%
as well as for covariants, such as
\begin{small}
\begin{align}
\HS_\mathbf{3}^{SU(3),\, \mathbf{3} \oplus \mathbf{\bar 3}} &= \int \dd\mu_{SU(3)}^{} (z_1, z_2)\, \chi_\mathbf{3}^*(z_1, z_2)\, \frac{1}{\textstyle\det_{\mathbf{3} \oplus \mathbf{\bar 3}}^{} (q, z_1,z_2)} \notag\\[8pt]
&= \oint_{\left|z_1\right|=1} \frac{\dd z_1}{2\pi i\, z_1} \oint_{\left|z_2\right|=1} \frac{\dd z_2}{2\pi i\, z_2} \left( 1 - z_1 z_2 \right) \left( 1 - \frac{z_1^2}{z_2} \right) \left( 1 - \frac{z_2^2}{z_1} \right)
\left( \frac{1}{z_1} + z_2 + \frac{z_1}{z_2} \right)
\notag\\[5pt]
&\hspace{20pt}
\times \frac{1}{\left( 1 - q\, z_1 \right) \left( 1 - q\, z_2^{-1} \right) \left( 1 - q\, z_1^{-1} z_2 \right)}
\frac{1}{\left( 1 - q\, z_1^{-1} \right) \left( 1 - q\, z_2 \right) \left( 1 - q\, z_1 z_2^{-1} \right)}
\notag\\[8pt]
&= \frac{q}{1-q^2} \,.
\end{align}
\end{small}%
The Hilbert series with more copies of $\mathbf{3}+\mathbf{\bar 3}$ as the building blocks can be obtained by similar expressions. For $m$ copies, \ie\ $\RBB = m\times \left( \mathbf{3}+\mathbf{\bar 3} \right)$, we have
\begin{equation}
\textstyle\det_{m\times \left( \mathbf{3}+\mathbf{\bar 3} \right)}^{} (q, z_1,z_2) = 
\left[ \textstyle\det_{\mathbf{3} \oplus \mathbf{\bar 3}}^{} (q, z_1,z_2) \right]^m \,,
\end{equation}
which leads us to 
\begin{equation}
\HS_R^{SU(3),\, m\times\left(\mathbf{3} \oplus \mathbf{\bar 3}\right)} (q) = \int \dd\mu_{SU(3)}^{} (z_1, z_2)\, \chi_R^*(z_1, z_2)\, \frac{1}{\left[ \textstyle\det_{\mathbf{3} \oplus \mathbf{\bar 3}}^{} (q, z_1,z_2) \right]^m} \,,
\end{equation}
with the Haar measure $\dd\mu_{SU(3)}^{} (z_1, z_2)$ given in \cref{eqn:HaarSU3}, the characters $\chi_R^{}(z_1, z_2)$ listed in \cref{eqn:chiSU3}, and $\det_{\mathbf{3} \oplus \mathbf{\bar 3}}^{} (q, z_1,z_2)$ given in \cref{eqn:detSU3ffbar}. In \cref{tab:SU3ffbar}, we list the results of these Hilbert series for $m=1,2,3,4$ and $R=\I, \mathbf{3}, \mathbf{6}, \mathbf{8}, \mathbf{10}, \mathbf{27}$.

Now let us move on to the case of having adjoint reps as the building blocks. For a single adjoint rep $\RBB = \mathbf{8}$, the representation matrix and the determinant are
\begin{align}
g_\mathbf{8}^{} &= \text{diag} \left( 1\,,\, 1\,,\, z_1 z_2\,,\, z_1^{-1} z_2^{-1}\,,\, z_1 z_2^{-2}\,,\,
z_1^{-2} z_2\,,\, z_1^2 z_2^{-1}\,,\, z_1^{-1} z_2^2 \right) \,, \\[10pt]
\textstyle\det_\mathbf{8} (q, z_1,z_2) &= \left( 1 - q \right)^2
\left( 1 - q\, z_1 z_2 \right) \left( 1 - q\, z_1^{-1} z_2^{-1} \right) 
\left( 1 - q\, z_1 z_2^{-2} \right) \left( 1 - q\, z_1^{-2} z_2 \right)
\notag\\[5pt]
&\hspace{40pt}
\times \left( 1 - q\, z_1^2 z_2^{-1} \right) \left( 1 - q\, z_1^{-1} z_2^2 \right) \,.
\label{eqn:detSU3adj}
\end{align}
For $m$ copies of adjoint reps as the building blocks $\RBB = m\times (\mathbf{8})$, we have
\begin{equation}
\textstyle\det_{m\times(\mathbf{8})} (q, z_1,z_2) = \left[ \textstyle\det_\mathbf{8} (q, z_1,z_2) \right]^m \,.
\end{equation}
Therefore, we get the Hilbert series
\begin{equation}
\HS_R^{SU(3),\, m\times(\mathbf{8})} (q) = \int \dd\mu_{SU(3)}^{} (z_1, z_2)\, \chi_R^*(z_1, z_2)\, \frac{1}{\left[ \textstyle\det_{\mathbf{8}}^{} (q, z_1,z_2) \right]^m} \,,
\end{equation}
where $\dd\mu_{SU(3)}^{} (z_1, z_2)$, $\chi_R^{}(z_1, z_2)$ and $\det_{\mathbf{8}}^{} (q, z_1,z_2)$ are given in \cref{eqn:HaarSU3,eqn:chiSU3,eqn:detSU3adj}, respectively. In \cref{tab:SU3adj}, we list the results of these Hilbert series for $m=1,2,3$ and $R=\I, \mathbf{3}, \mathbf{6}, \mathbf{8}, \mathbf{10}, \mathbf{27}$.

\subsection*{Example Hilbert series for MFV covariants}

Now let us show explicit calculations of the Hilbert series for some MFV covariants discussed in \cref{sec:MFV}. In these cases, the symmetry group is the product group
\begin{equation}
G_f = U(3)_q \times U(3)_u \times U(3)_d \,,
\label{eqn:GFapp}
\end{equation}
and the building blocks are the Yukawa matrices $Y_u$, $Y_d$ that transform under $G_f$ as
\begin{subequations}\label{eqn:YuYdTransapp}
\begin{align}
Y_u &\quad\longrightarrow\quad
U_q\, Y_u\, U_u^\dagger \,, \\[5pt]
Y_d &\quad\longrightarrow\quad
U_q\, Y_d\, U_d^\dagger \,.
\end{align}
\end{subequations}
Here $U_q$, $U_u$, and $U_d$ are group elements of $U(3)_q$, $U(3)_u$, and $U(3)_d$ respectively. The hermitian conjugates of these Yukawa matrices are also building blocks, so the full set of building blocks are $\BB=\{Y_u, Y_d, Y_u^\dagger, Y_d^\dagger\}$.

To label the representations of the symmetry group $G_f$, we decompose it as
\begin{equation}
G_f = G_S \times G_U \,,
\end{equation}
with $G_S$ collecting the $SU(3)$ factors and $G_U$ collecting the $U(1)$ factors:
\begin{subequations}
\begin{align}
G_S &\equiv SU(3)_q \times SU(3)_u \times SU(3)_d \,, \label{eqn:GS} \\[5pt]
G_U &\equiv  U(1)_q \times  U(1)_u \times  U(1)_d \,. \label{eqn:GU}
\end{align}
\end{subequations}
In this decomposition, the building blocks $\BB$ are in the following reps
\begin{subequations}\label{eqn:RYuRYd}
\begin{align}
R_{Y_u}^{} &= \left( \mathbf{3}, \mathbf{\bar 3}, \mathbf{1} \right)^{G_S} \otimes \left( 1, -1, 0 \right)^{G_U} \,, \\[5pt]
R_{Y_d}^{} &= \left( \mathbf{3}, \mathbf{1}, \mathbf{\bar 3} \right)^{G_S} \otimes \left( 1, 0, -1 \right)^{G_U} \,, \\[5pt]
R_{Y_u^\dagger} &= \left( \mathbf{\bar 3}, \mathbf{3}, \mathbf{1} \right)^{G_S} \otimes \left( -1, 1, 0 \right)^{G_U} \,, \\[5pt]
R_{Y_d^\dagger} &= \left( \mathbf{\bar 3}, \mathbf{1}, \mathbf{3} \right)^{G_S} \otimes \left( -1, 0, 1 \right)^{G_U} \,.
\end{align}
\end{subequations}
Sometimes it is also convenient to consider the following combinations of the building blocks
\begin{subequations}\label{eqn:huhdapp}
\begin{align}
h_u &\equiv Y_u\, Y_u^\dagger
\quad\longrightarrow\quad
U_q\, h_u\, U_q^\dagger \,, \\[5pt]
h_d &\equiv Y_d\, Y_d^\dagger
\quad\longrightarrow\quad
U_q\, h_d\, U_q^\dagger \,.
\end{align}
\end{subequations}
They transform in the same way --- only under the subgroup $SU(3)_q$, as an invariant $\mathbf{1}$ plus an adjoint $\mathbf{8}$:
\begin{equation}
R_{h_u} = R_{h_d} = \left( \mathbf{1} \oplus \mathbf{8}, \mathbf{1}, \mathbf{1} \right)^{G_S}
\otimes \left( 0, 0, 0 \right)^{G_U} \,.
\label{eqn:RhuRhd}
\end{equation}

When discussing $G_f$ representations $R=R^{G_S} \otimes R^{G_U}$, we will often omit the $R^{G_U}$ part and focus on the more nontrivial part $R^{G_S}$. The charges for the $U(1)$ factors can often be inferred from the context. For example, we will demonstrate in below the calculations of the Hilbert series for the following goal reps of $G_f$:
\begin{equation}
\mathbf{(8,1,1)} \,,\quad
\mathbf{(1,8,1)} \,,\quad
\mathbf{(1,3,\bar{3})} \,.
\label{eqn:GoalRepsS}
\end{equation}
The omitted $U(1)$ charges for these reps are
\begin{equation}
\left( 0, 0, 0 \right) \,,\quad
\left( 0, 0, 0 \right) \,,\quad
\left( 0, 1, -1 \right) \,.
\label{eqn:GoalRepsU}
\end{equation}
Below, we show explicit expressions of using the Molien-Weyl formula in \cref{eqn:MolienR} to compute the Hilbert series for these MFV covariants.

Let us begin with the invariants, \ie\ the $\mathbf{(1,1,1)}$ representation. In this case, one can simplify the calculation by using $h_u, h_d$ in \cref{eqn:huhdapp} as the building blocks, because to make $U(3)_u$ invariants out of $\BB$, $Y_u$ and $Y_u^\dagger$ must come together in the form of $h_u$ to cancel the $U(3)_u$ transformation (\cf\ \cref{eqn:YuYdTransapp}); the same is true for $Y_d$ and $Y_d^\dagger$ to make $U(3)_d$ invariants. Therefore, everything is made out of $h_u$ and $h_d$. This significantly simplifies the calculation, because one only needs to integrate over the group $SU(3)_q$, instead of the three unitary groups (\cf\ \cref{eqn:RhuRhd}). Applying the Molien-Weyl formula to compute the Hilbert series, we get
\begin{equation}
\HS_\mathbf{(1,1,1)}(q) = \int \dd\mu_{SU(3)_q}^{} (z_1,z_2)\;
\frac{1}{\det_{h_u}(q, z_1, z_2)}\, \frac{1}{\det_{h_d}(q, z_1, z_2)} \,.
\label{eqn:HS111Explicit}
\end{equation}
Here the Haar measure $\dd\mu_{SU(3)_q}^{}$ is given in \cref{eqn:HaarSU3}. The determinant factors from the building blocks $h_u$ and $h_d$ are the same (as they form the same type of rep):
\begin{equation}
\textstyle\det_{h_u}(q, z_1, z_2) = \textstyle\det_{h_d}(q, z_1, z_2)
= \textstyle\det_{\mathbf{8\oplus 1}}^{} (q^2, z_1,z_2) \,.
\end{equation}
Note that the first argument of the last expression is set to $q^2$, because our new building blocks $h_u$, $h_d$ are quadratic in the Yukawa matrices --- the grading variable $q$ is tracking the powers of the Yukawas. The determinant function in the last expression above can be computed following the rep decomposition:
\begin{equation}
\textstyle\det_{\mathbf{8\oplus 1}}^{} (q, z_1,z_2)
= \textstyle\det_{\mathbf{8}}^{} (q, z_1,z_2)\, \textstyle\det_{\mathbf{1}}^{} (q, z_1,z_2) \,,
\end{equation}
where the determinant for the adjoint rep is already given in \cref{eqn:detSU3adj}, and the determinant for the singlet rep is simply
\begin{equation}
\textstyle\det_{\mathbf{1}}^{} (q, z_1,z_2) = 1-q \,.
\label{eqn:detSU3Inv}
\end{equation}
Plugging all these ingredients into \cref{eqn:HS111Explicit} leads us to the result given in \cref{eqn:HS111}.

Now we move on to the $\mathbf{(8,1,1)}$-covariants. In this case, the same simplification for the $\mathbf{(1,1,1)}$ rep works --- everything is again made out of $h_u, h_d$, and we only need to integrate over the group $SU(3)_q$. Applying the Molien-Weyl formula to compute the Hilbert series, we get
\begin{equation}
\HS_\mathbf{(8,1,1)}(q) = \int \dd\mu_{SU(3)_q}^{} (z_1,z_2)\;
\chi_\mathbf{8}^*(z_1,z_2)\; \frac{1}{\det_{h_u}(q, z_1, z_2)}\, \frac{1}{\det_{h_d}(q, z_1, z_2)} \,.
\label{eqn:HS811Explicit}
\end{equation}
Here the character $\chi_\mathbf{8}^*(z_1,z_2)=\chi_\mathbf{8}(z_1,z_2)$ can be found in \cref{eqn:chiSU3}, and other ingredients are the same as in \cref{eqn:HS111Explicit}. Plugging in these ingredients leads us to the result given in \cref{eqn:HS811}.

Next, we move on to the $\mathbf{(1,8,1)}$-covariants. As we are selecting $U(3)_d$ invariants, $Y_d$ and $Y_d^\dagger$ must still come together in the form of $h_d = Y_d\, Y_d^\dagger$. Therefore, we can use $\BB = \{ Y_u, Y_u^\dagger, h_d \}$ as our building blocks. Checking their representations in \cref{eqn:RYuRYd,eqn:RhuRhd}, we see that it is necessary to integrate over two groups, $SU(3)_q$ and $U(3)_u$:
\begin{align}
\HS_\mathbf{(1,8,1)}(q) &= \int\dd\mu_{SU(3)_q}^{} (z_1, z_2) \int \dd\mu_{SU(3)_u}^{} (w_1,w_2)
\int \dd\mu_{U(1)_u}^{}(y)\, \chi_\mathbf{8}^*(w_1,w_2)
\notag\\[5pt]
&\hspace{-30pt} \times
\frac{1}{\det_{h_d}(q, z_1, z_2)}\,
\frac{1}{\textstyle\det_{Y_u} (q, z_1, z_2, w_1, w_2, y)} \,
\frac{1}{\textstyle\det_{Y_u^\dagger} (q, z_1, z_2, w_1, w_2, y)} \,.
\label{eqn:HS181Explicit}
\end{align}
The Haar measure and characters have been explicitly provided before, so is the determinant factor $\det_{h_d}(q, z_1, z_2)$. The determinant factors from $Y_u$ and $Y_u^\dagger$ can be worked out from their representations under $G_f$, which we recall from \cref{eqn:RYuRYd} as
\begin{subequations}
\begin{align}
R_{Y_u}^{} &= \left( \mathbf{3}, \mathbf{\bar 3}, \mathbf{1} \right)^{G_S} \otimes \left( 1, -1, 0 \right)^{G_U} \,, \\[5pt]
R_{Y_u^\dagger} &= \left( \mathbf{\bar 3}, \mathbf{3}, \mathbf{1} \right)^{G_S} \otimes \left( -1, 1, 0 \right)^{G_U} \,.
\end{align}
\end{subequations}
This tells us, for example, the representation matrix for the building block $Y_u$ is
\begin{align}
g_{Y_u}^{}(z_1, z_2, w_1, w_2, y)
&= g_\mathbf{3}^{}(z_1, z_2) \otimes g_\mathbf{\bar 3}^{}(w_1, w_2) \otimes y^{-1}
\notag\\[5pt]
&= \mqty( z_1 & & \\ & z_2^{-1} & \\ & & z_1^{-1} z_2 )
\otimes \mqty( w_1^{-1} & & \\ & w_2 & \\ & & w_1 w_2^{-1} ) \otimes y^{-1} \,,
\label{eqn:gYu}
\end{align}
where the representation matrices $g_\mathbf{3}^{}$ and $g_\mathbf{\bar 3}^{}$ were provided in \cref{eqn:g3SU3}. Computing the determinant factor $\textstyle\det_{Y_u}(q, z_1, z_2, w_1, w_2, y)$ from \cref{eqn:gYu} is straightforward:
\begin{align}
\textstyle\det_{Y_u}(q, z_1, z_2, w_1, w_2, y) &= \det\Big[ 1 - q\, g_{Y_u}^{}(z_1, z_2, w_1, w_2, y) \Big]
\notag\\[5pt]
&\hspace{-105pt} =
\left( 1 - q\, z_1\, w_1^{-1}\,     y^{-1} \right)
\left( 1 - q\, z_1\, w_2\,          y^{-1} \right)
\left( 1 - q\, z_1\, w_1 w_2^{-1}\, y^{-1} \right) 
\notag\\[5pt] 
&\hspace{-105pt} \times
\left( 1 - q\, z_2^{-1}\, w_1^{-1}\,     y^{-1} \right)
\left( 1 - q\, z_2^{-1}\, w_2\,          y^{-1} \right)
\left( 1 - q\, z_2^{-1}\, w_1 w_2^{-1}\, y^{-1} \right) 
\notag\\[5pt] 
&\hspace{-105pt} \times
\left( 1 - q\, z_1^{-1} z_2\, w_1^{-1}\,     y^{-1} \right)
\left( 1 - q\, z_1^{-1} z_2\, w_2\,          y^{-1} \right)
\left( 1 - q\, z_1^{-1} z_2\, w_1 w_2^{-1}\, y^{-1} \right) \,.
\label{eqn:detYu}
\end{align}
As $Y_u^\dagger$ forms the conjugate representation of $Y_u$
\begin{equation}
g_{Y_u^\dagger}^{}(z_1, z_2, w_1, w_2, y) = g_{Y_u}^{}(z_1^{-1}, z_2^{-1}, w_1^{-1}, w_2^{-1}, y^{-1}) \,,
\end{equation}
its determinant factor can be computed from \cref{eqn:detYu} as
\begin{equation}
\textstyle\det_{Y_u^\dagger}(q, z_1, z_2, w_1, w_2, y)
= \textstyle\det_{Y_u}(q, z_1^{-1}, z_2^{-1}, w_1^{-1}, w_2^{-1}, y^{-1}) \,.
\label{eqn:detYudag}
\end{equation}
With these ingredients, we can evaluate \cref{eqn:HS181Explicit} and obtain the result in \cref{eqn:HS181}.

Finally, let us compute the Hilbert series for the $\mathbf{(1,3,\overline 3)}$-covariants. In this case, we have to use the original building blocks --- the Yukawa matrices $Y_u, Y_d, Y_u^\dagger, Y_d^\dagger$. From their representations listed in \cref{eqn:RYuRYd}, we see that it is necessary to integrate over three non-abelian groups, $SU(3)_q \times U(3)_u \times U(3)_d$. Note that it is enough to integrate over two $U(1)$ factors, since the Yukawa matrices are invariant under the baryon number rotations $U(1)_{\text{BN}}$, which is the diagonal component of $G_U$ in \cref{eqn:GU}. The Molien-Weyl integral then reads
\begin{align}
\HS_\mathbf{(1,3,\overline 3)}(q) &= \int \dd\mu_{SU(3)_q}^{} (z_1, z_2)
\int \dd\mu_{SU(3)_u}^{} (w_1, w_2)
\int \dd\mu_{SU(3)_d}^{} (v_1, v_2)
\notag\\[5pt]
&\hspace{0pt} \times
\int \dd\mu_{U(1)_u}^{}(y)
\int \dd\mu_{U(1)_d}^{}(x)\; \chi_{\mathbf{(1,3,\overline 3)}_{+1,-1}}^*(w_1, w_2, v_1, v_2, y, x)
\notag\\[5pt]
&\hspace{0pt} \times
\frac{1}{\textstyle\det_{Y_u} (q, z_1, z_2, w_1, w_2, y)} \,
\frac{1}{\textstyle\det_{Y_u^\dagger} (q, z_1, z_2, w_1, w_2, y)}
\notag\\[5pt]
&\hspace{0pt} \times
\frac{1}{\textstyle\det_{Y_d} (q, z_1, z_2, v_1, v_2, x)} \,
\frac{1}{\textstyle\det_{Y_d^\dagger} (q, z_1, z_2, v_1, v_2, x)} \,,
\label{eqn:HS133barExplicit}
\end{align}
where the notation $\mathbf{(1,3,\overline 3)}_{+1,-1}$ indicates that the $U(1)_u\times U(1)_d$ charges are $(+1,-1)$ and thus the full character reads
\begin{align}
\chi_{\mathbf{(1,3,\overline 3)}_{+1,-1}} (w_1, w_2, v_1, v_2, y, x)
&= \chi_\mathbf{3}^{SU(3)}(w_1, w_2)\, \chi_\mathbf{\bar 3}^{SU(3)}(v_1, v_2)\, \chi_{+1}^{U(1)}(y) \, \chi_{-1}^{U(1)}(x)
\notag\\[5pt] 
&\hspace{-50pt}
= \left( w_1 + w_2^{-1} + w_1^{-1} w_2 \right) 
\left( v_1^{-1} + v_2 + v_1 v_2^{-1} \right) y\, x^{-1} \,.
\end{align}
The determinant factors from $Y_d$ and $Y_d^\dagger$ can be obtained from \cref{eqn:detYu,eqn:detYudag} by a proper substitution of variables:
\begin{subequations}
\begin{align}
\textstyle\det_{Y_d}(q, z_1, z_2, v_1, v_2, x)
&= \textstyle\det_{Y_u}(q, z_1, z_2, w_1=v_1, w_2=v_2, y=x) \,, \\[5pt]
\textstyle\det_{Y_d^\dagger}(q, z_1, z_2, v_1, v_2, x)
&= \textstyle\det_{Y_u^\dagger}(q, z_1, z_2, w_1=v_1, w_2=v_2, y=x) \,.
\end{align}
\end{subequations}
Evaluating \cref{eqn:HS133barExplicit} gives us the result in \cref{eqn:HS133bar}.

\begin{table}[t]
\renewcommand{\arraystretch}{1.5}
\setlength{\arrayrulewidth}{.2mm}
\setlength{\tabcolsep}{0.8em}
\centering\footnotesize
\begin{tabular}{ccccc}
\toprule
$U(1)$            & $(+1, -1)$
                  & $2 \times (+1, -1)$
                  & $3 \times (+1, -1)$
                  & $4 \times (+1, -1)$ \\
\midrule
\vspace{3mm}
$H$               & $\Ge$ & $\Ge$ & $\Ge$ & $\Ge$ \\
\vspace{5mm}
$\dfrac{1}{\Den}$ & $\dfrac{1}{1-q^2}$
                  & $\dfrac{1}{(1-q^2)^3}$
                  & $\dfrac{1}{(1-q^2)^5}$
                  & $\dfrac{1}{(1-q^2)^7}$ \\
\vspace{5mm}
$\Num_\I$       & $1$
                & $\begin{tiny} \mqty[ 1 + q^2 ] \to 2 \end{tiny}$
                & $\begin{tiny} \mqty[ 1 + 4q^2 +q^4 ] \to 6 \end{tiny}$
                & $\begin{tiny} \mqty[ 1 +9q^2 +9q^4 +q^6 ] \to 20 \end{tiny}$\\
\vspace{5mm}
$\Num_{\pm 1}$  & $\begin{tiny} \mqty{ q } \to \Sat{1} \end{tiny}$
                & $\begin{tiny} \mqty{ 2q } \to \Sat{2} \end{tiny}$
                & $\begin{tiny} \mqty{ 3q } \mqty[ 1 +q^2 ] \to \Sat{6} \end{tiny}$
                & $\begin{tiny} \mqty{ 4q } \mqty[ 1 +3q^2 +q^4 ] \to \Sat{20} \end{tiny}$ \\
\vspace{5mm}
$\Num_{\pm 2}$  & $\begin{tiny} \mqty{ q^2 } \to \Sat{1} \end{tiny}$
                & $\begin{tiny} \mqty{ q^2 }
                \mqty[ 3 \\ -q^2 ] \to \Sat{2} \end{tiny}$
                & $\begin{tiny} \mqty{ 6q^2 } \to \Sat{6} \end{tiny}$
                & $\begin{tiny} \mqty{ 10q^2 } \mqty[ 1 \\ +q^2 ] \to \Sat{20} \end{tiny}$ \\
\vspace{5mm}
$\Num_{\pm 3}$  & $\begin{tiny} \mqty{ q^3 } \to \Sat{1} \end{tiny}$
                & $\begin{tiny} \mqty{ 2q^3 }
                \mqty[ 2 \\ -q^2 ] \to \Sat{2} \end{tiny}$
                & $\begin{tiny} \mqty{ q^3 }
                \mqty[ 10 \\ -5q^2 \\ +q^4 ] \to \Sat{6} \end{tiny}$
                & $\begin{tiny} \mqty[ 20q^3 ] \to \Sat{20} \end{tiny}$ \\
\vspace{5mm}
$\Num_{\pm 4}$  & $\begin{tiny} \mqty{ q^4 } \to \Sat{1} \end{tiny}$
                & $\begin{tiny} \mqty{ q^4 }
                \mqty[ 5 \\ -3q^2 ] \to \Sat{2} \end{tiny}$
                & $\begin{tiny} \mqty{ 3q^4 }
                \mqty[ 5 \\ -4q^2 \\ +q^4 ] \to \Sat{6} \end{tiny}$
                & $\begin{tiny} \mqty{ q^4 }
                \mqty[ 35 \\ -21q^2 \\ +7q^4 \\ -q^6 ] \to \Sat{20} \end{tiny}$ \\
\vspace{5mm}
$\Num_{\pm 5}$  & $\begin{tiny} \mqty{ q^5 } \to \Sat{1} \end{tiny}$
                & $\begin{tiny} \mqty{ 2q^5 }
                \mqty[ 3 \\ -2q^2 ] \to \Sat{2} \end{tiny}$
                & $\begin{tiny} \mqty{ 3q^5 }
                \mqty[ 7 \\ -7q^2 \\ +2q^4 ] \to \Sat{6} \end{tiny}$
                & $\begin{tiny} \mqty{ 4q^5 }
                \mqty[ 14 \\ -14q^2 \\ +6q^4 \\ -q^6 ] \to \Sat{20} \end{tiny}$ \\
\vspace{5mm}
$\Num_{\pm 10}$ & $\begin{tiny} \mqty{ q^{10} } \to \Sat{1} \end{tiny}$
                & $\begin{tiny} \mqty{ q^{10} }
                \mqty[ 11 \\ -9q^2 ] \to \Sat{2} \end{tiny}$
                & $\begin{tiny} \mqty{ 6q^{10} }
                \mqty[ 11 \\ -16q^2 \\ +6q^4 ] \to \Sat{6} \end{tiny}$
                & $\begin{tiny} \mqty{ 2q^{10} }
                \mqty[ 143 \\ -273q^2 \\ +182q^4 \\ -42q^6 ] \to \Sat{20} \end{tiny}$ \\
\vspace{5mm}
$\Num_{\pm 15}$ & $\begin{tiny} \mqty{ q^{15} } \to \Sat{1} \end{tiny}$
                & $\begin{tiny} \mqty{ 2q^{15} }
                \mqty[ 8 \\ -7q^2 ] \to \Sat{2} \end{tiny}$
                & $\begin{tiny} \mqty{ q^{15} }
                \mqty[ 136 \\ -221q^2 \\ +91q^4 ] \to \Sat{6} \end{tiny}$
                & $\begin{tiny} \mqty{ 4q^{15} }
                \mqty[ 204 \\ -459q^2 \\ +351q^4 \\ -91q^6 ] \to \Sat{20} \end{tiny}$ \\
\vspace{5mm}
$\Num_{\pm 20}$ & $\begin{tiny} \mqty{ q^{20} } \to \Sat{1} \end{tiny}$
                & $\begin{tiny} \mqty{ q^{20} }
                \mqty[ 21 \\ -19q^2 ] \to \Sat{2} \end{tiny}$
                & $\begin{tiny} \mqty{ 3q^{20} }
                \mqty[ 77 \\ -132q^2 \\ +57q^4 ] \to \Sat{6} \end{tiny}$
                & $\begin{tiny} \mqty{ q^{20} }
                \mqty[ 1771 \\ -4301q^2 \\ +3519q^4 \\ -969q^6 ] \to \Sat{20} \end{tiny}$ \\
\vspace{5mm}
$\Num_{\pm 25}$ & $\begin{tiny} \mqty{ q^{25} } \to \Sat{1} \end{tiny}$
                & $\begin{tiny} \mqty{ 2q^{25} }
                \mqty[ 13 \\ -12q^2 ] \to \Sat{2} \end{tiny}$
                & $\begin{tiny} \mqty{ 3q^{25} }
                \mqty[ 117 \\ -207q^2 \\ +92q^4 ] \to \Sat{6} \end{tiny}$
                & $\begin{tiny} \mqty{ 4q^{25} }
                \mqty[ 819 \\ -2079q^2 \\ +1771q^4 \\ -506q^6 ] \to \Sat{20} \end{tiny}$ \\
\bottomrule
\end{tabular}
\caption{The Hilbert series $\HS_R^{U(1),\, m\times(+1,-1)} = \frac{1}{D} N_R$ for $U(1)$ group covariants of rep $R=\I, \pm 1, \pm 2, \pm 3, \pm 4, \pm 5, \pm 10, \pm 15, \pm 20, \pm 25$, with $m=1,2,3,4$ pairs of $(+1, -1)$ as the building blocks. The arrows denote evaluating the expressions at $q=1$. One can verify that all cases satisfy the theorem in \cref{eqn:TheoremBrion}, where the unbroken subgroup $H$ under a generic vev of the building blocks is listed at the top of the table. When rank saturation occurs, the numbers are colored in \Sat{pink}.}
\label{tab:U1}
\end{table}
\begin{table}[t]
\renewcommand{\arraystretch}{1.6}
\setlength{\arrayrulewidth}{.2mm}
\setlength{\tabcolsep}{0.8em}
\centering\footnotesize
\begin{tabular}{ccccc}
\toprule
$SU(2)$            & $(\mathbf{3})$
                   & $2 \times (\mathbf{3})$
                   & $3 \times (\mathbf{3})$
                   & $4 \times (\mathbf{3})$ \\
\midrule
\vspace{3mm}
$H$                & $U(1)$ & $Z_2$ & $Z_2$ & $Z_2$ \\
\vspace{5mm}
$\dfrac{1}{\Den}$  & $\dfrac{1}{1-q^2}$
                   & $\dfrac{1}{(1-q^2)^3}$
                   & $\dfrac{1}{(1-q^2)^6}$
                   & $\dfrac{1}{(1-q^2)^9}$ \\
\vspace{5mm}
$\Num_\I$          & $1$
                   & $1$
                   & $\begin{tiny} \mqty[ 1 +q^3 ] \to 2 \end{tiny}$
                   & $\begin{tiny} \mqty[ 1 +q^2 \\ +4q^3 \\ +q^4 +q^6 ] \to 8 \end{tiny}$ \\
\vspace{5mm}
$\Num_\mathbf{2}$  & $0$
                   & $0$
                   & $0$
                   & $0$ \\
\vspace{5mm}
$\Num_\mathbf{3}$  & $\begin{tiny} \mqty[ q ] \to 1 \end{tiny}$
                   & $\begin{tiny} \mqty{ q }
                   \mqty[ 2 \\ +q ]
                   \to \Sat{3} \end{tiny}$
                   & $\begin{tiny} \mqty{ 3q }
                   \mqty[ 1 \\ +q ]
                   \to \Sat{2\times 3} \end{tiny}$
                   & $\begin{tiny} \mqty{ 2q }
                   \mqty[ 2 + 3q \\ +2q^2 \\ +3q^3 +2q^4 ]
                   \to \Sat{8\times 3} \end{tiny}$ \\
\vspace{5mm}
$\Num_\mathbf{4}$  & $0$
                   & $0$
                   & $0$
                   & $0$ \\
\vspace{5mm}
$\Num_\mathbf{5}$  & $\begin{tiny} \mqty[ q^2 ] \to 1 \end{tiny}$
                   & $\begin{tiny} \mqty{ q^2 }
                   \mqty[ 3 \\ +2q ]
                   \to \Sat{5} \end{tiny}$
                   & $\begin{tiny} \mqty{ q^2 }
                   \mqty[ 6 \\ +8q \\ -3q^3 \\ -q^4 ]
                   \to \Sat{2\times 5} \end{tiny}$
                   & $\begin{tiny} \mqty{ 10q^2 }
                   \mqty[ 1 +2q +q^2 ]
                   \to \Sat{8\times 5} \end{tiny}$ \\
\vspace{5mm}
$\Num_\mathbf{10}$ & $0$
                   & $0$
                   & $0$
                   & $0$ \\
\vspace{5mm}
$\Num_\mathbf{15}$ & $\begin{tiny} \mqty[ q^7 ] \to 1 \end{tiny}$
                   & $\begin{tiny} \mqty{ q^7 }
                   \mqty[ 8 \\ +7q ]
                   \to \Sat{15} \end{tiny}$
                   & $\begin{tiny} \mqty{ 3q^7 }
                   \mqty[ 12 \\ +21q \\ -16q^3 \\ -7q^4 ]
                   \to \Sat{2\times 15} \end{tiny}$
                   & $\begin{tiny} \mqty{ 5q^7 }
                   \mqty[ 24 +63q \\ +24q^2 -65q^3 \\ -60q^4 +7q^5 \\ +24q^6 +7q^7 ]
                   \to \Sat{8\times 15} \end{tiny}$ \\
\vspace{5mm}
$\Num_\mathbf{20}$ & $0$
                   & $0$
                   & $0$
                   & $0$ \\
\vspace{5mm}
$\Num_\mathbf{25}$ & $\begin{tiny} \mqty[ q^{12} ] \to 1 \end{tiny}$
                   & $\begin{tiny} \mqty{ q^{12} }
                   \mqty[ 13 \\ +12q ]
                   \to \Sat{25} \end{tiny}$
                   & $\begin{tiny} \mqty{ q^{12} }
                   \mqty[ 91 \\ +168q \\ -143q^3 \\ -66q^4 ]
                   \to \Sat{2\times 25} \end{tiny}$
                   & $\begin{tiny} \mqty{ 5q^{12} }
                   \mqty[ 91 +252q \\ +91q^2 -320q^3 \\ -305q^4 +44q^5 \\ +143q^6 +44q^7 ]
                   \to \Sat{8\times 25} \end{tiny}$ \\
\bottomrule
\end{tabular}
\caption{The Hilbert series $\HS_R^{SU(2),\, m\times(\mathbf{3})} = \frac{1}{D} N_R$ for $SU(2)$ group covariants of rep $R=\I, \mathbf{2}, \mathbf{3}, \mathbf{4}, \mathbf{5}, \mathbf{10}, \mathbf{15}, \mathbf{20}, \mathbf{25}$, with $m=1,2,3,4$ adjoint reps as the building blocks. The arrows denote evaluating the expressions at $q=1$. One can verify that all cases satisfy the theorem in \cref{eqn:TheoremBrion}, where the unbroken subgroup $H$ under a generic vev of the building blocks is listed at the top of the table. When rank saturation occurs, the numbers are colored in \Sat{pink}.}
\label{tab:SU2adj}
\end{table}
\begin{table}[t]
\renewcommand{\arraystretch}{1.6}
\setlength{\arrayrulewidth}{.2mm}
\setlength{\tabcolsep}{0.8em}
\centering\footnotesize
\begin{tabular}{ccccc}
\toprule
$SU(3)$            & $(\mathbf{3}+\mathbf{\bar{3}})$
                   & $2 \times (\mathbf{3}+\mathbf{\bar{3}})$
                   & $3 \times (\mathbf{3}+\mathbf{\bar{3}})$
                   & $4 \times (\mathbf{3}+\mathbf{\bar{3}})$ \\
\midrule
\vspace{3mm}
$H$                & $SU(2)$ & $\Ge$ & $\Ge$ & $\Ge$ \\
\vspace{5mm}
$\dfrac{1}{\Den}$  & $\dfrac{1}{1-q^2}$
                   & $\dfrac{1}{(1-q^2)^4}$
                   & $\dfrac{1}{(1-q^2)^9 (1-q^3)}$
                   & $\dfrac{1}{(1-q^2)^{12} (1-q^3)^4}$ \\
\vspace{5mm}
$\Num_\I$          & $1$
                   & $1$
                   & $\begin{tiny} \mqty[ 1 +q^3 ] \to 2 \end{tiny}$
                   & $\begin{tiny} \mqty[ 1 +4q^2 \\ +4q^3 +10q^4 \\ +8q^5 +14q^6 +8q^7 \\
                   +10q^8 +4q^9 \\ +4q^{10} +q^{12} ]
                   \to 68 \end{tiny}$ \\
\vspace{5mm}
$\Num_\mathbf{3}$  & $\begin{tiny} \mqty[ q ] \to 1 \end{tiny}$
                   & $\begin{tiny} \mqty{ q } \mqty[ 2 \\ +q ] \to \Sat{3} \end{tiny}$
                   & $\begin{tiny} \mqty{ 3q }
                   \mqty[ 1 +q ]
                   \to \Sat{2\times 3} \end{tiny}$
                   & $\begin{tiny} \mqty{ q }
                   \mqty[ 4 +6q \\ +16q^2 +23q^3 \\ +36q^4 +34q^5 +36q^6 \\ 
                   +23q^7 +16q^8 \\ +6q^9 +4q^{10} ]
                   \to \Sat{68\times 3} \end{tiny}$ \\
\vspace{5mm}
$\Num_\mathbf{6}$  & $\begin{tiny} \mqty[ q^2 ] \to 1 \end{tiny}$
                   & $\begin{tiny} \mqty{ q^2 }
                   \mqty[ 3 \\ +2q \\ +q^2 ]
                   \to \Sat{6} \end{tiny}$
                   & $\begin{tiny} \mqty{ 3q^2 }
                   \mqty[ 2 +3q \\ +2q^2 -q^3 \\ -3q^4 +q^5 ]
                   \to \Sat{2\times 6} \end{tiny}$
                   & $\begin{tiny} \mqty{ 2q^2 }
                   \mqty[ 5 + 12q \\ +30q^2 +34q^3 \\ +42q^4 \\ +34q^5 +30q^6 \\ +12q^7 +5q^8 ]
                   \to \Sat{68\times 6} \end{tiny}$ \\
\vspace{5mm}
$\Num_\mathbf{8}$  & $\begin{tiny} \mqty[ q^2 ] \to 1 \end{tiny}$
                   & $\begin{tiny} \mqty{ 4q^2 } \mqty[ 1 \\ +q ]
                   \to \Sat{8} \end{tiny}$
                   & $\begin{tiny} \mqty{ q^2 }
                   \mqty[ 9 +16q \\ -9q^3 -q^4 \\ +q^7 ]
                   \to \Sat{2\times 8} \end{tiny}$
                   & $\begin{tiny} \mqty{ 8q^2 }
                   \mqty[ 2 + 5q \\ +8q^2 +12q^3 \\ +14q^4 \\ +12q^5 +8q^6 \\ +5q^7 +2q^8 ]
                   \to \Sat{68\times 8} \end{tiny}$ \\
\vspace{5mm}
$\Num_\mathbf{10}$ & $\begin{tiny} \mqty[ q^3 ] \to 1 \end{tiny}$
                   & $\begin{tiny} \mqty{ q^3 }
                   \mqty[ 4 \\ +3q \\ +2q^2 \\ +q^3 ]
                   \to \Sat{10} \end{tiny}$
                   & $\begin{tiny} \mqty{ q^3 }
                   \mqty[ 10 +18q \\ +18q^2 +2q^3 \\ -27q^4 -18q^5 \\ +9q^6 +9q^7 \\ -q^9 ]
                   \to \Sat{2\times 10} \end{tiny}$
                   & $\begin{tiny} \mqty{ 20q^3 }
                   \mqty[ 1 +3q \\ +8q^2 \\ +10q^3 \\ +8q^4 \\ +3q^5 +q^6 ]
                   \to \Sat{68\times 10} \end{tiny}$ \\
\vspace{5mm}
$\Num_\mathbf{27}$ & $\begin{tiny} \mqty[ q^4 ] \to 1 \end{tiny}$
                   & $\begin{tiny} \mqty{ 3q^4 }
                   \mqty[ 3 \\ +4q \\ + 2q^2 ]
                   \to \Sat{27} \end{tiny}$
                   & $\begin{tiny} \mqty{ 9q^4}
                   \mqty[ 4 +10q \\ +6q^2 -8q^3 \\ -11 q^4 +4q^6 \\ +q^7 ]
                   \to \Sat{2\times 27} \end{tiny}$
                   & $\begin{tiny} \mqty{ q^4 }
                   \mqty[ 100 + 360q \\ +652q^2 +608q^3 \\ +343q^4 -16q^5 \\ -192q^6 -148q^7 \\
                   +24q^8 +104q^9 \\ +42q^{10} -16q^{11} \\ -26q^{12} -4q^{13} \\ +4q^{14} +q^{16} ]
                   \to \Sat{68\times 27} \end{tiny}$ \\
\bottomrule
\end{tabular}
\caption{The Hilbert series $\HS_R^{SU(3),\, m\times(\mathbf{3}+\mathbf{\bar 3})} = \frac{1}{D} N_R$ for $SU(3)$ group covariants of rep $R=\I, \mathbf{3}, \mathbf{6}, \mathbf{8}, \mathbf{10}, \mathbf{27}$, with $m=1,2,3, 4$ pairs of $\mathbf{3}+\mathbf{\bar 3}$ reps as the building blocks. The arrows denote evaluating the expressions at $q=1$. One can verify that all cases satisfy the theorem in \cref{eqn:TheoremBrion}, where the unbroken subgroup $H$ under a generic vev of the building blocks is listed at the top of the table. When rank saturation occurs, the numbers are colored in \Sat{pink}.}
\label{tab:SU3ffbar}
\end{table}

\begin{table}[t]
\renewcommand{\arraystretch}{1.6}
\setlength{\arrayrulewidth}{.2mm}
\setlength{\tabcolsep}{0.8em}
\centering\footnotesize
\begin{tabular}{cccc}
\toprule
$SU(3)$            & $(\mathbf{8})$ & $2 \times (\mathbf{8})$ & $3 \times (\mathbf{8})$ \\
\midrule
\vspace{3mm}
$H$                & $U(1)^2$ & $Z_3$ & $Z_3$ \\
\vspace{5mm}
$\dfrac{1}{\Den}$  & $\dfrac{1}{(1-q^2) (1-q^3)}$
                   & $\dfrac{1}{(1-q^2)^3 (1-q^3)^4 (1-q^4)}$
                   & $\dfrac{1}{(1-q^2)^6 (1-q^3)^7 (1-q^4)^3}$ \\
\vspace{3mm}
$\Num_\I$          & $1$ 
                   & $\begin{tiny} \mqty[ 1 +q^6 ] \to 2 \end{tiny}$ 
                   & $\begin{tiny} \mqty[ 1 +4q^3 + 6q^4 + 9q^5 \\ 
                   +20q^6 +21q^7 + 27q^8 \\
                   + 41q^9 +39q^{10} + 39q^{11} \\
                   + 41q^{12} + 27q^{13} + 21q^{14} \\ 
                   +20q^{15} + 9q^{16} + 6q^{17} \\
                   +4q^{18} + q^{21} ] \to 336 \end{tiny}$ \\
\vspace{2mm}
$\Num_\mathbf{3}$  & $0$ & $0$ & $0$ \\
\vspace{3mm}
$\Num_\mathbf{6}$  & $0$ & $0$ & $0$ \\
\vspace{5mm}
$\Num_\mathbf{8}$  & $\begin{tiny} \mqty{ q }
                   \mqty[ 1 \\ +q ]
                   \to 2 \end{tiny}$
                   & $\begin{tiny} \mqty{ 2q }
                   \mqty[ 1 +2q \\ +2q^2 \\ +2q^3 +q^4 ]
                   \to \Sat{2\times 8} \end{tiny}$
                   & $\begin{tiny} \mqty{ q }
                   \mqty[ 3 +9q +17q^2 \\ +39q^3 +75q^4 \\ +121q^5 +189q^6 \\ +255q^7 +300q^8 \\
                   +336q^9 +336q^{10} \\ +300q^{11} +255q^{12} \\ +189q^{13} +121q^{14} \\
                   +75q^{15} +39q^{16} \\ +17q^{17} +9q^{18} +3q^{19} ]
                   \to \Sat{336\times 8} \end{tiny}$ \\
\vspace{5mm}
$\Num_\mathbf{10}$ & $\begin{tiny} \mqty[ q^3 ] \to 1 \end{tiny}$
                   & $\begin{tiny} \mqty{ q^2 }
                   \mqty[ 1 +6q \\ +7q^2 +8q^3 \\ +4q^4 -3q^6 \\ -2q^7 -q^8 ]
                   \to \Sat{2\times 10} \end{tiny}$
                   & $\begin{tiny} \mqty{ q^2 }
                   \mqty[ 3 +19q \\ +42q^2 +84q^3 \\ +149q^4 +225q^5 \\ +312q^6 +402q^7 \\ +444q^8 +444q^9 \\
                   +402q^{10} +312q^{11} \\ +225q^{12} +149q^{13} \\ +84q^{14} +42q^{15} \\
                   +19q^{16} +3q^{17} ]
                   \to \Sat{336\times 10} \end{tiny}$ \\
\vspace{5mm}
$\Num_\mathbf{27}$ & $\begin{tiny} \mqty{ q^2 }
                   \mqty[ 1 \\ +q \\ +q^2 ] \to 3 \end{tiny}$
                   & $\begin{tiny} \mqty{ q^2 }
                   \mqty[ 3 +8q \\ +17q^2 +20q^3 \\ +19q^4 +8q^5 \\ -q^6 -8q^7 \\ -7q^8 -4q^9 \\ -q^{10} ]
                   \to \Sat{2\times 27} \end{tiny}$
                   & $\begin{tiny} \mqty{ 3q^2 }
                   \mqty[ 2 +9q \\ +29q^2 +67q^3 \\ +129q^4 +208q^5 \\ +294q^6 +366q^7 \\ +408q^8 +408q^9 \\ 
                   +366q^{10} +294q^{11} \\ +208q^{12} +129q^{13} \\ +67q^{14} +29q^{15} \\
                   +9q^{16} +2q^{17} ]
                   \to \Sat{336\times 27} \end{tiny}$ \\
\bottomrule
\end{tabular}
\caption{The Hilbert series $\HS_R^{SU(3),\, m\times(\mathbf{8})} = \frac{1}{D} N_R$ for $SU(3)$ group covariants of rep $R=\I, \mathbf{3}, \mathbf{6}, \mathbf{8}, \mathbf{10}, \mathbf{27}$, with $m=1,2,3$ adjoint reps as the building blocks. The arrows denote evaluating the expressions at $q=1$. One can verify that all cases satisfy the theorem in \cref{eqn:TheoremBrion}, where the unbroken subgroup $H$ under a generic vev of the building blocks is listed at the top of the table. When rank saturation occurs, the numbers are colored in \Sat{pink}.}
\label{tab:SU3adj}
\end{table}

\section{Cayley-Hamilton theorem}
\label{appsec:CayleyHamilton}

For any $n\times n$ square matrix $X$, one defines a characteristic polynomial as usual:
\begin{equation}
p_X (\lambda) \equiv \det \left( \lambda I_n - X \right) 
= \lambda^n + c_{n-1} \lambda^{n-1} + c_{n-2} \lambda^{n-2} + \cdots + c_1 \lambda + c_0 \,,
\label{eqn:pMdef}
\end{equation}
where $I_n$ is the $n\times n$ identity matrix and $\lambda$ is an ordinary number. The coefficients $c_k$ in this polynomial depend on the matrix $X$:
\begin{equation}
c_{n-k} = \frac{1}{k!}\, B_k (x_1, \cdots, x_k) \,,
\qquad\text{with}\qquad
x_r \equiv - \big[(r-1)!\big] \tr \big( X^r \big) \,,
\label{eqn:cinBk}
\end{equation}
where $B_k(x_1, \cdots, x_k)$ are the complete exponential Bell polynomials, defined as the coefficients in the following expansion
\begin{equation}
\exp \left( \sum_{r=1}^\infty x_r\, \frac{1}{r!}\, \alpha^r \right)
\equiv \sum_{k=0}^\infty B_k(x_1, \cdots, x_k)\, \frac{1}{k!}\, \alpha^k \,.
\label{eqn:Bkdef}
\end{equation}
To derive the relation in \cref{eqn:cinBk}, one can first consider the following expansion
\begin{align}
\det \left( I_n - \alpha X \right) &= \exp \Big[ \tr \log \left( I_n - \alpha X \right) \Big]
= \exp \left[ - \sum_{r=1}^\infty \frac{1}{r}\, \alpha^r \tr \big( X^r \big) \right]
\notag\\[5pt]
&= \sum_{k=0}^n B_k(x_1, \cdots, x_k)\, \frac{1}{k!}\, \alpha^k \,,
\label{eqn:detIalphaM}
\end{align}
where the $x_r$ in the second line are as defined in \cref{eqn:cinBk}, and we have used the definition of the complete exponential Bell polynomials in \cref{eqn:Bkdef}. Note that the sum truncates at $n$ here, because the determinant we are evaluating is a polynomial in $\alpha$ of at most degree $n$. With \cref{eqn:detIalphaM}, it is straightforward to compute the characteristic polynomial:
\begin{equation}
p_X (\lambda) = \lambda^n \det \left( I_n - \lambda^{-1} X \right)
= \sum_{k=0}^n B_k(x_1, \cdots, x_k)\, \frac{1}{k!}\, \lambda^{n-k} \,,
\end{equation}
which gives us \cref{eqn:cinBk}.

The Cayley-Hamilton theorem states that if one promotes the right-hand side of \cref{eqn:pMdef} to a matrix polynomial and plugs in $\lambda = X$, then it vanishes:
\begin{equation}
p_X (X) = X^n + c_{n-1}\, X^{n-1} + c_{n-2}\, X^{n-2} + \cdots + c_1\, X + c_0\, I_n = 0 \,,
\label{eqn:CHapp}
\end{equation}
which is of course anticipated from the left-hand side of \cref{eqn:pMdef}. Using the identity in \cref{eqn:CHapp}, one can rewrite $X^n$ into a polynomial in $X$ of lower degrees. Using it repeatedly, one can reduce any polynomial in $X$ to at most degree $n-1$.

For the purpose of this paper, we consider the case $n=3$, in which the characteristic polynomial reads
\begin{align}
p_X (\lambda) &\equiv \det \left( \lambda I_3 - X \right) \notag\\[5pt]
&= \lambda^3 - \left(\tr_X\right) \lambda^2 - \frac12 \left( \tr_{X^2} - \tr_X^2 \right) \lambda
- \frac16 \left( 2\tr_{X^3} - 3\tr_{X^2} \tr_X + \tr_X^3 \right) \,,
\end{align}
where we can also recognize the last piece as the determinant of the matrix:
\begin{equation}
\det(X) = \frac16 \left( 2\tr_{X^3} - 3\tr_{X^2} \tr_X + \tr_X^3 \right) \,.
\end{equation}
The Cayley-Hamilton identity in \cref{eqn:CHapp} then reads
\begin{equation}
X^3 = \left(\tr_X\right) X^2 + \frac12 \left( \tr_{X^2} - \tr_X^2 \right) X
+ \frac16 \left( 2\tr_{X^3} - 3\tr_{X^2} \tr_X + \tr_X^3 \right) I_3 \,,
\label{eqn:CH3app}
\end{equation}
with $I_3$ the $3\times 3$ identity matrix. This identity leads to many redundancy relations among polynomials of $3\times3$ matrices. To obtain their most general form, we can take $X$ to be $X = x\, A + y\, B + z\, C$,
and focus on the term proportional to $xyz$. This leads us to the following identity
\begin{align}
&\hspace{-10pt}
ABC + ACB + BAC + BCA + CAB + CBA
\notag\\[8pt]
&= \tr_C \Big( AB + BA \Big) + \tr_B \Big( AC + CA \Big) + \tr_A \Big( BC + CB \Big)
\notag\\[5pt]
&\quad
+ \Big( \tr_{BC} - \tr_B \tr_C \Big)\, A
+ \Big( \tr_{AC} - \tr_A \tr_C \Big)\, B
+ \Big( \tr_{AB} - \tr_A \tr_B \Big)\, C
\notag\\[5pt]
&\quad
+ \Big( \tr_{ABC} + \tr_{ACB} \Big)
- \Big(  \tr_{AB} \tr_C + \tr_{AC} \tr_B + \tr_{BC} \tr_A \Big)
+ \tr_A \tr_B \tr_C \,,
\label{eqn:CH3ABCapp}
\end{align}
which holds for any three $3\times 3$ matrices $A$, $B$, $C$. One can verify that taking the special case $A = B = C = X$ will reproduce the single matrix identity in \cref{eqn:CH3app}.

Let us show some explicit examples of using the identity in \cref{eqn:CH3ABCapp} to obtain linear redundancy relations among the matrix monomials in $h_u, h_d$. For example, taking $A=B=C=h_u$ gives
\begin{equation}
h_u^3 = \Big( \tr_{h_u} \Big)\, h_u^2 + \frac12\, \Big( \tr_{h_u^2} - \tr_{h_u}^2 \Big)\, h_u
+ \frac16\, \Big( 2 \tr_{h_u^3} - 3 \tr_{h_u^2} \tr_{h_u} + \tr_{h_u}^3 \Big) \,,
\label{eqn:hu3Reduce}
\end{equation}
which rewrites $h_u^3$ into lower order monomials $h_u^2$, $h_u$, and $I_3$. Taking $A=h_u$ and $B=C=h_d$ gives
\begin{align}
h_u h_d^2 + h_d h_u h_d + h_d^2 h_u &= \Big( \tr_{h_d} \Big)\, \big( h_u h_d + h_d h_u \big)
+ \Big( \tr_{h_u} \Big)\, h_d^2
\notag\\[5pt]
&\quad
+ \frac12\, \Big( \tr_{h_d^2} - \tr_{h_d}^2 \Big)\, h_u
+ \Big( \tr_{h_u h_d} - \tr_{h_u} \tr_{h_d} \Big)\, h_d
\notag\\[5pt]
&\quad
+ \Big( \tr_{h_u h_d^2} - \tfrac12 \tr_{h_d^2} \tr_{h_u}
- \tr_{h_u h_d} \tr_{h_d} + \tfrac12 \tr_{h_u} \tr_{h_d}^2 \Big) \,,
\label{eqn:huhdhdReduce}
\end{align}
which allows us to rewrite one linear combination of the $\O(q^6)$ monomials into that of $\O(q^4)$ and below. Similarly, taking $A=h_u^2$ and $B=C=h_d$ gives
\begin{align}
h_u^2 h_d^2 + h_d h_u^2 h_d + h_d^2 h_u^2 &= \Big( \tr_{h_d} \Big)\, \big( h_u^2 h_d + h_d h_u^2 \big)
+ \Big( \tr_{h_u^2} \Big)\, h_d^2
\notag\\[5pt]
&\quad
+ \frac12\, \Big( \tr_{h_d^2} - \tr_{h_d}^2 \Big)\, h_u^2
+ \Big( \tr_{h_u^2 h_d} - \tr_{h_u^2} \tr_{h_d} \Big)\, h_d
\notag\\[5pt]
&\quad
+ \Big( \tr_{h_u^2 h_d^2} - \tfrac12 \tr_{h_u^2} \tr_{h_d^2}
- \tr_{h_u^2 h_d} \tr_{h_d} + \tfrac12 \tr_{h_u^2} \tr_{h_d}^2 \Big) \,,
\label{eqn:hu2hdhdReduce}
\end{align}
and taking $A=h_u^2$, $B=h_u$, $C=h_d$ would lead to
\begin{align}
2 h_u^3 h_d + h_u^2 h_d h_u + h_u h_d h_u^2 + 2 h_d h_u^3 &= 2\, \Big( \tr_{h_d} \Big)\, h_u^3
+ \Big( \tr_{h_u} \Big)\, \big( h_u^2 h_d + h_d h_u^2 \big)
\notag\\[5pt]
&\hspace{-130pt}
+ \Big( \tr_{h_u^2} \Big)\, \big( h_u h_d + h_d h_u \big)
+ \Big( \tr_{h_u h_d} - \tr_{h_u} \tr_{h_d} \Big)\, h_u^2
\notag\\[5pt]
&\hspace{-130pt}
+ \Big( \tr_{h_u^2 h_d} - \tr_{h_u^2} \tr_{h_d} \Big)\, h_u
+ \Big( \tr_{h_u^3} - \tr_{h_u^2} \tr_{h_u} \Big)\, h_d
\notag\\[5pt]
&\hspace{-130pt}
+ \Big( 2 \tr_{h_u^3 h_d} - \tr_{h_u^3} \tr_{h_d} - \tr_{h_u^2 h_d} \tr_{h_u}
- \tr_{h_u^2} \tr_{h_u h_d} + \tr_{h_u^2} \tr_{h_u} \tr_{h_d} \Big) \,.
\label{eqn:hu2huhdReduce}
\end{align}

\bibliographystyle{JHEP}
\bibliography{ref}

\providecommand{\href}[2]{#2}\begingroup\raggedright\begin{thebibliography}{10}

\bibitem{Buchmuller:1985jz}
W.~Buchmuller and D.~Wyler, \emph{{Effective Lagrangian Analysis of New
  Interactions and Flavor Conservation}},
  \href{https://doi.org/10.1016/0550-3213(86)90262-2}{\emph{Nucl.Phys.}
  {\bfseries B268} (1986) 621}.

\bibitem{Grzadkowski:2010es}
B.~Grzadkowski, M.~Iskrzynski, M.~Misiak and J.~Rosiek, \emph{{Dimension-Six
  Terms in the Standard Model Lagrangian}},
  \href{https://doi.org/10.1007/JHEP10(2010)085}{\emph{JHEP} {\bfseries 1010}
  (2010) 085} [\href{https://arxiv.org/abs/1008.4884}{{\ttfamily 1008.4884}}].

\bibitem{DAmbrosio:2002ex}
G.~D'Ambrosio, G.~Giudice, G.~Isidori and A.~Strumia, \emph{{Minimal flavor
  violation: An Effective field theory approach}},
  \href{https://doi.org/10.1016/S0550-3213(02)00836-2}{\emph{Nucl.Phys.}
  {\bfseries B645} (2002) 155}
  [\href{https://arxiv.org/abs/hep-ph/0207036}{{\ttfamily hep-ph/0207036}}].

\bibitem{Cirigliano:2005ck}
V.~Cirigliano, B.~Grinstein, G.~Isidori and M.B.~Wise, \emph{{Minimal flavor
  violation in the lepton sector}},
  \href{https://doi.org/10.1016/j.nuclphysb.2005.08.037}{\emph{Nucl.Phys.}
  {\bfseries B728} (2005) 121}
  [\href{https://arxiv.org/abs/hep-ph/0507001}{{\ttfamily hep-ph/0507001}}].

\bibitem{Sturmfels:inv}
B.~Sturmfels, \emph{Algorithms in Invariant Theory}, Texts \& Monographs in
  Symbolic Computation, Springer Vienna (1993).

\bibitem{bruns1998cohen}
W.~Bruns and H.J.~Herzog, \emph{Cohen-Macaulay Rings}, no.~39 in Cambridge
  Studies in Advanced Mathematics, Cambridge University Press, 2~ed. (1998),
  \href{https://doi.org/10.1017/CBO9780511608681}{10.1017/CBO9780511608681}.

\bibitem{Popov1994}
V.L.~Popov and E.B.~Vinberg, \emph{Invariant theory},  in \emph{Algebraic
  Geometry IV: Linear Algebraic Groups Invariant Theory}, A.N.~Parshin and
  I.R.~Shafarevich, eds., (Berlin, Heidelberg), pp.~123--278, Springer Berlin
  Heidelberg (1994), \href{https://doi.org/10.1007/978-3-662-03073-8_2}{DOI}.

\bibitem{Benvenuti:2006qr}
S.~Benvenuti, B.~Feng, A.~Hanany and Y.-H.~He, \emph{{Counting BPS Operators in
  Gauge Theories: Quivers, Syzygies and Plethystics}},
  \href{https://doi.org/10.1088/1126-6708/2007/11/050}{\emph{JHEP} {\bfseries
  11} (2007) 050} [\href{https://arxiv.org/abs/hep-th/0608050}{{\ttfamily
  hep-th/0608050}}].

\bibitem{Feng:2007ur}
B.~Feng, A.~Hanany and Y.-H.~He, \emph{{Counting gauge invariants: The
  Plethystic program}},
  \href{https://doi.org/10.1088/1126-6708/2007/03/090}{\emph{JHEP} {\bfseries
  03} (2007) 090} [\href{https://arxiv.org/abs/hep-th/0701063}{{\ttfamily
  hep-th/0701063}}].

\bibitem{Gray:2008yu}
J.~Gray, A.~Hanany, Y.-H.~He, V.~Jejjala and N.~Mekareeya, \emph{{SQCD: A
  Geometric Apercu}},
  \href{https://doi.org/10.1088/1126-6708/2008/05/099}{\emph{JHEP} {\bfseries
  05} (2008) 099} [\href{https://arxiv.org/abs/0803.4257}{{\ttfamily
  0803.4257}}].

\bibitem{Jenkins:2009dy}
E.E.~Jenkins and A.V.~Manohar, \emph{{Algebraic Structure of Lepton and Quark
  Flavor Invariants and CP Violation}},
  \href{https://doi.org/10.1088/1126-6708/2009/10/094}{\emph{JHEP} {\bfseries
  10} (2009) 094} [\href{https://arxiv.org/abs/0907.4763}{{\ttfamily
  0907.4763}}].

\bibitem{Hanany:2010vu}
A.~Hanany, E.E.~Jenkins, A.V.~Manohar and G.~Torri, \emph{{Hilbert Series for
  Flavor Invariants of the Standard Model}},
  \href{https://doi.org/10.1007/JHEP03(2011)096}{\emph{JHEP} {\bfseries 03}
  (2011) 096} [\href{https://arxiv.org/abs/1010.3161}{{\ttfamily 1010.3161}}].

\bibitem{Lehman:2015via}
L.~Lehman and A.~Martin, \emph{{Hilbert Series for Constructing Lagrangians:
  expanding the phenomenologist's toolbox}},
  \href{https://doi.org/10.1103/PhysRevD.91.105014}{\emph{Phys. Rev. D}
  {\bfseries 91} (2015) 105014}
  [\href{https://arxiv.org/abs/1503.07537}{{\ttfamily 1503.07537}}].

\bibitem{Henning:2015daa}
B.~Henning, X.~Lu, T.~Melia and H.~Murayama, \emph{{Hilbert series and operator
  bases with derivatives in effective field theories}},
  \href{https://doi.org/10.1007/s00220-015-2518-2}{\emph{Commun. Math. Phys.}
  {\bfseries 347} (2016) 363}
  [\href{https://arxiv.org/abs/1507.07240}{{\ttfamily 1507.07240}}].

\bibitem{Lehman:2015coa}
L.~Lehman and A.~Martin, \emph{{Low-derivative operators of the Standard Model
  effective field theory via Hilbert series methods}},
  \href{https://doi.org/10.1007/JHEP02(2016)081}{\emph{JHEP} {\bfseries 02}
  (2016) 081} [\href{https://arxiv.org/abs/1510.00372}{{\ttfamily
  1510.00372}}].

\bibitem{Henning:2015alf}
B.~Henning, X.~Lu, T.~Melia and H.~Murayama, \emph{{2, 84, 30, 993, 560, 15456,
  11962, 261485, ...: Higher dimension operators in the SM EFT}},
  \href{https://doi.org/10.1007/JHEP08(2017)016}{\emph{JHEP} {\bfseries 08}
  (2017) 016} [\href{https://arxiv.org/abs/1512.03433}{{\ttfamily
  1512.03433}}].

\bibitem{Henning:2017fpj}
B.~Henning, X.~Lu, T.~Melia and H.~Murayama, \emph{{Operator bases,
  $S$-matrices, and their partition functions}},
  \href{https://doi.org/10.1007/JHEP10(2017)199}{\emph{JHEP} {\bfseries 10}
  (2017) 199} [\href{https://arxiv.org/abs/1706.08520}{{\ttfamily
  1706.08520}}].

\bibitem{Marinissen:2020jmb}
C.B.~Marinissen, R.~Rahn and W.J.~Waalewijn, \emph{{..., 83106786, 114382724,
  1509048322, 2343463290, 27410087742, ... efficient Hilbert series for
  effective theories}},
  \href{https://doi.org/10.1016/j.physletb.2020.135632}{\emph{Phys. Lett. B}
  {\bfseries 808} (2020) 135632}
  [\href{https://arxiv.org/abs/2004.09521}{{\ttfamily 2004.09521}}].

\bibitem{Bonnefoy:2021tbt}
Q.~Bonnefoy, E.~Gendy, C.~Grojean and J.T.~Ruderman, \emph{{Beyond Jarlskog:
  699 invariants for CP violation in SMEFT}},
  \href{https://doi.org/10.1007/JHEP08(2022)032}{\emph{JHEP} {\bfseries 08}
  (2022) 032} [\href{https://arxiv.org/abs/2112.03889}{{\ttfamily
  2112.03889}}].

\bibitem{Kondo:2022wcw}
D.~Kondo, H.~Murayama and R.~Okabe, \emph{{23, 381, 6242, 103268, 1743183,
  \textellipsis{} : Hilbert series for CP-violating operators in SMEFT}},
  \href{https://doi.org/10.1007/JHEP03(2023)107}{\emph{JHEP} {\bfseries 03}
  (2023) 107} [\href{https://arxiv.org/abs/2212.02413}{{\ttfamily
  2212.02413}}].

\bibitem{Bonnefoy:2023bzx}
Q.~Bonnefoy, E.~Gendy, C.~Grojean and J.T.~Ruderman, \emph{{Opportunistic CP
  violation}}, \href{https://doi.org/10.1007/JHEP06(2023)141}{\emph{JHEP}
  {\bfseries 06} (2023) 141}
  [\href{https://arxiv.org/abs/2302.07288}{{\ttfamily 2302.07288}}].

\bibitem{Ruhdorfer:2019qmk}
M.~Ruhdorfer, J.~Serra and A.~Weiler, \emph{{Effective Field Theory of Gravity
  to All Orders}}, \href{https://doi.org/10.1007/JHEP05(2020)083}{\emph{JHEP}
  {\bfseries 05} (2020) 083}
  [\href{https://arxiv.org/abs/1908.08050}{{\ttfamily 1908.08050}}].

\bibitem{Graf:2020yxt}
L.~Graf, B.~Henning, X.~Lu, T.~Melia and H.~Murayama, \emph{{2, 12, 117, 1959,
  45171, 1170086, \textellipsis{}: a Hilbert series for the QCD chiral
  Lagrangian}}, \href{https://doi.org/10.1007/JHEP01(2021)142}{\emph{JHEP}
  {\bfseries 01} (2021) 142}
  [\href{https://arxiv.org/abs/2009.01239}{{\ttfamily 2009.01239}}].

\bibitem{Bijnens:2022zqo}
J.~Bijnens, S.B.~Gudnason, J.~Yu and T.~Zhang, \emph{{Hilbert series and
  higher-order Lagrangians for the O(N) model}},
  \href{https://doi.org/10.1007/JHEP05(2023)061}{\emph{JHEP} {\bfseries 05}
  (2023) 061} [\href{https://arxiv.org/abs/2212.07901}{{\ttfamily
  2212.07901}}].

\bibitem{Graf:2022rco}
L.~Gr\'af, B.~Henning, X.~Lu, T.~Melia and H.~Murayama, \emph{{Hilbert series,
  the Higgs mechanism, and HEFT}},
  \href{https://doi.org/10.1007/JHEP02(2023)064}{\emph{JHEP} {\bfseries 02}
  (2023) 064} [\href{https://arxiv.org/abs/2211.06275}{{\ttfamily
  2211.06275}}].

\bibitem{Sun:2022aag}
H.~Sun, Y.-N.~Wang and J.-H.~Yu, \emph{{Hilbert Series and Operator Counting on
  the Higgs Effective Field Theory}},
  \href{https://arxiv.org/abs/2211.11598}{{\ttfamily 2211.11598}}.

\bibitem{Kobach:2017xkw}
A.~Kobach and S.~Pal, \emph{{Hilbert Series and Operator Basis for NRQED and
  NRQCD/HQET}},
  \href{https://doi.org/10.1016/j.physletb.2017.06.026}{\emph{Phys. Lett. B}
  {\bfseries 772} (2017) 225}
  [\href{https://arxiv.org/abs/1704.00008}{{\ttfamily 1704.00008}}].

\bibitem{Kobach:2018pie}
A.~Kobach and S.~Pal, \emph{{Reparameterization Invariant Operator Basis for
  NRQED and HQET}}, \href{https://doi.org/10.1007/JHEP11(2019)012}{\emph{JHEP}
  {\bfseries 11} (2019) 012}
  [\href{https://arxiv.org/abs/1810.02356}{{\ttfamily 1810.02356}}].

\bibitem{Delgado:2022bho}
A.~Delgado, A.~Martin and R.~Wang, \emph{{Constructing operator basis in
  supersymmetry: a Hilbert series approach}},
  \href{https://doi.org/10.1007/JHEP04(2023)097}{\emph{JHEP} {\bfseries 04}
  (2023) 097} [\href{https://arxiv.org/abs/2212.02551}{{\ttfamily
  2212.02551}}].

\bibitem{Delgado:2023ivp}
A.~Delgado, A.~Martin and R.~Wang, \emph{{Counting operators in N = 1
  supersymmetric gauge theories}},
  \href{https://doi.org/10.1007/JHEP07(2023)081}{\emph{JHEP} {\bfseries 07}
  (2023) 081} [\href{https://arxiv.org/abs/2305.01736}{{\ttfamily
  2305.01736}}].

\bibitem{Grojean:2023tsd}
C.~Grojean, J.~Kley and C.-Y.~Yao, \emph{{Hilbert series for ALP EFTs}},
  \href{https://arxiv.org/abs/2307.08563}{{\ttfamily 2307.08563}}.

\bibitem{Chang:2022crb}
S.~Chang, M.~Chen, D.~Liu and M.A.~Luty, \emph{{Primary observables for
  indirect searches at colliders}},
  \href{https://doi.org/10.1007/JHEP07(2023)030}{\emph{JHEP} {\bfseries 07}
  (2023) 030} [\href{https://arxiv.org/abs/2212.06215}{{\ttfamily
  2212.06215}}].

\bibitem{brion1993modules}
M.~Brion, \emph{Sur les modules de covariants}, {\emph{Annales scientifiques de
  l'École Normale Supérieure} {\bfseries 26} (1993) 1}.

\bibitem{broer1994hilbert}
B.~Broer, \emph{Hilbert series for modules of covariants}, {\emph{Algebraic
  Groups and Their Generalizations: Classical Methods (University Park, PA,
  1991)} {\bfseries 56} (1994) 321}.

\bibitem{Coleman:1969sm}
S.R.~Coleman, J.~Wess and B.~Zumino, \emph{{Structure of phenomenological
  Lagrangians. 1.}},
  \href{https://doi.org/10.1103/PhysRev.177.2239}{\emph{Phys. Rev.} {\bfseries
  177} (1969) 2239}.

\bibitem{Callan:1969sn}
C.G.~Callan, Jr., S.R.~Coleman, J.~Wess and B.~Zumino, \emph{{Structure of
  phenomenological Lagrangians. 2.}},
  \href{https://doi.org/10.1103/PhysRev.177.2247}{\emph{Phys. Rev.} {\bfseries
  177} (1969) 2247}.

\bibitem{Hochster74}
M.~Hochster and J.L.~Roberts, \emph{Rings of invariants of reductive groups
  acting on regular rings are cohen-macaulay},
  \href{https://doi.org/http://dx.doi.org/10.1016/0001-8708(74)90067-X}{\emph{Advances
  in Mathematics} {\bfseries 13} (1974) 115 }.

\bibitem{Stanley78}
R.P.~Stanley, \emph{Hilbert functions of graded algebras},
  \href{https://doi.org/http://dx.doi.org/10.1016/0001-8708(78)90045-2}{\emph{Advances
  in Mathematics} {\bfseries 28} (1978) 57 }.

\bibitem{Glashow:1970gm}
S.L.~Glashow, J.~Iliopoulos and L.~Maiani, \emph{{Weak Interactions with
  Lepton-Hadron Symmetry}},
  \href{https://doi.org/10.1103/PhysRevD.2.1285}{\emph{Phys. Rev.} {\bfseries
  D2} (1970) 1285}.

\bibitem{Dugan:1984qf}
M.~Dugan, B.~Grinstein and L.J.~Hall, \emph{{CP Violation in the Minimal N=1
  Supergravity Theory}},
  \href{https://doi.org/10.1016/0550-3213(85)90145-2}{\emph{Nucl. Phys.}
  {\bfseries B255} (1985) 413}.

\bibitem{Chivukula:1987py}
R.S.~Chivukula and H.~Georgi, \emph{{Composite Technicolor Standard Model}},
  \href{https://doi.org/10.1016/0370-2693(87)90713-1}{\emph{Phys.Lett.}
  {\bfseries B188} (1987) 99}.

\bibitem{Davidson:2006bd}
S.~Davidson and F.~Palorini, \emph{{Various definitions of Minimal Flavour
  Violation for Leptons}},
  \href{https://doi.org/10.1016/j.physletb.2006.09.016}{\emph{Phys.Lett.}
  {\bfseries B642} (2006) 72}
  [\href{https://arxiv.org/abs/hep-ph/0607329}{{\ttfamily hep-ph/0607329}}].

\bibitem{Gavela:2009cd}
M.~Gavela, T.~Hambye, D.~Hernandez and P.~Hernandez, \emph{{Minimal Flavour
  Seesaw Models}},
  \href{https://doi.org/10.1088/1126-6708/2009/09/038}{\emph{JHEP} {\bfseries
  0909} (2009) 038} [\href{https://arxiv.org/abs/0906.1461}{{\ttfamily
  0906.1461}}].

\bibitem{Alonso:2011jd}
R.~Alonso, G.~Isidori, L.~Merlo, L.A.~Munoz and E.~Nardi, \emph{{Minimal
  flavour violation extensions of the seesaw}},
  \href{https://doi.org/10.1007/JHEP06(2011)037}{\emph{JHEP} {\bfseries 1106}
  (2011) 037} [\href{https://arxiv.org/abs/1103.5461}{{\ttfamily 1103.5461}}].

\bibitem{Alonso:2013hga}
R.~Alonso, E.E.~Jenkins, A.V.~Manohar and M.~Trott, \emph{{Renormalization
  Group Evolution of the Standard Model Dimension Six Operators III: Gauge
  Coupling Dependence and Phenomenology}},
  \href{https://doi.org/10.1007/JHEP04(2014)159}{\emph{JHEP} {\bfseries 04}
  (2014) 159} [\href{https://arxiv.org/abs/1312.2014}{{\ttfamily 1312.2014}}].

\bibitem{Weinberg:1979sa}
S.~Weinberg, \emph{{Baryon and Lepton Nonconserving Processes}},
  \href{https://doi.org/10.1103/PhysRevLett.43.1566}{\emph{Phys. Rev. Lett.}
  {\bfseries 43} (1979) 1566}.

\bibitem{Arnold:2009ay}
J.M.~Arnold, M.~Pospelov, M.~Trott and M.B.~Wise, \emph{{Scalar Representations
  and Minimal Flavor Violation}},
  \href{https://doi.org/10.1007/JHEP01(2010)073}{\emph{JHEP} {\bfseries 01}
  (2010) 073} [\href{https://arxiv.org/abs/0911.2225}{{\ttfamily 0911.2225}}].

\bibitem{Alonso:2014zka}
R.~Alonso, H.-M.~Chang, E.E.~Jenkins, A.V.~Manohar and B.~Shotwell,
  \emph{{Renormalization group evolution of dimension-six baryon number
  violating operators}},
  \href{https://doi.org/10.1016/j.physletb.2014.05.065}{\emph{Phys. Lett. B}
  {\bfseries 734} (2014) 302}
  [\href{https://arxiv.org/abs/1405.0486}{{\ttfamily 1405.0486}}].

\bibitem{Colangelo:2008qp}
G.~Colangelo, E.~Nikolidakis and C.~Smith, \emph{{Supersymmetric models with
  minimal flavour violation and their running}},
  \href{https://doi.org/10.1140/epjc/s10052-008-0796-y}{\emph{Eur. Phys. J. C}
  {\bfseries 59} (2009) 75} [\href{https://arxiv.org/abs/0807.0801}{{\ttfamily
  0807.0801}}].

\bibitem{Mercolli:2009ns}
L.~Mercolli and C.~Smith, \emph{{Edm Constraints on Flavored Cp-Violating
  Phases}}, \href{https://doi.org/10.1016/j.nuclphysb.2009.03.010}{\emph{Nucl.
  Phys.} {\bfseries B817} (2009) 1}
  [\href{https://arxiv.org/abs/0902.1949}{{\ttfamily 0902.1949}}].

\bibitem{teranishi_1986}
Y.~Teranishi, \emph{The ring of invariants of matrices},
  \href{https://doi.org/10.1017/S0027763000022728}{\emph{Nagoya Mathematical
  Journal} {\bfseries 104} (1986) 149–161}.

\bibitem{Jarlskog:1985cw}
C.~Jarlskog, \emph{{A Basis Independent Formulation of the Connection Between
  Quark Mass Matrices, CP Violation and Experiment}},
  \href{https://doi.org/10.1007/BF01565198}{\emph{Z. Phys.} {\bfseries C29}
  (1985) 491}.

\bibitem{Jarlskog:1985ht}
C.~Jarlskog, \emph{{Commutator of the Quark Mass Matrices in the Standard
  Electroweak Model and a Measure of Maximal CP Violation}},
  \href{https://doi.org/10.1103/PhysRevLett.55.1039}{\emph{Phys. Rev. Lett.}
  {\bfseries 55} (1985) 1039}.

\bibitem{Isidori:2012ts}
G.~Isidori and D.M.~Straub, \emph{{Minimal Flavour Violation and Beyond}},
  \href{https://doi.org/10.1140/epjc/s10052-012-2103-1}{\emph{Eur. Phys. J. C}
  {\bfseries 72} (2012) 2103}
  [\href{https://arxiv.org/abs/1202.0464}{{\ttfamily 1202.0464}}].

\bibitem{Bartocci:2023nvp}
R.~Bartocci, A.~Biek\"otter and T.~Hurth, \emph{{A global analysis of the SMEFT
  under the minimal MFV assumption}},
  \href{https://arxiv.org/abs/2311.04963}{{\ttfamily 2311.04963}}.

\bibitem{Hill:1994hp}
C.T.~Hill, \emph{{Topcolor assisted technicolor}},
  \href{https://doi.org/10.1016/0370-2693(94)01660-5}{\emph{Phys. Lett. B}
  {\bfseries 345} (1995) 483}
  [\href{https://arxiv.org/abs/hep-ph/9411426}{{\ttfamily hep-ph/9411426}}].

\bibitem{Grinstein:2010ve}
B.~Grinstein, M.~Redi and G.~Villadoro, \emph{{Low Scale Flavor Gauge
  Symmetries}}, \href{https://doi.org/10.1007/JHEP11(2010)067}{\emph{JHEP}
  {\bfseries 1011} (2010) 067}
  [\href{https://arxiv.org/abs/1009.2049}{{\ttfamily 1009.2049}}].

\bibitem{Feldmann:2010yp}
T.~Feldmann, \emph{{See-Saw Masses for Quarks and Leptons in SU(5)}},
  \href{https://doi.org/10.1007/JHEP04(2011)043}{\emph{JHEP} {\bfseries 1104}
  (2011) 043} [\href{https://arxiv.org/abs/1010.2116}{{\ttfamily 1010.2116}}].

\bibitem{Guadagnoli:2011id}
D.~Guadagnoli, R.N.~Mohapatra and I.~Sung, \emph{{Gauged Flavor Group with
  Left-Right Symmetry}},
  \href{https://doi.org/10.1007/JHEP04(2011)093}{\emph{JHEP} {\bfseries 04}
  (2011) 093} [\href{https://arxiv.org/abs/1103.4170}{{\ttfamily 1103.4170}}].

\bibitem{Buras:2011wi}
A.J.~Buras, M.V.~Carlucci, L.~Merlo and E.~Stamou, \emph{{Phenomenology of a
  Gauged $SU(3)^3$ Flavour Model}},
  \href{https://doi.org/10.1007/JHEP03(2012)088}{\emph{JHEP} {\bfseries 1203}
  (2012) 088} [\href{https://arxiv.org/abs/1112.4477}{{\ttfamily 1112.4477}}].

\bibitem{Alonso:2016onw}
R.~Alonso, E.~Fernandez~Martinez, M.B.~Gavela, B.~Grinstein, L.~Merlo and
  P.~Quilez, \emph{{Gauged Lepton Flavour}},
  \href{https://doi.org/10.1007/JHEP12(2016)119}{\emph{JHEP} {\bfseries 12}
  (2016) 119} [\href{https://arxiv.org/abs/1609.05902}{{\ttfamily
  1609.05902}}].

\bibitem{Paradisi:2009ey}
P.~Paradisi and D.M.~Straub, \emph{{The SUSY CP Problem and the Mfv
  Principle}},
  \href{https://doi.org/10.1016/j.physletb.2009.12.054}{\emph{Phys. Lett.}
  {\bfseries B684} (2010) 147}
  [\href{https://arxiv.org/abs/0906.4551}{{\ttfamily 0906.4551}}].

\bibitem{Bobeth:2007dw}
C.~Bobeth, G.~Hiller and G.~Piranishvili, \emph{{Angular distributions of
  $\bar{B} \to \bar{K} \ell^+\ell^-$ decays}},
  \href{https://doi.org/10.1088/1126-6708/2007/12/040}{\emph{JHEP} {\bfseries
  12} (2007) 040} [\href{https://arxiv.org/abs/0709.4174}{{\ttfamily
  0709.4174}}].

\bibitem{Grinstein:1988me}
B.~Grinstein, M.J.~Savage and M.B.~Wise, \emph{{B ---\ensuremath{>} X(s) e+ e-
  in the Six Quark Model}},
  \href{https://doi.org/10.1016/0550-3213(89)90078-3}{\emph{Nucl. Phys. B}
  {\bfseries 319} (1989) 271}.

\bibitem{Ali:2013zfa}
A.~Ali, A.Y.~Parkhomenko and A.V.~Rusov, \emph{{Precise Calculation of the
  Dilepton Invariant-Mass Spectrum and the Decay Rate in $B^\pm \to \pi^\pm
  \mu^+ \mu^-$ in the SM}},
  \href{https://doi.org/10.1103/PhysRevD.89.094021}{\emph{Phys. Rev. D}
  {\bfseries 89} (2014) 094021}
  [\href{https://arxiv.org/abs/1312.2523}{{\ttfamily 1312.2523}}].

\bibitem{Bause:2021cna}
R.~Bause, H.~Gisbert, M.~Golz and G.~Hiller, \emph{{Interplay of dineutrino
  modes with semileptonic rare B-decays}},
  \href{https://doi.org/10.1007/JHEP12(2021)061}{\emph{JHEP} {\bfseries 12}
  (2021) 061} [\href{https://arxiv.org/abs/2109.01675}{{\ttfamily
  2109.01675}}].

\bibitem{Parrott:2022zte}
{\scshape HPQCD} collaboration, \emph{{Standard Model predictions for
  B\textrightarrow{}K\ensuremath{\ell}+\ensuremath{\ell}-,
  B\textrightarrow{}K\ensuremath{\ell}1-\ensuremath{\ell}2+ and
  B\textrightarrow{}K\ensuremath{\nu}\ensuremath{\nu}\textasciimacron{} using
  form factors from Nf=2+1+1 lattice QCD}},
  \href{https://doi.org/10.1103/PhysRevD.107.014511}{\emph{Phys. Rev. D}
  {\bfseries 107} (2023) 014511}
  [\href{https://arxiv.org/abs/2207.13371}{{\ttfamily 2207.13371}}].

\bibitem{Beneke:2019slt}
M.~Beneke, C.~Bobeth and R.~Szafron, \emph{{Power-enhanced leading-logarithmic
  QED corrections to $B_q \to \mu^+\mu^-$}},
  \href{https://doi.org/10.1007/JHEP10(2019)232}{\emph{JHEP} {\bfseries 10}
  (2019) 232} [\href{https://arxiv.org/abs/1908.07011}{{\ttfamily
  1908.07011}}].

\bibitem{LHCb:2020pcv}
{\scshape LHCb} collaboration, \emph{{Search for the Rare Decays $B^0_s\to
  e^+e^-$ and $B^0\to e^+e^-$}},
  \href{https://doi.org/10.1103/PhysRevLett.124.211802}{\emph{Phys. Rev. Lett.}
  {\bfseries 124} (2020) 211802}
  [\href{https://arxiv.org/abs/2003.03999}{{\ttfamily 2003.03999}}].

\bibitem{Bause:2022rrs}
R.~Bause, H.~Gisbert, M.~Golz and G.~Hiller, \emph{{Model-independent analysis
  of $b \rightarrow d$ processes}},
  \href{https://doi.org/10.1140/epjc/s10052-023-11586-9}{\emph{Eur. Phys. J. C}
  {\bfseries 83} (2023) 419}
  [\href{https://arxiv.org/abs/2209.04457}{{\ttfamily 2209.04457}}].

\bibitem{LHCb:2021awg}
{\scshape LHCb} collaboration, \emph{{Measurement of the $B^0_s\to\mu^+\mu^-$
  decay properties and search for the $B^0\to\mu^+\mu^-$ and
  $B^0_s\to\mu^+\mu^-\gamma$ decays}},
  \href{https://doi.org/10.1103/PhysRevD.105.012010}{\emph{Phys. Rev. D}
  {\bfseries 105} (2022) 012010}
  [\href{https://arxiv.org/abs/2108.09283}{{\ttfamily 2108.09283}}].

\bibitem{LHCb:2021vsc}
{\scshape LHCb} collaboration, \emph{{Analysis of Neutral B-Meson Decays into
  Two Muons}},
  \href{https://doi.org/10.1103/PhysRevLett.128.041801}{\emph{Phys. Rev. Lett.}
  {\bfseries 128} (2022) 041801}
  [\href{https://arxiv.org/abs/2108.09284}{{\ttfamily 2108.09284}}].

\bibitem{Fleischer:2017ltw}
R.~Fleischer, R.~Jaarsma and G.~Tetlalmatzi-Xolocotzi, \emph{{In Pursuit of New
  Physics with $B^0_{s,d}\to\ell^+\ell^-$}},
  \href{https://doi.org/10.1007/JHEP05(2017)156}{\emph{JHEP} {\bfseries 05}
  (2017) 156} [\href{https://arxiv.org/abs/1703.10160}{{\ttfamily
  1703.10160}}].

\bibitem{ParticleDataGroup:2022pth}
{\scshape Particle Data Group} collaboration, \emph{{Review of Particle
  Physics}}, \href{https://doi.org/10.1093/ptep/ptac097}{\emph{PTEP} {\bfseries
  2022} (2022) 083C01}.

\bibitem{LHCb:2015hsa}
{\scshape LHCb} collaboration, \emph{{First measurement of the differential
  branching fraction and $C\!P$ asymmetry of the $B^\pm\to\pi^\pm\mu^+\mu^-$
  decay}}, \href{https://doi.org/10.1007/JHEP10(2015)034}{\emph{JHEP}
  {\bfseries 10} (2015) 034}
  [\href{https://arxiv.org/abs/1509.00414}{{\ttfamily 1509.00414}}].

\bibitem{BELLE:2019xld}
{\scshape BELLE} collaboration, \emph{{Test of lepton flavor universality and
  search for lepton flavor violation in $B \rightarrow K\ell \ell$ decays}},
  \href{https://doi.org/10.1007/JHEP03(2021)105}{\emph{JHEP} {\bfseries 03}
  (2021) 105} [\href{https://arxiv.org/abs/1908.01848}{{\ttfamily
  1908.01848}}].

\bibitem{LHCb:2021trn}
{\scshape LHCb} collaboration, \emph{{Test of lepton universality in
  beauty-quark decays}},
  \href{https://doi.org/10.1038/s41567-023-02095-3}{\emph{Nature Phys.}
  {\bfseries 18} (2022) 277}
  [\href{https://arxiv.org/abs/2103.11769}{{\ttfamily 2103.11769}}].

\bibitem{LHCb:2014cxe}
{\scshape LHCb} collaboration, \emph{{Differential branching fractions and
  isospin asymmetries of $B \to K^{(*)} \mu^+ \mu^-$ decays}},
  \href{https://doi.org/10.1007/JHEP06(2014)133}{\emph{JHEP} {\bfseries 06}
  (2014) 133} [\href{https://arxiv.org/abs/1403.8044}{{\ttfamily 1403.8044}}].

\bibitem{Belle-II:2023esi}
{\scshape Belle-II} collaboration, \emph{{Evidence for $B^{+}\to
  K^{+}\nu\bar{\nu}$ Decays}},
  \href{https://arxiv.org/abs/2311.14647}{{\ttfamily 2311.14647}}.

\bibitem{Hatton:2021syc}
D.~Hatton, C.T.H.~Davies, J.~Koponen, G.P.~Lepage and A.T.~Lytle,
  \emph{{Determination of $\overline{m}_b/\overline{m}_c$ and $\overline{m}_b$
  from $n_f=4$ lattice QCD$+$QED}},
  \href{https://doi.org/10.1103/PhysRevD.103.114508}{\emph{Phys. Rev. D}
  {\bfseries 103} (2021) 114508}
  [\href{https://arxiv.org/abs/2102.09609}{{\ttfamily 2102.09609}}].

\bibitem{Blake:2016olu}
T.~Blake, G.~Lanfranchi and D.M.~Straub, \emph{{Rare $B$ Decays as Tests of the
  Standard Model}},
  \href{https://doi.org/10.1016/j.ppnp.2016.10.001}{\emph{Prog. Part. Nucl.
  Phys.} {\bfseries 92} (2017) 50}
  [\href{https://arxiv.org/abs/1606.00916}{{\ttfamily 1606.00916}}].

\bibitem{brocker2013representations}
T.~Br{\"o}cker and T.~Tom~Dieck, \emph{Representations of compact Lie groups},
  vol.~98, Springer Science \& Business Media (2013).

\end{thebibliography}\endgroup

\end{spacing}

\end{document}